\documentclass[a4paper,12pt]{article}
\usepackage[utf8]{inputenc}

\usepackage[a4paper]{geometry}
\usepackage{jheppub,
amsmath,amssymb,amsfonts,amsxtra,mathrsfs,graphics,graphicx,amsthm,epsfig,
bm,longtable,float,color,tikz,mathtools,xfrac,footnote,rotating,lscape}
\restylefloat{table}
\usepackage{dsfont}
\usepackage{multicol}
\usepackage{lipsum}
\usepackage{textgreek}
\usetikzlibrary{decorations.pathmorphing}
\usetikzlibrary{decorations.markings}
\usetikzlibrary{quotes,arrows.meta}
\usetikzlibrary{arrows, decorations.markings, calc, fadings, decorations.pathreplacing, patterns, decorations.pathmorphing, positioning}
\usepackage{tikz-cd}

\usepackage{scalerel}
\newlength\bshft
\def\fakebold#1{\ThisStyle{\ooalign{$\SavedStyle#1$\cr%
  \kern-\bshft$\SavedStyle#1$\cr%
  \kern\bshft$\SavedStyle#1$}}}

\usetikzlibrary{positioning,shapes}
\usetikzlibrary{chains}
\usetikzlibrary{arrows,fit,decorations.pathreplacing}
\tikzstyle{every picture}+=[remember picture]
\tikzstyle{na} = [baseline=-.5ex]

\usepackage{empheq}
\usepackage{multirow}
\usepackage{booktabs}
\usepackage[american]{babel}

\usepackage[utf8]{inputenc}

\usepackage{array,booktabs}

\makeatletter
\newcommand{\vast}{\bBigg@{1}}
\newcommand{\Vast}{\bBigg@{5}}
\makeatother

\usepackage{hyperref}

\newcommand{\nn}{\nonumber}

\newcommand{\be}{\begin{equation}} \newcommand{\ee}{\end{equation}}
\newcommand{\bea}{\begin{equation} \begin{aligned}} \newcommand{\eea}{\end{aligned} \end{equation}}

\newcommand{\ex}{{\mathrm{e}}}

\newcommand{\vol}{\mathrm{vol}}
\newcommand{\Vol}{\mathrm{Vol}}

\newcommand{\wb}{\overline}

\DeclareMathOperator{\sign}{sign}

\newcommand{\cA}{\mathcal{A}}

\newcommand{\cC}{\mathcal{C}}

\newcommand{\cE}{\mathcal{E}}

\newcommand{\cP}{\mathcal{P}}

\newcommand{\bC}{\mathbb{C}}

\newcommand{\bF}{\mathbb{F}}

\newcommand{\bM}{\mathbb{M}}

\newcommand{\bP}{\mathbb{P}}

\newcommand{\bR}{\mathbb{R}}

\newcommand{\bT}{\mathbb{T}}

\newcommand{\bV}{\mathbb{V}}

\newcommand{\bZ}{\mathbb{Z}}

\newcommand{\fkt}{\mathfrak{t}}

\usepackage{tikz}
\usetikzlibrary{arrows,calc}

\usepackage{dsfont}
\usepackage[bbgreekl]{mathbbol}
\usepackage{mathrsfs}
\usepackage{caption}
\usepackage{subcaption}

\usepackage{bbold}
\DeclareSymbolFontAlphabet{\mathbb}{AMSb}
\DeclareSymbolFontAlphabet{\mathbbl}{bbold}

\DeclareMathAlphabet{\mathpzc}{OT1}{pzc}{m}{it}

\newcommand{\ie}{\textit{i.e.}}

\newcommand{\ZZ}{\mathbb{Z}}
\newcommand{\RR}{\mathbb{R}}
\newcommand{\CC}{\mathbb{C}}
\newcommand{\dd}{\mathrm{d}}

\newcommand{\AdS}{\mathrm{AdS}}
\newcommand{\KE}{\mathrm{KE}}
\newcommand{\spindle}{\mathbbl{\Sigma}}
\newcommand{\Morb}{\mathbb{M}}

\newcommand{\flt}{\mathfrak{t}}
\newcommand{\flp}{\mathfrak{p}}

\newcommand{\evol}{\mathds{V}}
\newcommand{\Fext}{F}

\newcommand{\fan}{d}
\newcommand{\fix}{n}

\definecolor{cadmiumgreen}{rgb}{0.0, 0.42, 0.24}

\usepackage[Symbol]{upgreek}
\usepackage{bm}

\usepackage{calligra}
\DeclareMathAlphabet{\mathcalligra}{T1}{calligra}{m}{n}

\theoremstyle{plain}

  \theoremstyle{definition}

\providecommand{\examplename}{Example}
\providecommand{\theoremname}{Theorem}

\makeatletter
\g@addto@macro\bfseries{\boldmath}
\makeatother

\makeatletter
\newcommand*{\rom}[1]{\expandafter\@slowromancap\romannumeral #1@}
\makeatother

\begin{document}
\begin{titlepage}
\begin{center}

\today

\vskip 2.5cm

{\Large \bf  Equivariant volume extremization and holography}

\vskip 2cm
\hskip -0.15truecm {Edoardo Colombo$^{\mathrm{a,b}}$, Federico Faedo$^{\mathrm{c,d}}$, Dario Martelli$^{\mathrm{c,d}}$ and Alberto Zaffaroni$^{\mathrm{a,b}}$}

\vskip 0.8cm

${}^{\mathrm{a}}$\textit{Dipartimento di Fisica, Universit\`a di Milano-Bicocca\\
Piazza della Scienza 3, 20126 Milano, Italy}

\vskip 0.2cm

${}^{\mathrm{b}}$\textit{INFN, Sezione di Milano-Bicocca, Piazza della Scienza 3, 20126 Milano, Italy}

\vskip 0.2cm

${}^{\mathrm{c}}$\textit{Dipartimento di Matematica ``Giuseppe Peano'', Universit\`a di Torino,\\
Via Carlo Alberto 10, 10123 Torino, Italy}

\vskip 0.2cm

${}^{\mathrm{d}}$\textit{INFN, Sezione di Torino,  Via Pietro Giuria 1, 10125 Torino, Italy}

\end{center}

\vskip 2 cm

\begin{abstract}

\noindent
In a previous paper two of us (D.M. and A.Z.) proposed that a vast class of gravitational extremization problems in holography can be formulated in terms of the equivariant volume of the internal geometry, or of the cone over it. We substantiate this claim by analysing supergravity solutions corresponding to branes partially or totally wrapped on a four-dimensional orbifold, both in M-theory as well as in type II supergravities.
We show that our approach recovers the relevant gravitational central charges/free energies of several known supergravity solutions and can be used to compute these also for solutions that are not known explicitly.
Moreover, we demonstrate the validity of previously conjectured gravitational block formulas for M5 and D4 branes.
In the case of M5 branes we make contact with a recent approach based on localization of equivariant forms, constructed with Killing spinor bilinears.

\end{abstract}

\end{titlepage}

\setcounter{tocdepth}{2}

\tableofcontents

%
%

\section{Introduction}

In this paper we propose a general prescription to write extremal functions for supergravity solutions with a holographic dual. The extremal functions depend on equivariant parameters for the expected abelian isometries of the background and a set of parameters describing the geometry. The extremization with respect to the parameters gives the free energy of the supergravity solution, that is holographically equal to the central charge or free energy of the dual conformal field theory. On the quantum field theory side, this construction is the gravitational  dual of extremizing   the central charge (in even dimensions) or the sphere partition function (in odd dimensions)  in order to find the exact R-symmetry. There is a huge literature about extremal functions for black hole, and more generally for holographic solutions. Extremal functions of known black holes and black strings can be expressed in terms of gravitational blocks \cite{Hosseini:2019iad} and strongly suggest that some equivariant localization is at work. Following   \cite{Martelli:2023oqk} we will indeed express the extremal functions in terms of a universal geometrical quantity, the equivariant volume of the internal supergravity geometry.

Given a symplectic orbifold $\Morb_{2m}$ of real dimension $2m$ with a toric\footnote{The toric assumption is not essential, but is made for two reasons. Firstly, if a geometry has a symmetry group that contains $\mathbb{T}^m=U(1)^m$, we need to extremize over the corresponding $m-1$ equivariant parameters not fixed by supersymmetry, otherwise the critical point found would not be a \emph{bona fide} extremum of the gravitational action. Secondly, in this case the fixed point theorem simplifies to a sum of contributions at isolated fixed points. More generally, it would be straightforward to proceed assuming a $\mathbb{T}^k=U(1)^k$ Hamiltonian action, with $1<k < m$.}
 action of $\mathbb{T}^m=U(1)^m$ generated by the Hamiltonian $H$, we can define the equivariant volume
\be \evol =(-1)^m\int_{\Morb_{2m}} \ex^{-\frac{\omega^\mathbb{T}}{2\pi}} \, ,\ee
where $\omega^\mathbb{T}=\omega +2\pi H$ is the equivariant K\"ahler form. In addition to the vectors of the fan $V^A$, it depends on
the $m$ equivariant parameters $\epsilon_I$ for the torus $\mathbb{T}^m$ action and on the K\"ahler parameter $\lambda_A$ of the geometry. The latter enter in the expansion of the K\"ahler class in a sum of Chern classes of toric line bundles
\be -\frac{[\omega]}{2\pi}=\sum_A \lambda_A c_1(L_A) \, .\ee
 The equivariant volume of toric orbifolds is a basic topological object. It can be computed using a fixed point formula and it is only sensitive to the degenerations of the torus  $\mathbb{T}^m$ near the fixed points. In the applications to holography one encounters metrics that are not K\"ahler and not even symplectic, but with underlying spaces that are in fact symplectic toric orbifolds and one can nevertheless define $\evol$ and use it to compute topological quantities that ultimately will not depend on the metric. In many examples when the underlying geometry is  not strictly symplectic or toric  we can also  define a natural generalization of $\evol$ by a sort of analytical continuation.\footnote{This happens for geometries where the fan is not strictly convex or geometries involving $S^4$.} Given these properties, the equivariant volume is the gravitational analogue of quantum field theory quantities like 't Hooft anomalies and supersymmetric indices that are invariant under small deformations of the theory once symmetries and matter content are fixed.%
 \footnote{It was already anticipated in the literature that we can extract 't Hooft anomalies from supergravity using equivariant co-homology. See for example~\cite{Bah:2019rgq,Bah:2019vmq,Bah:2020jas,Hosseini:2020vgl}.}
 All these quantities are insensitive to the UV behaviour and allow to compute IR observables by extremization principles.
 In \cite{Martelli:2023oqk} it was argued therefore that all extremization problems in gravity can be reformulated in terms of the equivariant volume. It was shown that this is true for volume minimization \cite{Martelli:2005tp,Martelli:2006yb} (dual to $a$ \cite{Intriligator:2003jj} and $F$-maximization \cite{Jafferis:2010un}) and the formalism of GK geometry \cite{Couzens:2018wnk,Gauntlett:2018dpc} (dual to $c$ \cite{Benini:2012cz} and ${\cal I}$-extremization \cite{Benini:2015eyy}). It was proposed that this should be true more generally.

As a partial check of this proposal, it has been shown in \cite{Martelli:2023oqk} that all known extremization problems for branes wrapped over a sphere or a spindle in type II and M theory can be reformulated in terms of an extremal function
\be
F\, =\, \evol^{(\alpha)}(\lambda_A, \epsilon_I)\, ,
\ee
 subject to a set of flux constraints
\be
\label{firstMA}
\nu M_A =-\frac{\partial \evol^{(\beta)}}{\partial \lambda_A} \, ,
\ee
where $M_A$ are the integer fluxes of the relevant RR or M theory antisymmetric form,
obeying
\bea
\label{firstsuMA}
 \sum_A V^A_I M_A\, = \, 0\, .
 \eea
$\nu$ is a normalization constant\footnote{\emph{A priori} there is also an overall normalization constant in the definition of  $F$, again depending on the
 type of brane and the dimension of the internal geometry, however this can always be absorbed in a rescaling of the $\lambda_A$, using the homogeneity of $\evol^{(\gamma)}$. For simplicity, in the examples we will indicate only the type of brane as a subscript in $\nu$, omitting the dependence on the dimension of the internal geometry.} that depends on the type of brane and the dimension of the internal geometry, and    $\evol^{(\gamma)}$ is the homogeneous piece of $\evol$ of degree $\gamma$ in $\lambda_A$.
 We also note that from (\ref{firstMA}) and (\ref{firstsuMA}) it follows, using the properties of $\evol$, that the constraint
 \be
\label{nowtopconstr}
 \evol^{(\beta-1)} \, =\, 0 \,
\ee
must be satisfied. Although it is not an independent relation, one can regard this as a topological constraint necessary in order to impose the flux quantization, analogously to the GK formalism \cite{Couzens:2018wnk}.
 The integers $\alpha$ and $\beta$ depend on the type of brane. By a simple scaling argument, it was found that
\bea
	&\text{D3  branes in type IIB:}  \qquad  & \alpha &= 2 \,, & \beta &= 2\\
	&\text{M2 branes in M theory:}   \qquad  & \alpha &= 3 \,, & \beta &= 3\\
	&\text{M5 branes in M theory:}   \qquad  & \alpha &= 3 \,, & \beta &= 2\\
	&\text{D4 branes in massive type IIA:}  \qquad  & \alpha &= 5 \,, & \beta &= 3\\
	&\text{D2 branes in massive type IIA:}  \qquad  & \alpha &= 5 \,, & \beta &= 4\,.
\eea
 The extremal function $F$ can be normalized such that its extremum reproduces the central charge of the dual field theory in even dimensions and the logarithm of the sphere partition function in odd dimensions and we will use this convention in the following.

In this paper we show that this construction also holds for known extremization problems for branes (partially or totally)  wrapped over four-dimensional toric orbifolds. We need to generalize our   construction by introducing {\it higher times} in the equivariant volume
\be\label{highertimes} \evol =(-1)^m\int_{\Morb_{2m}} \ex^{-\frac{\omega^\mathbb{T}}{2\pi} + \sum_{k=2}  \lambda_{A_1\ldots A_k} c_1^\mathbb{T}(L_{A_1}) \ldots c_1^\mathbb{T}(L_{A_k}) }\, ,\ee
where $\lambda_{A_1\ldots A_k}$ are symmetric tensors and a sum over repeated indices $A_i$ is understood. Higher times have appeared only recently in the literature \cite{Nekrasov:2021ked} and are still poorly studied. The previous expression has a large gauge invariance and many parameters are redundant.
As we will see, the  equivariant volume with higher times contains all the information  needed to fully capture the topological properties and the quantization of fluxes for
a very large class of
supergravity solutions.

The above construction relies on even-dimensional toric orbifolds. For supergravity backgrounds AdS$_d \times M_k$ with odd-dimensional internal space $M_k$ the geometry to consider is the cone over $M_k$, as familiar from holography. This cone is often a non-compact toric Calabi-Yau, or, in the case of supersymmetry preserved with anti-twist, a non-convex generalization.\footnote{See for example \cite{Boido:2022mbe}.} When $M_k$ is even-dimensional, we consider the equivariant volume of the compact $M_k$ itself. Some M5 brane solutions have a $\mathbb{Z}_2$ symmetry that allows to cut into half the number of fixed point and consider an equivalent problem for a non-compact Calabi-Yau (half of the manifold). This was done in  \cite{Martelli:2023oqk} for M5 branes wrapped on a spindle.

Our approach naturally incorporates the GMS construction based  on GK geometry \cite{Couzens:2018wnk,Gauntlett:2018dpc} as well as the recent localization technique based on Killing spinor bilinears in  M theory \cite{BenettiGenolini:2023kxp}. Indeed, we will show that, for M5 solutions
with even-dimensional $M_6$ or $M_8$, our approach is effectively equivalent  to  the one in \cite{BenettiGenolini:2023kxp}. In particular, all the geometrical constraints that must be imposed on a case-by-case analysis in order to find the free energy in \cite{BenettiGenolini:2023kxp} appear naturally in our construction as an extremization with respect to all the parameters that are not fixed by the flux quantization conditions. On the one hand, this is a nice confirmation of our prescription. On the other hand,  our approach for the toric case is more general, it covers in a simple and universal way the even and odd-dimensional cases, it naturally extends to massive type IIA solutions, which are not yet covered by the previous techniques, and expresses everything in terms of the extremization of a universal quantity, the equivariant volume of the associated geometry, without referring to supergravity quantities. We are confident that when the explicit case-by-case supergravity analysis will be performed for the missing backgrounds it will confirm our general prescription.

 The paper is organized as follows. In section \ref{sec:equiv} we define the equivariant volume of a general toric orbifold and we review some of its basic properties following \cite{Martelli:2023oqk}. We also introduce the concept of higher times, which are necessary to parameterize all the fluxes supported by a given geometry. In section \ref{Mtheory} we
analyse  M theory solutions with M5 brane flux. In section \ref{AdS3xM8} we  consider  solutions associated with M5 branes wrapped over a four-dimensional orbifold $\Morb_4$. We show  that the free energy can
 be obtained by extremizing the appropriate term in the equivariant volume and that the result agrees with the field theory computation in \cite{Martelli:2023oqk}, obtained by integrating the anomaly of the M5 brane theory over $\Morb_4$. In section \ref{sec:M5} we consider  solutions that are potentially related to M5 branes wrapped on a  two-cycle in $\Morb_4$. By extremizing the appropriate term in the equivariant volume, we reproduce known results in the literature and extend them to predictions for solutions still to be found.
  In section \ref{GGS} we compare our prescription with the recent approach based on Killing spinor bilinears in  M theory \cite{BenettiGenolini:2023kxp}.  In section \ref{II theory} we consider solutions in type II string theory with geometries that are fibrations over a four-dimensional orbifold $\Morb_4$. In
section \ref{AdS2xM8} we consider  massive type IIA solutions associated with D4 branes wrapped around a  four-dimensional toric orbifold $\Morb_4$ and derive the free energy proposed in \cite{Faedo:2022rqx}.
In section \ref{subsec:AdS4xM6} we consider massive type IIA solutions associated with a system of D4/D8 branes, with the former wrapped on a two-cycle in $\Morb_4$. Extremizing the appropriate term in the equivariant volume we are able to reproduce the gravitational free energy computed from the explicit solution.
In section \ref{sec:D3} we consider type IIB solutions with D3 flux associated with $S^3/\bZ_p$ fibrations over $\Morb_4$, which could potentially arise as the near-horizon limit of a system of D3 branes wrapped on a two-cycle of the four-dimensional orbifold $\Morb_4$. This example can be covered by the formalism of GK geometry, that we here extend to the case of fibrations over  orbifolds, using the equivariance with respect to the full four-torus $\bT^4$. In this and other previous examples with M5 branes, we observe that, in order to obtain the correct critical point, one should allow all the equivariant parameters not fixed by symmetries to vary, thus rectifying some previous results in the literature. We conclude with a discussion of open problems and future perspectives. Three appendices contain technical aspects of some computations.

\section{Equivariant volume with higher times}\label{sec:equiv}

In this section we review and generalize some basic facts about the equivariant volume of general toric orbifolds that will be used in the following. We adopt the conventions of
\cite{Martelli:2023oqk}, to which we refer for more details and a review of equivariant localization.

We consider a toric orbifold $\Morb_{2m}$ with an action $\mathbb{T}^m$  generated by the $m$ vector fields $\partial_{\phi_I}$. We introduce  $m$ equivariant parameters $\epsilon_I$, with $I=1,\ldots,m$, and the vector field $\xi= \epsilon_I \partial_{\phi_I}$ and consider equivariantly closed forms  $\alpha^{\mathbb{T}}$
satisfying
\be  (\dd + 2\pi  i_{\xi}) \alpha^{\mathbb{T}}=0 \, .\ee
We will be dealing with varieties and orbifolds of different dimension and, when needed, we will also write $\alpha^{\mathbb{T}^m}$ to specify the dimension.
Each toric orbifold comes equipped with a fan, a collection of integer vectors $V^A$ that exhibits  $\Morb_{2m}$ as a $\mathbb{T}^m$ fibration over a convex polytope
\be
	\cP=\{y_I\in\bR^m\:|\:y_IV_I^A\geq\lambda_A\}\:.
\ee
On each of the facets $y_I V^A_I =\lambda_A$ a particular circle in $\mathbb{T}^m$ degenerates, as familiar from toric geometry.\footnote{Notice that for toric varieties the vectors $V^A$ are primitive, while in the case of generic symplectic toric orbifolds they are not. We can define for each $V^A$ a primitive vector $\hat V^A$ and a positive integer $n_A$ such that $v^A = n_A \hat V^A$. The symplectic divisor $D_A$ has a local $\mathbb{Z}_{n_A}$ singularity. For more details we refer to \cite{Martelli:2023oqk}.} Each facet defines a toric divisor $D_A$, with associated line bundle $L_A$, and an  equivariant Chern class
\be \label{Cherneq} c_1^{\mathbb{T}}(L_A) = c_1(L_A) + 2\pi \epsilon_I \mu_I^A = \dd ( \mu_I^A \dd \phi_I) + 2\pi \epsilon_I \mu_I^A\, \, .\ee
The functions $\mu^A$ play the role of moment maps. We will not need their explicit expression, which is discussed in \cite{Martelli:2023oqk}, but we will frequently use the relation
\be\label{equivChern} \sum_A V^A_I c_1^{\mathbb{T}}(L_A) =-\epsilon_I \, ,\ee
which is the equivariant version of the co-homological statement $ \sum_A V^A_I c_1(L_A)=0$ following from the toric equivalence relations among divisors
$ \sum_A V^A_I D_A=0$.

The equivariant volume with higher times \eqref{highertimes} can be computed with a fixed point formula that can also be taken as an operative definition
\be  \evol(\lambda_{A_1\ldots A_K},\epsilon_I) =(-1)^m \sum_{\alpha=1}^\fix  \frac{ \ex^{ \tau^\mathbb{T} |_{y_\alpha}  } } {d_\alpha\, e^{\mathbb{T}}|_{y_\alpha}}\, ,\ee
where $y_\alpha$ are the fixed points of  the $\mathbb{T}^m$ action, $d_\alpha$  the order of the orbifold singularity at $y_\alpha$, $e^{\mathbb{T}}$ the equivariant Euler class of the tangent bundle at $y_\alpha$ and
\be\label{timeform} \tau^\mathbb{T} = \sum_{k=1}  \lambda_{A_1\ldots A_k} c_1^\mathbb{T}(L_{A_1}) \ldots c_1^\mathbb{T}(L_{A_k}) \, .\ee

To use the localization formula we assume that $\Morb_{2m}$  has only isolated orbifold singularities. This is the case if the fan
is  the union of $m$-dimensional cones $\{V^{A_1},\ldots ,V^{A_m}\}$.\footnote{We can always resolve the fan by adding  vectors if this condition is not met.} Each cone $\alpha=(A_1,\ldots, A_m)$
corresponds to an isolated fixed point  $y_\alpha$ with a local orbifold singularity of order
\be d_\alpha = \bigl|\det (V^{A_1}, \ldots ,V^{A_m})\bigr| \, .\ee
The restriction of the Euler class is given by
\be e^{\mathbb{T}}\bigr|_{y_\alpha} =\prod_{A_i \in \alpha} c^{\mathbb{T}}_1(L_{A_i}) \bigr|_{y_\alpha} \, .\ee
The restriction of the Chern classes at the fixed points $y_\alpha$ are computed as follows~\cite{Martelli:2023oqk}
  \be\label{rest3}  c^{\mathbb{T}}_1(L_A) \bigr|_{y_\alpha} =   \begin{cases} -\frac{ \epsilon_I U_I^{A}}{d_\alpha} \, \qquad &{\rm if}\,  A\in \alpha  \\
0 \, \qquad &{\rm if} \,  A\notin \alpha \end{cases}\, ,\ee
where $U_I^{A}$ are the inward normal vectors to the facets of the cone $\alpha$ defined and normalized by the relation
\be U^{A_i}_\alpha\cdot V^{A_j} = d_\alpha \delta_{ij}\, .\ee

Now the fixed point formula gives
\be \label{fp} \evol(\lambda_{A_1\ldots A_k},\epsilon_I)=\sum_{\alpha=(A_1,\ldots, A_m)} \frac{\ex^{\tau_\alpha}}{d_\alpha \prod_{i=1}^m \frac{\epsilon \cdot U^{A_i}_\alpha}{d_\alpha} } \, ,\ee
where $\tau_\alpha$ is the restriction of the equivariant form \eqref{timeform} to the fixed point $y_\alpha$ and explicitly reads
\be \tau_\alpha= - \sum_{i=1}^m \lambda_{A_i}  \Bigl(\frac{\epsilon \cdot U^{A_i}_\alpha} {d_{\alpha} }\Bigr) + \sum_{i,j=1}^m \lambda_{A_iA_j}  \Bigl(\frac{\epsilon \cdot U^{A_i}_\alpha}{d_{\alpha}}\Bigr)\Bigl(\frac{\epsilon \cdot U^{A_j}_\alpha}{d_{\alpha}}\Bigr) +\ldots
 \, \ee
 The equivariant volume can be expanded in power series of the higher times,
 \be   \evol(\lambda_{A_1\ldots A_k},\epsilon_I)=\sum_{n=0}^\infty  \evol^{(n)}(\lambda_{A_1\ldots A_k},\epsilon_I)\, ,\ee
 where we denote with $\evol^{(n)}$ the homogeneous component of degree $n$ in the set of higher times $\lambda_{A_1\ldots A_k}$ for all $k$. $\evol^{(n)}$ is a polynomial in $\epsilon_I$ in the compact case, while it can be a rational function  of $\epsilon_I$ when $\Morb_{2m}$ is non-compact.

 In the examples in \cite{Martelli:2023oqk} only single times ($\lambda_A$) were used. In this paper we will use single and double times ($\lambda_A$ and $\lambda_{AB}$).
 As a general rule, to fully capture the parameters of the supergravity solution, we need  a number of independent parameters at least equal to the number of fixed points.
 Functionally, indeed $\evol$ is a function of $\epsilon_I$ and $\tau_\alpha$ only.
Notice that there is a large redundancy in the description with higher times. Due to the relation \eqref{equivChern}, $\tau^\mathbb{T}$ is invariant under the {\it gauge transformations}
\bea\label{gauge}
\lambda_{A_1\ldots A_{k+1}}\rightarrow \lambda_{A_1\ldots A_{k+1}} + \beta_I^{(A_1\ldots A_k} V^{A_{k+1})}_I \, , \qquad  \lambda_{A_1\ldots A_{k}}\rightarrow \lambda_{A_1\ldots A_{k}} + \epsilon_I \beta_I^{A_1\ldots A_k} \, ,
\eea
where $\beta_I^{A_1\ldots A_k}$ is  symmetric in the indices  $A_1\ldots A_k$.
Notice that the  subgroup with  $\epsilon_I \beta_I^{A_1\ldots A_k}=0$ acts only on $\lambda_{A_1\ldots A_{k+1}}$ without mixing times of different degree and it is the only transformation allowed for single times.
In the Calabi-Yau case, where the vectors in the fan lie on a plane identified by the direction $I=CY$, say $V^A_{CY}=1$,\footnote{In this paper $\epsilon_{CY}$ will be identified with $\epsilon_3$ in all examples.} this subgroup can also be written as \be\label{gaugeCY} \lambda_{A_1\ldots A_{k}}\rightarrow \lambda_{A_1\ldots A_{k}}+(\epsilon_{CY} V^{(A_1}_{I} - \epsilon_{I} ) \gamma^{A_2\ldots A_k)}_{I}\, ,\ee
 generalizing the results in  \cite{Martelli:2023oqk}.
Many times can be therefore gauge-fixed to zero.

As an example we give some more explicit expressions about the four-dimensional case. These will be heavily used in the following since
in this paper we will mostly consider geometries that are fibrations over four-dimensional compact orbifolds $\Morb_4$. For clarity of notation, we will use capital letters $(V^A_I,A,I)$ for the higher-dimensional geometry and lower-case letters $(v^a_i,a,i)$  for $\Morb_4$. The fan of a  four-dimensional compact orbifold is just a collection of two-dimensional integer vectors $v^a$, $a=1,\dots, \fan$, that define a convex polygon in the plane. The fixed points are associated with the cones $(v^a,v^{a+1})$, where we take a counter-clockwise order for the vector and identify cyclically $v^{a+d}=v^a$. Notice that in the compact four-dimensional case the number of fixed points is equal to the number of vectors in the fan and we can use the index~$a$ to label both. With the notations of \cite{Martelli:2023oqk} we define the quantities
\begin{equation} \label{epsilon}
	\epsilon_1^a = \frac{\epsilon \cdot u^a_1}{d_{a,a+1}} \,,  \qquad
	\epsilon_2^a = \frac{\epsilon \cdot u^a_2}{d_{a,a+1}} \,,
\end{equation}
where $u^a_1$ and $u^a_2$ are the inward normals to the  cones $(v^a,v^{a+1})$. Explicitly
\begin{equation}
  \epsilon^{a}_1 = -\frac{\det(v^{a+1}, \epsilon)}{\det( v^a , v^{a+1} )} \,,  \qquad
	\epsilon^{a}_2 = \frac{\det (v^a, \epsilon )}{\det (v^a ,v^{a+1} )}\,,
\end{equation}
 where $\epsilon \equiv (\epsilon_1,\epsilon_2)$.   In particular, the equivariant Euler class of the tangent bundle at a fixed point $y_a$  reads
\be
e^{\mathbb{T}}\bigr|_{y_a} =  \epsilon^{a}_1  \, \epsilon^{a}_2\, ,
\ee
 and the order of the local orbifold singularity is
 \be d_{a,a+1}=\det (v^a,v^{a+1}) \, .\ee
The restriction to the fixed points of the equivariant Chern classes $c_1^{\mathbb{T}}(L_a)$   can be written as
\begin{equation}\label{res2d}
 c_1^{\mathbb{T}}(L_a)\bigr|_{y_b}  =  -( \delta_{a,b} \epsilon_1^b +  \delta_{a,b+1}\epsilon_2^b )\, .
 \end{equation}
 The fixed point formula \eqref{fp} for the equivariant volume specializes to the expression\footnote{Notice that there is no summation on $a$ in the exponent.}
\bea\label{eqvolume4d}
 \evol(\lambda_{a_1\ldots a_k},\epsilon_i)  &
=\sum_{a=1}^d \frac{\ex^{ -\lambda_a \epsilon^a_1 -\lambda_{a+1} \epsilon^a_2 +\lambda_{a,a} (\epsilon^a_1)^2 +2\lambda_{a,a+1} \epsilon^a_1\epsilon^a_2+\lambda_{a+1,a+1} (\epsilon^a_2)^2 +\ldots}}{d_{a,a+1}\,\epsilon_1^{a}\epsilon_2^{a} }\, .
 \eea

In the following, we will also need the intersections matrix of divisors, which is independent of the equivariant parameters $\epsilon_1,\epsilon_2$ \cite{Martelli:2023oqk}:
\begin{equation}\label{divisors}
D_a\cdot D_b= D_{ab} = \int_{\Morb_{4}} c_1^{\mathbb{T}}(L_{a}) c_1^{\mathbb{T}}(L_{b}) =
\begin{cases}
\frac{1}{d_{a-1,a}} & \text{if $b=a- 1$}\, ,\\
\frac{1}{d_{a,a+1}} & \text{if $b=a +1$}\, ,\\
 -\frac{d_{a-1,a+1}}{d_{a-1,a}d_{a,a+1}} & \text{if $b=a$}\, ,\\
 0 & \text{otherwise}\, .
\end{cases}
\end{equation}

\section{$\AdS_3$ and $\AdS_5$  solutions in M theory}\label{Mtheory}

We start by analysing M theory solutions with M5 brane flux  and show that the free energy can be obtained by extremizing the appropriate term in the equivariant volume. The case of M5 branes wrapped on a spindle have been already studied in \cite{Martelli:2023oqk}. Here we focus on geometries that are fibrations over a four-dimensional toric orbifold $\Morb_4$.

\subsection{$\AdS_3\times M_8$ solutions}
\label{AdS3xM8}

In this section we consider $\AdS_3\times M_8$ solutions in M theory,
where\footnote{In general, $M_8$ is itself an orbifold.}
$M_8$ is an $S^4$ fibration over the four-dimensional orbifold $\Morb_4$.
Examples of this form have been found  in
\cite{Cheung:2022ilc} and further discussed in \cite{Faedo:2022rqx,Couzens:2022lvg,Bomans:2023ouw}.
They are obtained by uplifting
$\AdS_3\times\Morb_4$ solutions of $D=7$ maximal gauged supergravity to eleven dimensions.
These $\AdS_3\times M_8$ solutions can be interpreted as the near-horizon geometry of a system of M5 branes wrapped around $\Morb_4$.

We need first to identify the topological structure of the underlying geometry. We will focus on the case of toric $\Morb_4$. The  eight-dimensional geometry $M_8$ is not strictly toric, but it admits an action of $\bT^4=U(1)^4$.
If $\fan$ is the dimension of the fan of $\Morb_4$, there are $2 d$ fixed points of the torus action obtained  by selecting
a fixed point on $\Morb_4$ and combining it with the North and South pole of $S^4$. We will assume that there is a $\mathbb{Z}_2$ symmetry of the fibration that identifies the
North and South pole contributions to the fixed point formula. In this situation we can consider half of the geometry, a $\bC^2$ fibration over $\Morb_4$ with the geometry of a non-compact toric CY$_4$. One can understand the appearance of the fibre  $\bC^2$ from
the transverse geometry of the brane system, which is $\bC^2\times\bR$, with $S^4$ embedded inside.
 We then consider a CY$_4$ with fan generated by the vectors
 \be
\label{C2overMorb_4_fan}
	V^a=(v^a,1,\fkt_a)\, ,\qquad V^{d+1} =(0,0,1,0)\, ,\qquad V^{d+2}=(0,0,1,1)\,,
\ee
where $v^a$, $a=1,\dots,d$, are the vectors of the fan of $\Morb_4$ and $\fkt_a$ are integers specifying the twisting of $\CC^2$ over $\Morb_4$.
When supersymmetry is preserved with anti-twist \cite{Ferrero:2021etw}, the toric diagram is not convex and it does not strictly define a toric geometry. We will nevertheless proceed also in this case, considering it as an extrapolation from the twist case. The non-convex case is obtained from the formulas in this paper by sending $v^a\rightarrow \sigma^a v^a$, where $\sigma^a=\pm1$.

In addition to the metric, the supergravity solution is specified by the integer fluxes of the M theory four-form along all the non-trivial four-cycles. The toric four-cycles of the geometry  are $\Morb_4$ itself, the sphere $S^4$ and $\bP^1$ fibrations over the toric two-cycles $\Sigma_a\subset \Morb_4$.
In our half-geometry, the sphere $S^4$ and $\bP^1\subset S^4$ are replaced with copies of  $\bC^2$ and $\bC$. All together,  the toric four-cycles  correspond  to  all the possible intersections of the toric divisors $D_A\cap D_B$ and we can therefore introduce a matrix of fluxes $M_{AB}$. As usual, not all toric divisors are inequivalent in co-homology. The
relations $\sum_A V^A_I D_A=0$ imply that the matrix of fluxes satisfy
\be \sum_A V^A_I M_{AB}=0\, .\ee

We are now ready to formulate our prescription for  the  extremal function. For M5 branes in M theory, as discussed in \cite{Martelli:2023oqk} and in the introduction,
we define the free energy to extremize as
\be F=\evol^{(3)}(\lambda_A,\lambda_{AB},\epsilon_I) \, ,\ee
and impose the flux constraints\footnote{We put a bar on top of $\nu_{M5}$ to stress that we are using a half-geometry. To have the correct normalization of the free energy when using half of the geometry, the parameter $\nu_{M5}$ must be rescaled as in formula \eqref{half-full}, as we will discuss more extensively in section \ref{sec:M5}.}
\be
\label{AdS3xM8_prescription}
	\wb\nu_{M5}\,(2-\delta_{AB})\,M_{AB}=-
		\frac{\partial}{\partial\lambda_{AB}}\evol^{(2)}(\lambda_A,\lambda_{AB},\epsilon_I)\, .
\ee
Here the index $A=1,\ldots,d+2$ runs over all the vectors of the fan of the CY$_4$, whereas we reserve the lower-case index $a=1,\ldots,d$
for the vectors of the fan of the base $\Morb_4$.
On the other hand, the index $I=1,2,3,4$ runs over the equivariant parameters of the CY$_4$ and we will use $i=1,2$ for the directions inside $\Morb_4$.
We have added a $(2-\delta_{AB})$ factor in the equation for the fluxes for convenience.
It is easy to see using \eqref{fp} that this equation can be equivalently rewritten as
\be
	\wb\nu_{M5}\,M_{AB}=-\frac{\partial^2}{\partial\lambda_A\partial\lambda_B}\evol^{(3)}(\lambda_A,\lambda_{AB},\epsilon_I)\: ,
\ee
and one may wonder if we really need higher times. The answer is yes. As we will discuss later, with only single times the previous equation cannot be solved.\footnote{One would need to restrict the $\fkt_a$ in order to find solutions.}

In the rest of this section we will show that $F$ reproduces the expected extremal function and its factorization in  gravitational blocks
discussed in \cite{Faedo:2022rqx,Martelli:2023oqk}.

\subsubsection{The equivariant volume with double times}
The $\bT^4$ torus action on the CY$_4$ has $d$ fixed points, each one corresponding to a cone in the fan with generators $(V^a,V^{a+1},V^{d+1},V^{d+2})$,
$a=1,\ldots,d$.
In particular, there is a one-to-one  correspondence between these fixed points and the ones of the base orbifold $\Morb_4$;
for the latter the fixed points correspond to two-dimensional cones of the form $(v^a,v^{a+1})$ and they can be labelled by the index $a$.
The order of the orbifold singularities associated with the fixed points of CY$_4$ and $\Morb_4$ also match:
\be
	d_{a,a+1,d+1,d+2}=\bigl|\det(V^a,V^{a+1},V^{d+1},V^{d+2})\bigr|=\bigl|\det(v^a,v^{a+1})\bigr|=d_{a,a+1}\:.
\ee
Therefore, the fixed point formula for the equivariant volume with higher times of CY$_4$ takes the following form:
\be
\label{f.p.fmla_AdS3xM8}
	\evol(\lambda_A,\lambda_{AB},\epsilon_I)=
		\sum_a\frac{\ex^{\tau_a}}{d_{a,a+1}\:e^{\bT^4}|_{a}}\:.
\ee
Here, $\tau_a$ is the restriction to the fixed point $a$ of the form \eqref{timeform}%
\be
	\tau_a=\left(\sum_A\lambda_A\,c_1^{\bT^4}(L_A)+\sum_{A,B}\lambda_{AB}\,c_1^{\bT^4}(L_A)\,c_1^{\bT^4}(L_B)\right)\Bigg|_{a}\:,
\ee
while at the denominator we have the restriction of the Euler class $e^{\bT^4}$
\be
	e^{\bT^4}\bigr|_{a}=\Big(c_1^{\bT^4}(L_a)\,c_1^{\bT^4}(L_{a+1})\,c_1^{\bT^4}(L_{d+1})\,c_1^{\bT^4}(L_{d+2})\Big)
		\Big|_{a}\:.
\ee
The restrictions of the Chern classes
can be computed using \eqref{rest3}.
The inward normals to the faces of the cone generated by $(V_a,V_{a+1},V_{d+1},V_{d+2})$ are
 \bea
	&U^a=(u^{a}_{1},0,0)\, ,\\
	&U^{a+1}=(u^{a}_{2},0,0)\,,\\
	&U^{d+1}=\big((\fkt_a-1) u^{a}_{1}+(\fkt_{a+1}-1) u^{a}_{2}\,, d_{a,a+1}, -d_{a,a+1}\big)\,,\\
	&U^{d+2}=(-\fkt_a  u^{a}_{1}-\fkt_{a+1} u^{a}_2\,,0,d_{a,a+1})\,,
\eea
where $u^a_1$ and $u^a_2$ are the two-dimensional normals to the cone $(v^a,v^{a+1})$. Using the notations introduced in \eqref{epsilon} we find
\begin{align}\nn
	&c_1^{\bT^4}(L_a)\bigr|_{a}=-\frac{\epsilon_i\,(u^{a}_{1})_i}{d_{a,a+1}}=-\epsilon_1^a\,,\\\nn
	&c_1^{\bT^4}(L_{a+1})\bigr|_{a}=-\frac{\epsilon_i\,(u^{a}_{2})_i}{d_{a,a+1}}=-\epsilon_2^a\,,\\[2mm]
	&c_1^{\bT^4}(L_b)\bigr|_{a}=0\,,\qquad b\ne a,a+1\,,\\[4mm]\nn
	&c_1^{\bT^4}(L_{d+1})\bigr|_{a}=-(\fkt_a-1)\epsilon_1^a-(\fkt_{a+1}-1)\epsilon_2^a-\epsilon_3+\epsilon_4\, , \\[4mm]\nn
	&c_1^{\bT^4}(L_{d+2})\bigr|_{a}=\fkt_a\epsilon_1^a+\fkt_{a+1}\epsilon_2^a-\epsilon_4\,,
\end{align}
where for simplicity we have used Einstein notation for the sums over the index $i=1,2$.

We can write the equivariant volume  of the CY$_4$ as an integral over the base orbifold $\Morb_4$
of four-dimensional equivariant forms  with $\epsilon_3$ and $\epsilon_4$ as parameters.
Let us denote with $\bT$ the two-dimensional torus associated with $\epsilon_1$ and $\epsilon_2$,
and let $c_1^{\bT}(L_a)$ be the equivariant Chern classes associated to the restrictions of the line bundles $L_a$ to the base $\Morb_4$.
We can then take advantage of the one-to-one correspondence between fixed point of the CY$_4$ and fixed points of $\Morb_4$
and, using \eqref{res2d}, we can rewrite \eqref{f.p.fmla_AdS3xM8} as
\be
	\evol(\lambda_A,\lambda_{AB},\epsilon_I)=\int_{\Morb_4}\frac{\ex^{\tau^\bT}}{\cC_{d+1}\,\cC_{d+2}}\:,
\ee
where
\bea
\label{AdS3xM8_equivariant_forms}
	&\tau^\bT=\sum_A\lambda_A\:\cC_A+\sum_{A,B}\lambda_{AB}\:\cC_A\:\cC_B\, , \\
	&\cC_a\,=\,c_1^\bT(L_a)\:,\qquad a=1,\ldots,d\, , \\[3mm]
	&\cC_{d+1}\,=\,-\epsilon_3+\epsilon_4+\sum_a(\fkt_a-1)c_1^\bT(L_a)\, , \\
	&\cC_{d+2}\,=\,-\epsilon_4-\sum_a\fkt_ac_1^\bT(L_a)\:.
\eea
Notice the relations $\sum_av^a_i\,c_1^\bT(L_a)=-\epsilon_i$
and $\sum_AV_I^A\,\cC_A=-\epsilon_I$, following from \eqref{equivChern}.\footnote{The second relation, which can be checked by direct computation, is obviously
the restriction of  $\sum_AV^A_I\,c_1^{\bT^4}(L_A)=-\epsilon_I$ to $\Morb_4$.}

The homogeneous component of degree $\alpha$ of the equivariant volume with higher times can
be expressed as
\be
\label{AdS3xM8_homogeneous_components}
	\evol^{(\alpha)}(\lambda_A,\lambda_{AB},\epsilon_I)=\int_{\Morb_4}\frac{(\tau^\bT)^\alpha}{\alpha!\:\cC_{d+1}\,\cC_{d+2}}=
		\sum_a\frac{B^{(\alpha)}_a}{d_{a,a+1}\:\epsilon_1^a\:\epsilon_2^a}\:,
\ee
where we have defined $B^{(\alpha)}_a$ to be the restriction over the $a$-th fixed point of $\Morb_4$ of the following equivariant form:
\be
\label{Bdef}
	B^{(\alpha)}=\frac{(\tau^\bT)^\alpha}{\alpha!\:\cC_{d+1}\,\cC_{d+2}}\:.
\ee
For later reference we derive the relation between $B^{(\alpha)}_a$ and $B^{(\beta)}_a$
\begin{align} \label{Bchange}
	B^{(\beta)}_a=&\,\frac{(\tau_a)^\beta}{\beta!\,\big(\cC_{d+1}\,\cC_{d+2}\big)|_{a}}=
		\frac{(\alpha!)^{\frac\beta\alpha}}{\beta!}
		\left[\frac{(\tau_a)^\alpha}{\alpha!\,\big(\cC_{d+1}\,\cC_{d+2}\big)|_{a}}\right]^{\frac\beta\alpha}
		\Big[\big(\cC_{d+1}\,\cC_{d+2}\big)|_{a}\Big]^{\frac\beta\alpha-1}\\
	=&\,\frac{(\alpha!)^{\frac\beta\alpha}}{\beta!} \big(B_a^{(\alpha)}\big)^{\frac\beta\alpha}
	\big((1-\fkt_a)\epsilon_1^a+(1-\fkt_{a+1})\epsilon_2^a-\epsilon_3+\epsilon_4\big)^{\frac\beta\alpha-1}
		\big(\fkt_a\epsilon_1^a+\fkt_{a+1}\epsilon_2^a-\epsilon_4\big)^{\frac\beta\alpha-1}. \nonumber
\end{align}
When $\alpha$ is even this formula holds in terms of absolute values and the signs must be fixed separately.
This will not be the case for the computation of this section, so we postpone the discussion about the signs to section \ref{AdS2xM8}.
\subsubsection{Solving the flux constraints}
The flux constraints \eqref{AdS3xM8_prescription} reads
\be
\label{AdS3xM8_flux_equation}
	\wb\nu_{M5}\,(2-\delta_{AB})\,M_{AB}=-\frac{\partial\evol^{(2)}}{\partial\lambda_{AB}}=
		-(2-\delta_{AB})\int_{\Morb_4}\frac{\cC_A\,\cC_B\,\tau^\bT}{\cC_{d+1}\,\cC_{d+2}}\:,
\ee
or, equivalently
\be
\label{AdS3xM8_flux_equation2}
	\wb\nu_{M5}\,M_{AB}=-\int_{\Morb_4}\frac{\cC_A\,\cC_B\,\tau^\bT}{\cC_{d+1}\,\cC_{d+2}}=
		-\sum_a\frac{B^{(1)}_a\cdot\big(\cC_A\,\cC_B\big)|_{a}}{d_{a,a+1}\:\epsilon_1^a\:\epsilon_2^a}\:.
\ee
Let us focus on the $A,B\in\{1,\ldots,d\}$ sector. Using \eqref{res2d} we find
\bea
\label{M_ab_and_B_a_equations}
	&\wb\nu_{M5}\,M_{a,a+1}=-\frac{B^{(1)}_a}{d_{a,a+1}}\, , \\
	&\wb\nu_{M5}\,M_{a,a}=-\frac{B^{(1)}_a\:\epsilon_1^a}{d_{a,a+1}\:\epsilon_2^a}
		-\frac{B^{(1)}_{a-1}\:\epsilon_2^{a-1}}{d_{a-1,a}\:\epsilon_1^{a-1}}\, , \\[2mm]
	&\wb\nu_{M5}\,M_{ab}=0\qquad\text{when }b\ne a,\,a+1,\,a-1\:.
\eea
These equations give  constraints on the fluxes but they have a very simple solution
\bea
\label{AdS3xM8_flux_equations_solution}
	&B^{(1)}_a=-\wb\nu_{M5}\,N\, , \\
	&M_{ab}=N\,D_{ab}\:,
\eea
where $D_{ab}$ is the intersection matrix of divisors \eqref{divisors} and $N$ is any integer that is a multiple of all the products $d_{a-1,a}\,d_{a,a+1}$.

This can be seen as follows. By combining the first two equations we obtain
\be
	M_{a,a}=M_{a,a+1}\:\frac{\epsilon_1^a}{\epsilon_2^a}+
		M_{a,a-1}\:\frac{\epsilon_2^{a-1}}{\epsilon_1^{a-1}}\:,
\ee
and using the
relation \cite{Martelli:2023oqk}
\be
	\frac{\epsilon_1^a}{d_{a,a+1}\:\epsilon_2^a}+\frac{\epsilon_2^{a-1}}{d_{a-1,a}\:\epsilon_1^{a-1}}\,=\,
		-\:\frac{d_{a-1,a+1}}{d_{a-1,a}\:\:d_{a,a+1}}\, ,
\ee
we can rewrite this as
\be
\label{epsilon_dependence_LHSvsRHS}
	M_{a,a}\,d_{a,a+1}+M_{a,a-1}\,d_{a-1,a+1}=
		\frac{\epsilon_1^a}{\epsilon_2^a}\,\big(M_{a,a+1}\,d_{a,a+1}-M_{a,a-1}\,d_{a-1,a}\big)\:.
\ee
Given that the fluxes $M_{AB}$ and the orders of the orbifold singularity $d_{a,a+1}$ are just integers,
the only way that this equation can be true for general values of $\epsilon$ is for both sides to vanish.
This implies that $M_{ab}$ is proportional to the intersections $D_{ab}$ given in \eqref{divisors}.
We can then conclude that the only solution to equations (\ref{M_ab_and_B_a_equations}) is \eqref{AdS3xM8_flux_equations_solution}.
Notice that there is just one independent flux associated with the $M_{ab}$ components of the flux matrix. This was to be expected since this corresponds
to the M theory four-form flux on $S^4$.

The values of the remaining entries of the matrix of fluxes $M_{AB}$ are related to the fibration parameters. By substituting $B^{(1)}_a=-\wb\nu_{M5}\,N$ in (\ref{AdS3xM8_flux_equation2}) we find
\be
\label{AdS3xM8_fluxes_result}
	M_{AB}=N\sum_a\frac{\big(\cC_A\,\cC_B\big)|_{a}}{d_{a,a+1}\:\epsilon_1^a\:\epsilon_2^a}
		=N\int_{\Morb_4}\cC_A\,\cC_B=N\sum_{c,d}\fkt_A^c\,\fkt_B^d\,D_{cd}\:.
\ee
In the last step we have used \eqref{divisors} and for convenience we have defined $\fkt_A^c$ as
\be
\label{fktAc}
	\fkt_A^c=
	\begin{cases}
		\delta^c_A\qquad	&A\in\{1,\ldots,d\}\\
		\fkt_c-1\qquad		&A=d+1\\
		-\fkt_c\qquad		&A=d+2
	\end{cases}\:.
\ee
Given that the $\fkt_a$ are integers,  the fluxes $M_{AB}$ in (\ref{AdS3xM8_fluxes_result}) are all integers.

We note that the expression (\ref{AdS3xM8_fluxes_result})  for $M_{AB}$ satisfies the relation required to be considered
a matrix of fluxes,
\be
	\sum_AV_I^AM_{AB}=0\:.
\ee
This can easily be verified by noting that
\be
	\sum_AV_I^A\,\fkt^c_A=
		\begin{cases}
			v_i^c\quad	&I\equiv i=1,2\\
			0\quad		&I=3,4
		\end{cases}
		\:,\qquad\sum_av^c_iD_{cd}=0\:.
\ee

The simplest solution to the equations
\be
	B^{(1)}_a\equiv\frac{\tau_a(\lambda_A(\epsilon_I),\lambda_{AB}(\epsilon_I),\epsilon_I)}
		{\big(\cC_{d+1}\,\cC_{d+2}\big)|_{a}}=-\wb\nu_{M5}\,N
\ee
is to set $\lambda_{d+1,d+2}=-\frac12\,\wb\nu_{M5}N$ while setting all the other $\lambda_A$ and $\lambda_{AB}$ to zero.
We note that in general there exist no solutions to these equations with $\lambda_{AB}=0$ for all $A,B$,
meaning that the inclusion of the higher times to the equivariant volume is necessary.
This stems from the fact that when $\lambda_{AB}=0$ only $d-1$ of the  $\tau_a$ are independent:
using the gauge invariance \eqref{gaugeCY}
\be \lambda_A \rightarrow \lambda_A+\sum_{I=1}^4 \gamma_I (\epsilon_3 V^A_I -\epsilon_I)\, ,\ee
three out of the $d+2$ K\"ahler moduli $\lambda_A$ can be set to zero.

\subsubsection{The extremal function and $c$-extremization}
We are now ready to compute the extremal function
\be F(\epsilon_I)=\evol^{(3)}(\lambda_A(\epsilon_I),\lambda_{AB}(\epsilon_I),\epsilon_I)\, .\ee
The dual field theory is supposed to be the two-dimensional SCFT obtained compactifying on $\Morb_4$ the $(2,0)$ theory living on a stack of $N$ M5 branes. The gravitational extremization problem should correspond to $c$-extremization in the dual two-dimensional SCFT.

A general comment that applies to all the examples in this paper is the following.
The free energy must be extremized with respect to {\it all but one} of the parameters $\epsilon_I$ in order to find the critical point. The value of the remaining parameter must be instead  fixed by requiring the correct scaling of the supercharge under the R-symmetry vector field $\xi$. This is familiar from the constructions in \cite{Martelli:2005tp,Martelli:2006yb,Couzens:2018wnk,Gauntlett:2018dpc}.  In our case,
we extremize with respect to $\epsilon_4$, $\epsilon_1$ and $\epsilon_2$  with $\epsilon_3$ fixed to a canonical value.\footnote{Notice that the free energy is homogeneous of degree two in the parameters $\epsilon_I$, so it makes no sense to extremize with respect to all parameters. The specific numerical value of the equivariant parameter fixed by supersymmetry depends on the setup considered as well as on conventions. In this paper we will not fix the numerical values of this parameter from first principles, but rather we will show that this can be absorbed  by the parameter $\nu$.}

Using relations (\ref{AdS3xM8_homogeneous_components}) and (\ref{Bchange}) we find
\bea\label{AdS3xM8_prescription_result0}
	&F=\sum_a\frac{B^{(3)}_a}{d_{a,a+1}\:\epsilon_1^a\:\epsilon_2^a}\:,\\
	&B^{(3)}_a=\frac16(-\wb\nu_{M5}\,N)^3\,\big((1-\fkt_a)\epsilon_1^a+(1-\fkt_{a+1})\epsilon_2^a-\epsilon_3+\epsilon_4\big)^2
		\big(\fkt_a\epsilon_1^a+\fkt_{a+1}\epsilon_2^a-\epsilon_4\big)^2 \, ,
\eea
which matches the form of  the conjectured formula of \cite{Faedo:2022rqx} in terms of gravitational blocks \cite{Hosseini:2019iad}.\footnote{The convention for the sign of the free energy in \cite{Faedo:2022rqx} is the opposite of ours.}

To  make contact with the dual field theory, we can also write our result in terms of an integral of equivariant forms over the base $\Morb_4$ as follows:
\bea
\label{AdS3xM8_prescription_result}
	F&=-\frac16\,\wb\nu_{M5}^3\,N^3\int_{\Morb_4}\cC_{d+1}^2\,\cC_{d+2}^2 \\
	&=-\frac16\,\wb\nu_{M5}^3\,N^3\int_{\Morb_4}\Big(\epsilon_3-\epsilon_4+\sum_a(1-\fkt_a)c_1^\bT(L_a)\Big)^2
		\Big(\epsilon_4+\sum_a\fkt_ac_1^\bT(L_a)\Big)^2\:.
\eea
This expression correctly reproduces the M5 brane anomaly polynomial integrated over the four-dimensional orbifold $\Morb_4$ as computed in
\cite{Martelli:2023oqk}.\footnote{Attention must be paid when performing the comparison since the symbol  $F$ refers to the central charge here, while it refers to  the integral of the anomaly polynomial in~\cite{Martelli:2023oqk} (see also~\eqref{2danomeqintegralmorb4}).}

Let us briefly review the comparison with field theory, referring to \cite{Martelli:2023oqk} for details.  The anomaly polynomial of the 2d SCFT is obtained by integrating
the eight-form anomaly polynomial of the six-dimensional theory over $\Morb_4$, which, at large $N$, gives
 \begin{equation}\label{intanompolspindleM5}
{\cal A}_{2{\rm d}} =  \int_{\Morb_4}{\cal A}_{6{\rm d}} = \frac{N^3}{24} \int_{\Morb_4} c_1(F_1)^2 c_1(F_2)^2 \, ,
\end{equation}
where $F_I$ are the generators of the $U(1)\times U(1)\subset SO(5)_R$ Cartan subgroup of the $(2,0)$ theory R-symmetry.
The $c_1(F_I)$ can be decomposed as
\begin{equation}
\label{c1FImorb4}
c_1(F_I) = \Delta_I c_1(F_R^{2{\rm d}}) -  \flp_I^a  \Big ( c_1({L}_a) + 2\pi \mu_{a}^i c_1({\cal J}_i)\Big )\, ,
\end{equation}
where  $F_R^{2{\rm d}}$, ${\cal J}_1, {\cal J}_2$ are  line bundles associated with the $2{\rm d}$ R-symmetry and the two global symmetries coming from the isometries of $\Morb_4$. They correspond to background fields for the two-dimensional theory with no legs along $\Morb_4$.
Substituting (\ref{c1FImorb4}) in (\ref{intanompolspindleM5})
and setting  $c_1({\cal J}_i) = \epsilon_i c_1(F_R^{2{\rm d}})$,
leads
to the  equivariant integral
\begin{equation}
\label{2danomeqintegralmorb4}
{\cal A}_{2{\rm d}}  =  \frac{c_r}{6} c_1(F_R^{2{\rm d}})^2= \frac{N^3}{24}   c_1(F_R^{2{\rm d}})^2 \int_{\Morb_4} (\Delta_1 - \flp_1^a  c_1^{\mathbb{T}}(L_a)  )^2  (\Delta_2  - \flp_2^a  c_1^{\mathbb{T}}(L_a) )^2 \, .
\end{equation}
 Preserving supersymmetry with a twist requires $c_1(F_1)+ c_1(F_2)= 2 c_1(F_R^{2{\rm d}}) -\sum_a c_1(L_a)$
which gives \cite{Martelli:2023oqk}
  \be
\Delta_1 + \Delta_2 = 2  +\det (W,\epsilon)\, , \qquad \flp_1^a + \flp_2^a = 1 + \det (W,v^a)   \, ,
\ee
 where $\epsilon = (\epsilon_1,\epsilon_2)$ and $W\in \RR^2$ is a two-dimensional constant vector.\footnote{$W$ can be gauged away, see \cite{Martelli:2023oqk}.}  The two-dimensional central charge $c_r$ is extracted  from \eqref{2danomeqintegralmorb4} and should be extremized with respect to $\epsilon_i$ and $\Delta_I$ subject to the previous constraint. We then see that the extremization of the gravitational free energy is equivalent to $c$-extremization under the identifications
  \be \Delta_1 = \epsilon_4\, ,\qquad \Delta_2 = \epsilon_3-\epsilon_4\, ,\qquad  \flp_1^a= \flt_a\,,\qquad  \flp_2^a=1- \flt_a\, ,\qquad W=0 \, ,\ee
  where we set $\epsilon_3=2$ for convenience. The free energy $F$ is actually homogeneous of degree two in $\epsilon_I$. To match the free energy with the central charge
  we have to set  $\epsilon_3^2\wb\nu_{M5}^3=-6$.   The case of anti-twist is similar and can be discussed by taking a non-convex fan for $\Morb_4$. The most general supersymmetry condition is now
 $c_1(F_1)+ c_1(F_2)= 2 c_1(F_R^{2{\rm d}}) -\sum_a \sigma^a c_1(L_a)$  where $\sigma_a=\pm 1$ as discussed in \cite{Faedo:2022rqx} and requires
   \be
\Delta_1 + \Delta_2 = 2  +\det (W,\epsilon)\, , \qquad \flp_1^a + \flp_2^a = \sigma_a + \det (W,v^a)   \, .
\ee
This case can be just obtained by {\it formally} sending $v^a\rightarrow \sigma^a v^a$ everywhere, implying $\epsilon_1^a\rightarrow \sigma^a \epsilon_1^a$ and
$\epsilon_2^a\rightarrow \sigma^{a+1} \epsilon_2^a$.

\subsection{AdS$_5\times M_6$ solutions}\label{sec:M5}

In this section we consider a generalization of the family  of M theory solutions found in \cite{Gauntlett:2004zh} and further studied in \cite{Gauntlett:2006ai}. Their geometry  is AdS$_5\times M_6$ where $M_6$ is a manifold obtained as a
 $\mathbb{P}^1$ bundle over
a four-dimensional compact manifold $B_4$, that can be either a K\"ahler-Einstein  manifold ($B_4=\mathrm{KE}_4$) or the product of two KE$_2$  ($B_4=\Sigma_1\times \Sigma_2)$. The bundle is the projectivization of the
 canonical bundle over $B_4$,  $\mathbb{P}( K \oplus \cal O)$.
Here we consider the  case where  $B_4$ is replaced by a generic four-dimensional toric orbifold $\Morb_4$. Notice that generically $M_6$ can be an orbifold,\footnote{Using our formalism,
we could easily study the case that $M_6$ is a generic toric six-dimensional orbifold. It would be interesting to understand what kinds of orbifold admit an holographic interpretation.} like in the solutions discussed in \cite{Ferrero:2021wvk}.
In addition to recovering the gravitational central charges of the existing solutions, we give a prediction for these more general backgrounds that  are still to be found.  These solutions are potentially interpreted as M5 branes wrapped over a two-cycle in $\Morb_4$ (see for example \cite{Bah:2019rgq,Bah:2019vmq}).

The topological structure of the underlying geometry can be encoded in the fan
\bea\label{fanM6} V^a=(v^a,1)\, ,\qquad  V^{d+1}=(0,0,1)\, ,\qquad  V^{d+2}=(0,0,-1)\, , \qquad a=1, \ldots, d \, ,
\eea
where $v^a$ are the two-dimensional vectors in the fan of $\Morb_4$.
We will use a capital  index $A$ to run over $a=1,\ldots, d$, $d+1$ and $d+2$.
That this is the right  geometry   can be seen  looking at the symplectic reduction presentation
$\bC^{d+2}//G$ of $M_6$. Here $G$ is the subgroup of the torus $\bT^{d+2}=U(1)^{d+2}$ generated
by the GLSM charges
\be \sum_A Q_A^k V^A_I=0 \, ,\qquad \qquad k=1,\ldots,d-1 \, . \ee
We can choose the following basis of GLSM charges
\be (q_a^p, -\sum_a q_a^p,0) \, , \qquad (0,\ldots, 0, 1,1) \, ,\ee
where $q_a^p$ are the $d-2$ charges for $\Morb_4$, $\sum_a q_a^p v^a_i=0$. The first $d-2$ vectors define the canonical bundle $K$ of $\Morb_4$ with an extra copy of $\mathbb{C}$. The final charge vector projectivizes it and gives indeed the geometry we are interested in:
\be \mathbb{P}( K \oplus \cal O) \, . \ee

We need also to specify  the integer fluxes of the M theory four-form along all the non-trivial four-cycles. There are $d+2$ toric four-cycles in the geometry, associated with the divisors $D_A$. The divisors $D_a$  are $\bP^1$ fibrations over the toric two-cycles $\Sigma_a\subset \Morb_4$, while $D_{d+1}$ and $D_{d+2}$ are copies of $\Morb_4$ sitting at the North and South pole of $\bP^1$, respectively.  All together,  they define a vector of fluxes $M_A$.   The
relations $\sum_A V^A_I D_A=0$ imply that not all toric divisors are inequivalent and that the vector of fluxes satisfies
\be \label{fluxc}\sum_A V^A_I M_{A}=0\, .\ee

Since we are dealing with M5 branes in M theory,
we define the free energy  as in section \ref{AdS3xM8}
\be F=\evol^{(3)}(\lambda_A,\lambda_{AB},\epsilon_I) \, ,\ee
and, since now we have a vector of fluxes, we  impose the flux constraints
\be
\label{AdS5xM6_prescription}
	\nu_{M5}\,M_{A}=-
		\frac{\partial}{\partial\lambda_{A}}\evol^{(2)}(\lambda_A,\lambda_{AB},\epsilon_I)\, .
\ee

Differently from the case discussed in the previous section, for these geometries there is no general field theory result for the central charge of the dual four-dimensional SCFTs. Our results here can therefore be seen as a prediction for the general form
of the off-shell central charge, which presumably can be obtained by integrating the M5 brane anomaly polynomial on a suitable two-cycle inside $M_6$, or using the method of \cite{Bah:2019rgq}. In order to compare with the existing literature, we will therefore consider in some detail a number of explicit examples of $\Morb_4$, including KE$_4$ and $\Sigma_1\times \Sigma_2$, but also other examples for which there is no known supergravity solution, nor field-theoretic understanding. The equations to be solved in the extremization problem typically lead
 to finding the zeroes
 of simultaneous polynomials of high degree and are therefore not manageable. For this reason, we will proceed by making different technical assumptions to simplify the algebra. One such general assumption is the existence of a $\ZZ_2$ symmetry acting on the $\mathbb{P}^1$ fibre, as we discuss below. Furthermore, we will occasionally restrict to non-generic fluxes in order to simplify the otherwise unwieldy expressions.

If we restrict to a class of geometries with a $\mathbb{Z}_2$ symmetry that exchanges the North and South poles of $\bP^1$, we can consider a simplified geometry obtained by cutting $\bP^1$ into half. We thus obtain a non-compact  Calabi-Yau geometry given by the canonical bundle over  $\Morb_4$. The corresponding  fan is obtained by dropping $V^{d+2}$:
\bea\label{halffan} V^a=(v^a,1)\, ,\qquad  V^{d+1}=(0,0,1)\ \, , \qquad a=1,\ldots,d \,.
\eea
Notice that this is a (partial) resolution of a CY$_3$ cone where  $V^{d+1}$ is associated with a compact divisor. Supergravity solutions with $\mathbb{Z}_2$ symmetry have been considered in \cite{Gauntlett:2004zh,Gauntlett:2006ai} where they correspond to set the parameter called $c$ to zero. Effectively, the $\mathbb{Z}_2$ symmetry reduces by one the number of independent fluxes we can turn on, thus simplifying the calculations. Notice that the on-shell equivariant volume $\evol$ for the half-geometry is half of the one for the total geometry.
The relation between the parameters to use in the two cases, in order to have the same normalization for the free energy, is the following
\be\label{half-full} \wb\nu_{M5} = 2^{-2/3} \nu_{M5} \, ,\ee
where $\wb\nu_{M5}$ is the correct one for half-geometries.

Notice that we introduced single and double times in \eqref{AdS5xM6_prescription}. We can immediately understand the need for higher times. In a compact geometry, $\evol^{(2)}(\lambda_A)$ with only single times would vanish identically.\footnote{For a compact geometry $\evol^{(2)}(\lambda_A)=-\frac 12 \sum_{AB} \lambda_A\lambda_B\int_{M_6} c_1^\bT(L_A) c_1^\bT(L_B) =0$ since it is the integral of a four-form at most on a six-dimensional manifold. In the non-compact case, this condition is evaded and $\evol^{(2)}(\lambda_A)$ is a rational function of $\epsilon_I$. See \cite{Martelli:2023oqk} for details.}
As we will discuss later, the double times are generically necessary also when imposing the $\mathbb{Z}_2$ symmetry in order to have enough parameters to solve the equations.\footnote{In the case of compactification on a spindle they are not necessary~\cite{Martelli:2023oqk}.}

\subsubsection{Geometries with $\mathbb{Z}_2$ symmetry}\label{Z2}

We consider first  geometries with $\mathbb{Z}_2$ symmetry. Cutting $M_6$ into half we consider the non-compact CY$_3$ specified by the fan \eqref{halffan}.
The $I=3$ condition in \eqref{fluxc} gives
\be  M_{d+1} = - \sum_a M_a \, ,\ee
thus fixing the flux along $\Morb_4$ in terms of the other fluxes. The $I=1,2$ conditions in \eqref{fluxc}  give two linear relations among the $M_a$, leaving a total number $d-2$ of independent fluxes. Notice that geometries without $\mathbb{Z}_2$ symmetry
have one additional independent flux, as we discuss later.

The fan is the union of $d$ cones $(V^a,V^{a+1},V^{d+1})$ and we see that the number of fixed points is the same of that of the base $\Morb_4$. It is then easy to write the equivariant volume with higher times as a sum over the fixed points of $\Morb_4$
\be \evol = \sum_a \frac{\ex^{\tau_a}}{d_{a,a+1}\epsilon^a_1 \epsilon^a_2(\epsilon_3-\epsilon^a_1-\epsilon^a_2)} \, ,\ee
where
\be \tau_a= \left(\sum_A \lambda_A\,c_1^{\mathbb{T}^3}(L_A)+\sum_{A,B} \lambda_{AB}\, c_1^{\mathbb{T}^3}(L_A)\,c_1^{\mathbb{T}^3}(L_B) \right )\Bigg|_{ a} \, \ee
or, more explicitly,
\bea \tau_a= &-\lambda_a \epsilon_1^a -\lambda_{a+1} \epsilon_2^a -\lambda_{d+1} (\epsilon_3-\epsilon^a_1-\epsilon^a_2) +\lambda_{aa} (\epsilon_1^a)^2 + 2 \lambda_{a,a+1} \epsilon_1^a \epsilon_2^a + \lambda_{a+1,a+1} (\epsilon_2^a)^2 \\
&+2 (\lambda_{a,d+1} \epsilon_1^a +\lambda_{a+1,d+1} \epsilon_2^a)(\epsilon_3-\epsilon^a_1-\epsilon^a_2) +\lambda_{d+1,d+1}(\epsilon_3-\epsilon^a_1-\epsilon^a_2)^2 \, .\eea

Notice that the equations \eqref{AdS5xM6_prescription} are not solvable with only single times.  $M_{d+1}=-\sum_a M_a \ne 0$ while, for $\lambda_{AB}=0$,
\bea -\frac{\partial \evol^{(2)}}{\partial \lambda_{d+1}}&= \sum_a \frac{-\lambda_a \epsilon_1^a -\lambda_{a+1} \epsilon_2^a -\lambda_{d+1} (\epsilon_3-\epsilon^a_1-\epsilon^a_2)}{d_{a,a+1}\epsilon^a_1 \epsilon^a_2} \\ &=  \int_{\Morb_4} \left(\sum_a \lambda_a c_1^{\mathbb{T}}(L_a) -\lambda_{d+1} \Big(\epsilon_3 +\sum_a  c_1^{\mathbb{T}}(L_a)\Big)\right) =0\, ,\eea
being the integral of a two-form at most.

The equations \eqref{AdS5xM6_prescription} explicitly read
\begin{align}
	(I) \qquad  &\wb\nu_{M5} M_a = \frac{ \epsilon_1^a \tau_a}{d_{a,a+1}\epsilon^a_1 \epsilon^a_2(\epsilon_3-\epsilon^a_1-\epsilon^a_2)} +\frac{ \epsilon_2^{a-1} \tau_{a-1}}{d_{a-1,a}\epsilon^{a-1}_1 \epsilon^{a-1}_2(\epsilon_3-\epsilon^{a-1}_1-\epsilon^{a-1}_2)} \,, \nn \\
	(II) \qquad &- \wb\nu_{M5}\sum_a M_a = \sum_a \frac{ \tau_a}{d_{a,a+1}\epsilon^a_1 \epsilon^a_2} \,.
\end{align}
These equations are not independent. In particular, $(II)$ follows from $(I)$.\footnote{Using $\sum_a v^a M_a=0$ and the vector identity $v^a_i \epsilon^a_1+v^{a+1}_i \epsilon^a_2=\epsilon_i$, one derives $\epsilon_i \sum_a \frac{\tau_a}{d_{a,a+1}\epsilon^a_1 \epsilon^a_2(\epsilon_3-\epsilon^a_1-\epsilon^a_2)}=0$ from $(I)$. Then, summing over $a$ in $(I)$, and using the previous identity:
\be \wb\nu_{M5} \sum_a M_a = \sum_a \frac{(\epsilon^a_1+\epsilon^a_2)\tau_a}{d_{a,a+1}\epsilon^a_1 \epsilon^a_2(\epsilon_3-\epsilon^a_1-\epsilon^a_2)} = - \sum_a \frac{\tau_a}{d_{a,a+1}\epsilon^a_1 \epsilon^a_2}   \, ,\ee
valid for $\epsilon_i\ne 0$. For $\epsilon_i= 0$ one should  pay more attention and we will see  in section \ref{subsec:AdS4xM6} one instance where a similar subtlety is important. In the present case we will check explicitly that both $(I)$ and $(II)$ are valid.} The equations $(I)$ can be written as
\be
	B_{a-1}^{(1)}-B_{a}^{(1)} = d_{a,a+1} \epsilon_2^a \wb\nu_{M5}  M_a\, ,  \qquad\qquad  B_a^{(1)}=-\frac{\tau_a}{\epsilon_3-\epsilon^a_1-\epsilon^a_2} \, .
\ee

It is then clear that these equations can be solved for $\tau_a$, but one ``time'', say $\tau_1$, remains undetermined.
Our prescription is  to {\it extremize} the free energy with respect to {\it all} parameters that are left undetermined after imposing the flux constraints. In this case then we extremize
\be \evol^{(3)} (\epsilon_i, \tau_1) \ee
with respect to $\epsilon_1$, $\epsilon_2$ and $\tau_1$, with $\epsilon_3$ set to some canonical value, fixed by the scaling of the supercharge under the R-symmetry vector field.
In the next subsection we will parameterize the free energy in a more convenient way.

\subsubsection{The extremal function for geometries with $\bZ_2$ symmetry}\label{genorb}

We can write the general form of the  extremal function for geometries with $\bZ_2$ symmetry.  Let us define
\be \tau^{\mathbb{T}^3}_{CY_3}=\sum_A \lambda_A\,c_1^{\mathbb{T}^3}(L_A)+\sum_{A,B} \lambda_{AB}\,c_1^{\mathbb{T}^3}(L_A)\,c_1^{\mathbb{T}^3}(L_B) \, ,\ee
the equivariant form with restriction $\tau_a$ at the fixed points. By  restricting the form to $\Morb_4$ and  considering $\epsilon_3$ as a parameter, we obtain
\bea
\label{tau_for_CY3}
	&\tau^\bT=\sum_A\lambda_A\:\cC_A+\sum_{A,B}\lambda_{AB}\:\cC_A\:\cC_B\:,\\
	&\cC_a\,=c_1^\bT(L_a)\:,\qquad a=1,\ldots,d\:,\\[2mm]
	&\cC_{d+1}\,=\,-\Bigl(\epsilon_3+\sum_ac_1^\bT(L_a)\Bigr)\:,
		\eea
where $c_1^{\mathbb{T}}(L_a)$ are the restrictions of the line bundles $L_a$ to the base $\Morb_4$ and $\mathbb{T}$ is the two-dimensional torus spanned by $\epsilon_1$ and $\epsilon_2$. From now on, unless explicitly said, all classes will refer to the base $\Morb_4$. In terms of $\tau^\bT$ the quadratic piece of the equivariant volume can be written as
\be \evol^{(2)}(\lambda_A,\lambda_{AB},\epsilon_I) = \frac 12 \int_{\Morb_4} \frac{(\tau^{\mathbb{T}})^2}{\epsilon_3 +\sum_a c_1^{\mathbb{T}}(L_a)}  \, .\ee
The flux constraints \eqref{AdS5xM6_prescription} give
\bea (I) \qquad  &- \wb\nu_{M5} M_a = \int_{\Morb_4} \frac{c_1^{\mathbb{T}}(L_a) \, \tau^{\mathbb{T}}}{\epsilon_3 +\sum_a c_1^{\mathbb{T}}(L_a)}\, , \\
(II) \qquad &- \wb\nu_{M5} \sum_a M_a = \int_{\Morb_4}\tau^{\mathbb{T}}\, .\eea
For a generic fan, using the gauge transformations \eqref{gauge} and \eqref{gaugeCY} we can set all $\lambda_a=\lambda_{a,a}=\lambda_{a,a+1}=0$.\footnote{For special symmetric fans, like $\bP^2$ and $\bP^1\times \bP^1$, and other simple examples with low $d$, one of the single times $\lambda_a$ remains unfixed. However, from the combined conditions $(I)+(II)$ we obtain $\sum_{ab}\lambda_a \int_{M_4} \frac{c_1^{\mathbb{T}}(L_a) \, c_1^{\mathbb{T}}(L_b)}{\epsilon_3 +\sum_a c_1^{\mathbb{T}}(L_a)}=0$ which implies that the remaining single time must vanish.}
We will show more formally in appendix \ref{app:addendum} that $\evol^{(3)}$ has a critical point at $\lambda_a=\lambda_{a,a}=\lambda_{a,a+1}=0$.
Then condition $(I)$ becomes
\begin{align}
	-\wb\nu_{M5} M_a &= \int_{\Morb_4} c_1^{\mathbb{T}}(L_a) \Bigl(- \lambda_{d+1} - 2 \sum_b \lambda_{b,d+1} c_1^{\mathbb{T}}(L_b)+\lambda_{d+1,d+1} \Bigl(\epsilon_3 +\sum_b c_1^{\mathbb{T}}(L_b)\Bigr)\Bigr) \nonumber \\
	&=  \sum_b D_{ab} (-2 \lambda_{b,d+1} +\lambda_{d+1,d+1}) \, .
\end{align}
We can similarly compute $(II)$ as an integral
\be - \wb\nu_{M5} \sum_a M_a = \int_{\Morb_4}\tau^{\mathbb{T}}=  \sum_{a,b} D_{ab} (-2 \lambda_{b,d+1} +\lambda_{d+1,d+1}) \, ,\ee
and see that it is automatically satisfied if $(I)$ is. Since  $\sum_a v^a_i M_a=0$, the flux constraints fix the $\lambda_{b,d+1}$ only up to the
 ambiguities
\bea \label{amb} &\lambda_{a,d+1}\rightarrow  \lambda_{a,d+1} +\sum_{i=1}^2 \delta_i v^a_i + \gamma\, ,
\\ &\lambda_{d+1,d+1}\rightarrow  \lambda_{d+1,d+1}+2\gamma \, ,
\eea
where $\delta_i$ and $\gamma$ are free parameters. However, these free parameters can be all reabsorbed in a redefinition of \be \lambda_{d+1}\rightarrow \lambda_{d+1} +2 \gamma \epsilon_3 +2 \sum_{i=1}^2 \delta_i \epsilon_i\, ,\ee
since $\sum_a v^a_i c_1^{\mathbb{T}}(L_a)=-\epsilon_i$ and they are not really independent.

The free energy is then given by
\be \evol^{(3)} =\frac16  \int_{\Morb_4}\frac{(\tau^{\mathbb{T}})^3}{\epsilon_3 +\sum_a c_1^{\mathbb{T}}(L_a)} \, ,\ee
which explicitly gives
\be\label{V3} \evol^{(3)} =\frac16  \int_{\Morb_4}\Bigl(\epsilon_3 +\sum_a c_1^{\mathbb{T}}(L_a)\Bigr)^2 \Bigl( \bar\lambda_{d+1} + \sum_b \bar \lambda_{b,d+1} c_1^{\mathbb{T}}(L_b)\Bigr)^3 \, ,\ee
where we defined
 \be \bar \lambda_{d+1}= -\lambda_{d+1}+\lambda_{d+1,d+1} \epsilon_3 \, , \qquad \bar\lambda_{a,d+1}=-2 \lambda_{a,d+1}+\lambda_{d+1,d+1} \, ,\ee
 which are subject to the constraints
 \be\label{constronlabar}
  -\wb\nu_{M5} M_a =   \sum_b D_{ab} \bar \lambda_{b,d+1} \, .\ee

Substituting the solution of the flux constraints, $\evol^{(3)}$ becomes a function of $\epsilon_i$ and the extra parameter  $\bar \lambda_{d+1}$. Indeed, as we have seen, the ambiguities \eqref{amb} can be reabsorbed in a redefinition of $\bar\lambda_{d+1}$.
A direct evaluation gives
 \begin{align} \label{genevol3Z2}
	 & 6 \evol^{(3)} = \bar \lambda_{d+1}^3 \sum_{ab} D_{ab} + \bar\lambda_{d+1}^2 \Bigl(6 \epsilon_3 \sum_{ab} D_{ab}\bar\lambda_{a,d+1}  + 3 \sum_{abc} D_{abc}\bar\lambda_{a,d+1} \Bigr) \nonumber \\
	 & \quad + 3 \bar \lambda_{d+1} \Bigl(\epsilon_3^2 \sum_{ab} D_{ab}\bar\lambda_{a,d+1} \bar\lambda_{b,d+1}  + 2 \epsilon_3 \! \sum_{abc} D_{abc}\bar\lambda_{a,d+1} \bar\lambda_{b,d+1} + \sum_{abcd} D_{abcd}\bar\lambda_{a,d+1}\bar\lambda_{b,d+1}\Bigr) \nonumber \\
	 & \quad + \Bigl(\epsilon_3^2 \sum_{abc} D_{abc}\bar\lambda_{a,d+1}\bar\lambda_{b,d+1} \bar\lambda_{c,d+1}+2 \epsilon_3 \sum_{abcd} D_{abcd}\bar\lambda_{a,d+1}\bar\lambda_{b,d+1} \bar\lambda_{c,d+1} \nonumber \\
	 & \quad +\sum_{abcde} D_{abcde}\bar\lambda_{a,d+1}\bar\lambda_{b,d+1} \bar\lambda_{c,d+1}\Bigr) \,,
\end{align}
where the generalized intersection numbers are defined by
\be\label{multint} D_{a_1\ldots a_p} =\int_{\Morb_4} c_1^{\mathbb{T}}(L_{a_1}) \ldots c_1^{\mathbb{T}}(L_{a_p}) \, .\ee
Notice that $D_{ab}$ is $\epsilon$-independent, while $D_{a_1\cdots a_p}$ is a homogeneous function of degree $p-2$ in $\epsilon_1$ and $\epsilon_2$.
$\evol^{(3)}$ need to be extremized with respect to $\epsilon_1$, $\epsilon_2$ and  $\bar \lambda_{d+1}$, with $\epsilon_3$ set to the canonical value.

The critical point is generically at a non-zero value of  $\epsilon_1$ and $\epsilon_2$.  We can expect a critical
 point\footnote{In the opposite direction, of course one would have as critical point
$\epsilon_1=\epsilon_2=0$ if the base $B_4$ has \emph{no} continuous symmetries. This is the case for examples for del Pezzo surfaces dP$_k$ with $k>3$, which we do not treat here. This would lead one to suspect that all KE$_4$ have $\epsilon_1=\epsilon_2=0$ as critical point, but this is actually incorrect, as the example of the toric dP$_3$ will show.}
  at $\epsilon_1=\epsilon_2=0$ only if the background and the fluxes have some extra symmetry, as for examples in the cases where all $U(1)$ isometries are enhanced to a non-abelian  group.
In these particular cases,  we can further simplify the expression
 \bea \label{extrKE} 6 \evol^{(3)} &= \bar \lambda_{d+1}^3 \sum_{ab} D_{ab} + 6 \epsilon_3 \bar\lambda_{d+1}^2  \sum_{ab} D_{ab}\bar\lambda_{a,d+1} +
3 \epsilon_3^2 \bar \lambda_{d+1}  \sum_{ab} D_{ab}\bar\lambda_{a,d+1} \bar\lambda_{b,d+1}  +O(\epsilon_i^2) \\
&= \bar \lambda_{d+1}^3 \sum_{ab} D_{ab} - 6\wb\nu_{M5} \sum_a M_a \epsilon_3 \bar\lambda_{d+1}^2  +
3 \epsilon_3^2 \bar \lambda_{d+1}  \sum_{ab} D_{ab}\bar\lambda_{a,d+1} \bar\lambda_{b,d+1}  +O(\epsilon_i^2)\, ,\eea
and extremize it with respect to $\bar\lambda_{d+1}$.

As a check of our expression, we can reproduce the central charge of the existing solutions with K\"ahler-Einstein metrics and fluxes all equal  \cite{Gauntlett:2006ai}. The only toric four-manifolds that are also KE are $\mathbb{P}^2$, $\mathbb{P}^1\times\mathbb{P}^1$ and dP$_3$, with fans
\bea  \label{fanKE} &\mathbb{P}^2: \,\,\,\, v^1=(1,1)\, ,\, v^2=(-1,0)\, ,\, v^3=(0,-1)\, ,\\
 &\mathbb{P}^1\times\mathbb{P}^1: \,\, v^1=(1,0)\, ,\, v^2=(0,1)\, ,\, v^3=(-1,0)\, , \,v^4=(0,-1)\, ,\\
&\text{dP}_3: \, v^1=(1,0)\, ,\, v^2=(1,1)\, ,\, v^3=(0,1)\, ,\, v^4=(-1,0) \, ,\, v^5=(-1,-1)\, ,\, v^6=(0,-1)\, ,\eea
and intersection matrices
\begin{equation} \label{intersKE}
	\begin{alignedat}{2}
		&\mathbb{P}^2:  && D_{ab}=1\, ,  \\
		&\mathbb{P}^1\times\mathbb{P}^1: \quad && D_{ab}=1 \; \text{if} \; |a-b|=1 \, (\text{mod}\, 2) \quad \text{and} \quad \text{zero otherwise}\,, \\
		&\mathrm{dP}_3:  && D_{aa}=-1\, , \quad D_{a,a\pm 1}=1 \quad \text{and} \quad \text{zero otherwise}\, ,
	\end{alignedat}
\end{equation}
where the indices are cyclically identified. To compare with the KE$_4$ solutions, we set all $M_a\equiv N$. We can then choose all $\bar\lambda_{a,d+1}$  equal and we find
\be \label{zero_epsilon_proof}\sum_{abc} D_{abc}\bar\lambda_{a,d+1}=\sum_{abc} D_{abc}\bar\lambda_{a,d+1}\bar\lambda_{b,d+1} =\sum_{abc} D_{abc}\bar\lambda_{a,d+1}\bar\lambda_{b,d+1} \bar\lambda_{c,d+1} =0 \, ,\ee
thus ensuring that the linear terms  in $\epsilon_1$ and $\epsilon_2$ in $\evol^{(3)} $ vanish, and that  there is indeed a critical point  at $\epsilon_1=\epsilon_2=0$. Extremizing
\eqref{extrKE} we get
\be \label{freeKE}\evol^{(3)}= \epsilon_3^3 \wb\nu_{M5}^3(-5+3 \sqrt{3})  N^3 \{ \frac19 , \frac13, 2 \} \, ,\ee
 for $\mathbb{P}^2$, $\mathbb{P}^1\times\mathbb{P}^1$ and dP$_3$, respectively, which agrees with (2.16)  in \cite{Gauntlett:2006ai} for $\epsilon_3 \wb\nu_{M5} =3$.\footnote{$N_{C_N}$ in \cite{Gauntlett:2006ai}  can be identified with $M_{d+1}=-\sum M_a$, so that $N_{there}=-h N_{C_N}/M = h \sum M_a/M$ where $(h,M)$ are defined in \cite{Gauntlett:2006ai} and they have value $(3,9),(4,8)$ and $(2,6)$ for $\bP^2$, $\bP^1\times \bP^1$ and dP$_3$, respectively.}

 For $\mathbb{P}^1\times \mathbb{P}^1$ we can introduce a second flux. The general solution to $\sum_A V^A_I M_A=0$ is indeed
\be
	M_A = (N_1,N_2,N_1,N_2,-2N_1-2N_2) \,.
\ee
The background has an expected $SU(2)\times SU(2)$ symmetry that it is realized in the supergravity solution  \cite{Gauntlett:2006ai}, that now is not in the KE class.
Using the gauge transformation \eqref{gauge} we can set $\bar\lambda_{a+2,d+1}=\bar\lambda_{a,d+1}$. A simple computation then shows
that the  free energy extremized has a critical point in  $\epsilon_1=\epsilon_2=0$, consistently with the non-abelian isometry of the solution, with critical value
\be\label{P12} \evol^{(3)}=\frac{\epsilon_3^3 \wb\nu_{M5}^3}{6}\left ( 2(N_1^2+N_1 N_2 + N_2^2)^{3/2} - (2 N_1^3 +3 N_1^2 N_2+ 3 N_1 N_2^2 +2 N_2^3)\right ) \, ,\ee
which should be compared with (2.29) in \cite{Gauntlett:2006ai} with $N_1=p N$ and $N_2= q N$. This looks superficially different, but it can be rewritten in the form above (\emph{cf}.\ for example (F.14) in \cite{Bah:2019rgq}).

 \subsubsection{Examples of geometries with non-zero critical $\epsilon$}
 \label{nozeroeps:sec}

 So far, in all the explicit examples we have discussed we found that  $\epsilon_1=\epsilon_2=0$ is a critical point. However, we have already pointed out that for generic toric $\Morb_4$
 and/or with generic fluxes  this will not be the case.
 In this subsection we will investigate situations in which at least one of  $\epsilon_1,\epsilon_2$ is different from zero at the critical point, by considering geometries with $SU(2)\times U(1)$ symmetry, as well as the case of dP$_3$ with generic fluxes.
 Interestingly, it turns out that for dP$_3$ there exist two  special configurations of  fluxes (different from the case where they are all equal)
   where  the critical point is again $\epsilon_1=\epsilon_2=0$, but the corresponding supergravity solutions are not known. For four independent generic
  fluxes, instead, $\epsilon_1=\epsilon_2=0$ is not a critical point.

\paragraph*{dP$_3$ with unequal fluxes.}

The symmetry of dP$_3$ is  just $U(1)\times U(1)$ and the existence of the critical point $\epsilon_1=\epsilon_2=0$ of the extremization problem
is not obviously implied by the fact that there exists a KE metric on dP$_3$. In the basis of the fan as in  (\ref{fanKE}),  the general assignment of fluxes
compatible with  $\sum_A V_I^A M_A=0$ can be parameterized as
\bea
 M_A=(N_1,N_2,N_3,N_4,N_5,N_6, -2 N_1 - 3 N_2-2N_3 +N_5)\, ,
\eea
where we choose $N_1,N_2,N_3,N_5$ as independent, with  $N_4 = N_1 + N_2 - N_5$  and $N_6 = N_2 + N_3 - N_5$. Upon setting
$\lambda_a = \lambda_{a,a} = \lambda_{a,a+1} = 0$ using the gauge freedom, as discussed before, the constraint
(\ref{constronlabar}) on 	$\bar{\lambda}_{a,d+1}$ can be  solved by taking, for example
\begin{equation} \label{dP3genlambdas}
	\begin{aligned}
		\bar{\lambda}_{1,d+1} & = -\bar{\nu}_{M5} \frac{N_2 + N_3}{2} \,,  \qquad  & \bar{\lambda}_{2,d+1} & = -\bar{\nu}_{M5} \frac{N_3 + N_1}{2} \,, \\
		\bar{\lambda}_{3,d+1} &= -\bar{\nu}_{M5} \frac{N_1 + N_2}{2} \,,  \qquad  & \bar{\lambda}_{4,d+1} &= -\bar{\nu}_{M5} \frac{N_2 + N_3}{2} \,, \\
		\bar{\lambda}_{5,d+1} &= -\bar{\nu}_{M5} \frac{N_3 + N_1 + 2(N_2 - N_5)}{2} \,,  \qquad  & \bar{\lambda}_{6,d+1} &= -\bar{\nu}_{M5} \frac{N_1 + N_2}{2} \,.
	\end{aligned}
\end{equation}

Writing out the free energy \eqref{genevol3Z2}, up to linear order in $\epsilon_1,\epsilon_2$, we have
\be
\evol^{(3)}  = \left. \evol^{(3)} \right|_{\epsilon_i=0}+  \left.\partial_{\epsilon_1}\evol^{(3)} \right|_{\epsilon_i=0} \epsilon_1+ \left. \partial_{\epsilon_2}\evol^{(3)} \right|_{\epsilon_i=0} \epsilon_2+ O(\epsilon_i^2)
\ee
where the constant term is not particularly interesting
and
\begin{equation}
\label{lineartermsinV3}
	\begin{split}
		 \left.\partial_{\epsilon_1}\evol^{(3)} \right|_{\epsilon_i=0} &= \bar{\nu}_{M5} \frac{N_5 - N_2}{2} \Bigl[ 6\bar{\lambda}_{d+1}^2 - 12(N_2 + N_3) \bar{\nu}_{M5} \epsilon_3 \bar{\lambda}_{d+1} \\
		 & + \bigl( N_2^2 - 2N_5^2 + 3N_3 (N_2 + N_5) + N_5 N_2 \bigr) \bar{\nu}_{M5}^2 \epsilon_3^2 \Bigr]  \,, \\
		 \left.\partial_{\epsilon_2}\evol^{(3)} \right|_{\epsilon_i=0} &= \bar{\nu}_{M5} \frac{N_5 - N_2}{2} \Bigl[ 6\bar{\lambda}_{d+1}^2 - 12(N_2 + N_1) \bar{\nu}_{M5} \epsilon_3 \bar{\lambda}_{d+1} \\
		 & + \bigl( N_2^2 - 2N_5^2 + 3N_1 (N_2 + N_5) + N_5 N_2 \bigr) \bar{\nu}_{M5}^2 \epsilon_3^2 \Bigr] \,.
	\end{split}
\end{equation}

We see that for generic values of the fluxes   the expressions above cannot  be zero simultaneously, implying that
$\epsilon_i=0$ is not a critical point of the extremization.
The  complete extremization equations
 are unwieldy, so in the following we will instead concentrate on two special configurations of fluxes, with enhanced symmetry, for which  $\epsilon_1=\epsilon_2=0$ turns out to be a critical point.

 The first  special value of fluxes is clearly obtained for $N_5=N_2$, that leaves three fluxes $N_1,N_2,N_3$ free.
In this case the parameters (\ref{dP3genlambdas}) acquire
the cyclic symmetry $\bar{\lambda}_{a,d+1} = \bar{\lambda}_{a+3,d+1}$, analogously to the  $\bP^1\times \bP^1$ discussed in the previous section
 and indeed the linear terms in $\evol^{(3)}$ manifestly vanish, so that $\epsilon_i=0$ is a critical point.
 The fluxes display an enhanced symmetry:
\bea
 M_A=(N_1,N_2,N_3,N_1,N_2,N_3, -2 N_1 - 2 N_2-2N_3)\, .
\eea
 Extremizing $\mathbb{V}^{(3)}$ with respect to $\bar{\lambda}_{d+1}$
yields
\begin{equation}
	\begin{split}
		\bar{\lambda}_{d+1}^* &= \frac{2\bar{\nu}_{M5} \epsilon_3}{3} (N_1 + N_2 + N_3) \\
		& - \frac{\bar{\nu}_{M5} \epsilon_3}{3} \sqrt{4 (N_1^2 + N_2^2 + N_3^2) + 5 (N_1 N_2 + N_2 N_3 + N_3 N_1)}  \,,
	\end{split}
\end{equation}
and the corresponding value of the on-shell central charge is
\begin{equation}
\label{central_dP3}
	\begin{split}
	\evol^{(3)} &= \frac{2\bar{\nu}_{M5}^3 \epsilon_3^3}{27} \Bigl[ \bigl( 4 (N_1^2 + N_2^2 + N_3^2) + 5 (N_1 N_2 + N_2 N_3 + N_3 N_1) \bigr)^{3/2} \\
		& - (N_1 + N_2 + N_3) \bigl( 8 (N_1^2 + N_2^2 + N_3^2)+ 7 (N_1 N_2 + N_2 N_3 + N_3 N_1) \bigr) \Bigr] \,.
	\end{split}
\end{equation}
It can be checked that this expression agrees precisely with  the central charge given in eq.~(3.79) of \cite{BenettiGenolini:2023ndb} and  it correctly reduces to (\ref{freeKE}) upon setting $N_1=N_2=N_3=N$.

Notice that while the expression of $\bar{\lambda}_{d+1}^*$ depends on the specific gauge chosen  for  the parameters  $\bar{\lambda}_{a,d+1}$, the critical values  $\epsilon_i^*=0$ and the central charge (\ref{central_dP3}) do not rely on this.

The second  special value of fluxes that we found is $N_1 = N_3 = N_5$, which implies $N_a = N_{a+2}$, so that the fluxes have again an enhanced symmetry:
 \bea
 M_A = (N_1,N_2,N_1,N_2,N_1,N_2, -3 N_1 - 3 N_2)\, .
\eea
In this case, notice that
 the two expressions in (\ref{lineartermsinV3}) coincide, so that it is possible that both linear terms vanish, for a particular value of $\bar\lambda_{d+1}^*$, despite $N_2\neq N_5$. However,
 the parameters in (\ref{dP3genlambdas}) do not enjoy this new symmetry, so it is better to look for  a different gauge, where the parameters respect the additional symmetry, namely
 $\bar{\lambda}_{a,d+1} = \bar{\lambda}_{a+2,d+1}$.
 This can be achieved choosing
\begin{equation}
	\bar{\lambda}_{1,d+1} = -\bar{\nu}_{M5} \frac{N_1 + 2N_2}{3} \,,  \qquad  \bar{\lambda}_{2,d+1} = -\bar{\nu}_{M5} \frac{2N_1 + N_2}{3} \,,
\end{equation}
and cyclic permutations. In this gauge, we can now check that
$\evol^{(3)}$ has no linear terms in $\epsilon_1$ and $\epsilon_2$.
Therefore, extremizing $\evol^{(3)}$
  with respect to $\bar{\lambda}_{d+1}$, $\epsilon_1$ and $\epsilon_2$, we obtain the critical  values $\epsilon_{1,2}^*=0$ and
\begin{equation}
	\bar{\lambda}_{d+1}^* = \frac{6 (N_1 + N_2) - \sqrt{6 (5N_1^2 + 8N_1 N_2 + 5N_2^2)}}{6} \, \bar{\nu}_{M5} \epsilon_3 \,,
\end{equation}
and the corresponding value of the on-shell central charge is
\begin{equation}
	\mathbb{V}^{(3)} = \frac{\bar{\nu}_{M5}^3 \epsilon_3^3}{4} \Bigl[ \frac{\bigl( 6 (5N_1^2 + 8N_1 N_2 + 5N_2^2) \bigr)^{3/2}}{27} - 2(N_1 + N_2) \bigl( 3N_1^2 + 4N_1 N_2 + 3N_2^2 \bigr) \Bigr] \,.
\end{equation}
It can be checked that this expression agrees precisely with  the central charge given in eq.~(3.79) of \cite{BenettiGenolini:2023ndb} and  it correctly reduces to (\ref{freeKE}) upon setting $N_1=N_2=N$.

It would be interesting to construct explicit supergravity solutions corresponding to the two special  configurations of fluxes we found. If they exist, they should lie outside the KE class considered in  \cite{Gauntlett:2004zh}.

\paragraph*{$\Morb_4= S^2 \ltimes \spindle$.}

We now consider the toric orbifold $\mathbb{M}_4 = S^2 \ltimes \spindle$, namely a spindle $\spindle =\mathbb{W}\bP^1_{[n_+,n_-]}$ fibred over a two-sphere, which is a case with only an $SU(1)\times U(1)$ symmetry.
We take the following fan
\begin{equation}
		v^1 = (n_-, 0) \,,  \qquad  v^2 = (-k, 1) \,,  \qquad  v^3 = (-n_+, 0) \,,  \qquad  v^4 = (0, -1) \, ,
\end{equation}
and refer to \cite{Faedo:2022rqx}  for more details about this orbifold.
The total fan is as in \eqref{halffan} and  the constraint \eqref{fluxc} is solved by
\begin{equation}
	M_a = \left( \frac{N_1}{n_-}, N_2, \frac{N_1 - k \, N_2}{n_+}, N_2 \right) \,,  \qquad  M_{d+1} = -\sum_a M_a \, ,
\end{equation}
where $N_1,N_2$ parameterize the two independent fluxes, and notice that $N_2=N_4$ is implied by the $SU(2)$ symmetry acting on the base $S^2$.
The constraint  (\ref{constronlabar}) on 	$\bar{\lambda}_{a,d+1}$ is solved by taking
\begin{equation} \label{S2spindleconst}
	\bar{\lambda}_{3,d+1} = -n_+ \biggl( \bar{\nu}_{M5} N_2 + \frac{\bar{\lambda}_{1,d+1}}{n_-} \biggr) \,,  \quad  \bar{\lambda}_{4,d+1} = -\biggl( \bar{\nu}_{M5} N_1 + \frac{k \, \bar{\lambda}_{1,d+1}}{n_-} + \bar{\lambda}_{2,d+1} \biggr) \,,
\end{equation}
and we can choose a gauge in which
\begin{equation}
\label{S2spindlegauge}	\bar{\lambda}_{2,d+1} = -\frac12 \biggl( \bar{\nu}_{M5} N_1 + \frac{k \, \bar{\lambda}_{1,d+1}}{n_-} \biggr) \,.
\end{equation}
After using  the remaining  gauge freedom to fix $\bar{\lambda}_{1,d+1}$,
we are then left to extremize $\evol^{(3)}$ with respect to $\epsilon_1,\epsilon_2,\bar\lambda_{d+1}$.
One can show that the combination
\begin{equation}
k \frac{\partial \evol^{(3)}}{\partial \epsilon_1} - 2 \frac{\partial \evol^{(3)}}{\partial \epsilon_2} = 0
\end{equation}
implies $\epsilon_2^*=0$, as expected from the $SU(2)$ symmetry, while generically
$\epsilon_1^*\neq0$. In particular, $\epsilon_1^*$ is determined solving a quartic equation, which takes about half a page to be written, so we will refrain from reporting this. The on-shell central charge
can then written in terms of the parameters $N_1,N_2,k,n_+,n_-$ and  $\epsilon_1^*$.  For simplicity we shall present the results in three  special cases, where the equations are qualitatively unchanged, but simpler to write.

Firstly, let us set $k=0$. This leads to the direct product $\Morb_4= S^2 \times \spindle$   and in this case, defining
\begin{equation}
	\chi \equiv \frac{n_+ + n_-}{n_+ n_-} \,,  \qquad  \mu \equiv \frac{n_+ - n_-}{n_+ + n_-} \,,
\end{equation}
it is convenient to use the remaining gauge freedom to set
\begin{equation}
	\bar{\lambda}_{1,d+1} = -\bar{\nu}_{M5} \frac{2(1-\mu) N_2 - \mu \chi N_1}{2 (1+\mu) \chi}\,.
\end{equation}
Upon extremizing  we find that indeed
 $\epsilon_2^*=0$ and
\begin{equation}
	\bar{\lambda}_{d+1}^* = \frac{\bar{\nu}_{M5} \epsilon_3}{4\chi} \bigl[ 2(\chi N_1 + 2 N_2) \pm \mathtt{s}_1^{1/2} \bigr] \,,
\end{equation}
where we defined the quantity
\begin{equation}
	\mathtt{s}_1 = N_1^2 \chi^2 (\mu \chi \hat{\epsilon}_1^* - 2)^2 - 2N_2 (\chi N_1 + 2N_2) \bigl((1-\mu) \chi \hat{\epsilon}_1^* + 2 \bigr) \bigl( (1+\mu) \chi \hat{\epsilon}_1^* - 2 \bigr) \,.
\end{equation}
Here, $\hat{\epsilon}_1^*$ is solution to the quartic equation
\begin{equation}
	\begin{split}
		& 3 \bigl[ N_1^2 \mu \chi^2 (2 - \mu\chi \hat{\epsilon}_1) + 2 N_2 (\chi N_1 + 2N_2) (2\mu + (1-\mu^2) \chi \hat{\epsilon}_1) \bigr] \mathtt{s}_1^{1/2} - 3N_1^3 \mu\chi^3 (2-\mu\chi \hat{\epsilon}_1)^2 \\
		& \qquad - N_2 (3\chi^2 N_1^2 + 6\chi N_1 N_2 + 8 N_2^2) \bigl(12\mu + 4(1-3\mu^2)\chi \hat{\epsilon}_1 - 3\mu (1-\mu^2) \chi^2 \hat{\epsilon}_1^2 \bigr) =0 \,, \\
	\end{split}
\end{equation}
the critical value of $\epsilon_1$ is given by $\epsilon_1^* = \hat{\epsilon}_1^* \epsilon_3$ and the on-shell central charge reads
\begin{equation}
	\begin{split}
	\mathbb{V}^{(3)} &= \frac{\bar{\nu}_{M5}^3 \epsilon_3^3}{48\chi^2} \Bigl\{ \mathtt{s}_1^{3/2} - (2-\mu\chi \hat{\epsilon}_1^*) \bigl[ N_1^3 \chi^3 (2-\mu\chi \hat{\epsilon}_1^*)^2 \\
	& + N_2 (3\chi^2 N_1^2 + 6\chi N_1 N_2 + 8N_2^2) \bigl(2 + (1-\mu) \chi \hat{\epsilon}_1^*\bigr) \bigl(2 - (1+\mu) \chi \hat{\epsilon}_1^*\bigr) \bigr] \Bigr\} \,.
	\end{split}
\end{equation}
Notice that setting $n_+=n_-=1$ in the above expressions we get $\epsilon_1^*=0$ and reproduce
 the expression (\ref{P12}) for the central charge of the $\bP^1\times \bP^1$ case.

Following the reasoning in  \cite{Bah:2019rgq}, the total space $M_6$ may be also viewed  as an $\mathbb{F}_2$ fibred over the spindle $\spindle$ and we therefore interpret the corresponding putative AdS$_5\times M_6$ solution
 as arising from a stack of M5 branes at $\CC^2/\ZZ_2$ singularity, further wrapped on the spindle $\spindle$.
It would be very interesting to reproduce the above central charge from an anomaly computation, or to construct the explicit AdS$_5\times M_6$ supergravity solution.

A second  sub-case is obtained setting $n_+=n_-=1$, with $k>0$, and corresponds to the Hirzebruch surfaces $\mathbb{F}_k$.
Using the remaining gauge freedom now we can  set
\begin{equation}
	\bar{\lambda}_{1,d+1} = -\bar{\nu}_{M5} \frac{(2-k) N_2}{4}
\end{equation}
and we find that the remaining two extremization equations are solved by
\begin{equation}
	\bar{\lambda}_{d+1}^* = \frac{\bar{\nu}_{M5} \epsilon_3}{4} \bigl[ (2N_1 + (2-k) N_2) \pm \mathtt{s}_2^{1/2} \bigr] \,,
\end{equation}
where
\begin{equation}
	\mathtt{s}_2 = 4N_1^2 - 4N_1 N_2 (\hat{\epsilon}_1^* + 1) (\hat{\epsilon}_1^* - 1+k) - N_2^2 (\hat{\epsilon}_1^* + 1) \bigl( (4-2k-k^2) \hat{\epsilon}_1^* - 4+2k-k^2 \bigr)
\end{equation}
and $\hat{\epsilon}_1^*$ is the solution to the quartic equation
\begin{align}
	& 12 N_1^2 (2\hat{\epsilon}_1 + k) - 6N_1 N_2 \bigl(3k \hat{\epsilon}_1^2 - 2(2-2k-k^2) \hat{\epsilon}_1 - k(1-2k) \bigr) \nonumber \\
	& \qquad - N_2^2 \bigl( 3k (2-3k-k^2) \hat{\epsilon}_1^2 - 2(8-6k+3k^2+3k^3) \hat{\epsilon}_1 - k (2-3k+3k^2) \bigr) \nonumber \\
	& \qquad - 3 \bigl[ 2 N_1 (2\hat{\epsilon}_1 + k) + N_2 \bigl( (4-2k-k^2) \hat{\epsilon}_1 - k^2 \bigr) \bigr] \mathtt{s}_2^{1/2} = 0 \,.
\end{align}
The critical value of $\epsilon_1$ is again given by $\epsilon_1^* = \hat{\epsilon}_1^* \epsilon_3$ and the on-shell central charge reads
\begin{align}
	\mathbb{V}^{(3)} &= \frac{\bar{\nu}_{M5}^3 \epsilon_3^3}{24} \Bigl\{ \mathtt{s}_2^{3/2} - 8N_1^3 + 12N_1^2 N_2 (\hat{\epsilon}_1^* + 1) (\hat{\epsilon}_1^* - 1 + k) \\
	& - 6 N_1 N_2^2 (\hat{\epsilon}_1^* + 1) \bigl(k \hat{\epsilon}_1^*{}^2 - (2-k-k^2) \hat{\epsilon}_1^* + 2 - 2k - k^2\bigr) \nonumber \\
	& - N_2^3 (\hat{\epsilon}_1^* + 1) \bigl( k (2-3k-k^2) \hat{\epsilon}_1^*{}^2 - 2(4-2k+k^3) \hat{\epsilon}_1^* + (8-6k+3k^2-k^3) \bigr) \Bigr\} \,. \nonumber
\end{align}
Again, setting $k=0$ in the above expressions we get $\epsilon_1^*=0$ and reproduce the expression (\ref{P12}) for the central charge of the $\bP^1\times \bP^1$ case.

This result is manifestly not in agreement with  the central charge given in eq.~(3.79) of \cite{BenettiGenolini:2023ndb}, where by construction $\epsilon_1^*=\epsilon_2^*=0$.
In fact, we can reproduce this result if we impose by hand that $\epsilon_1^*=\epsilon_2^*=0$ so that
\begin{equation}
	\evol^{(3)}(\bar{\lambda}_{d+1}) = \frac{\bar{\lambda}_{d+1}}{6} \bigl[ 8\bar{\lambda}_{d+1}^2 - 6(2N_1 + (2-k) N_2) \bar{\nu}_{M5} \epsilon_3 \bar{\lambda}_{d+1} + 3N_2(2N_1 - k \, N_2) \bar{\nu}_{M5}^2 \epsilon_3^2 \bigr]
\end{equation}
and then extremizing this with respect to the remaining parameter $\bar \lambda_{d+1}$ yields
\begin{equation}
	\bar{\lambda}_{d+1}^* = \frac{2N_1 + (2-k) N_2 -\sqrt{4N_1^2 + 4(1-k) N_1 N_2 + (4-2k+k^2) N_2^2}}{4} \, \bar{\nu}_{M5} \epsilon_3 \,,
\end{equation}
 giving the on-shell central charge
 \begin{align}
	\evol^{(3)}(\bar{\lambda}_{d+1}^*) &= \frac{\bar{\nu}_{M5}^3 \epsilon_3^3}{24} \Bigl[ \bigl( 4N_1^2 + 4(1-k) N_1 N_2 + (4-2k+k^2) N_2^2 \bigr)^{3/2} \\
		& - (2N_1 + (2-k) N_2) \bigl( 4N_1^2 + 2 (1-2k) N_1 N_2 + (4-k+k^2) N_2^2 \bigr) \Bigr] \nonumber \,,
\end{align}
coinciding with the expression given in eq.~(3.79) of \cite{BenettiGenolini:2023ndb}. This, however, does not correspond to a true extremum of  $\evol^{(3)}$and therefore it is unlikely that there exist corresponding supergravity solutions, nor dual SCFTs.

Finally, let us also present the results for the particular configuration of fluxes $N_1 = \frac{k}{2} N$, $N_2 = N$ implying that\footnote{We take $k$ to be even in this case.}
\begin{equation}
	M_a = \left( \frac{k\,N}{2n_-}, N, -\frac{k\,N}{2n_+}, N \right) \,,
\end{equation}
without any assumption on $k$ and $n_+,n_-$.
In order to simplify the expressions, we make use of the remaining gauge freedom to set
\begin{equation}
	\bar{\lambda}_{1,d+1} = -\bar{\nu}_{M5} \frac{((1-\mu) (4+k\mu\chi) - k\chi) N}{(1+\mu) \chi (4+k\mu\chi)} \,.
\end{equation}
The extremization problem is then solved by
\begin{equation}
	\bar{\lambda}_{d+1}^* = \frac{\bar{\nu}_{M5} \epsilon_3 N}{2\chi \xi} \bigl( 2\xi \pm \mathtt{s}_3^{1/2} \bigr) \,,
\end{equation}
where we defined $\xi = 4+k\mu\chi$ and
\begin{equation}
	\mathtt{s}_3 = 4\xi^2 - 4\mu \chi \xi^2 \hat{\epsilon}_1^* - \chi^2 (8(2+k\mu\chi)(1-\mu^2) - (1-\mu^2+\mu^4) k^2 \chi^2) \hat{\epsilon}_1^*{}^2 \,.
\end{equation}
Here, $\hat{\epsilon}_1^*$ is solution to the quartic equation
\begin{equation}
	\begin{split}
		& 2\xi^2 (k\chi + 6\mu \xi) + 4\chi \xi^2 (4 - 3\mu^2 \xi) \hat{\epsilon}_1 - 3\chi^2 \Bigl( \frac{\xi^2 (1-\mu^2)(k\chi + 2\mu\xi)}{2} - k^3 \chi^3 \Bigr) \hat{\epsilon}_1^2 \\
		& \qquad - 3 \bigl( 2\mu \xi^2 + \chi (8(2+k\mu\chi)(1-\mu^2) - (1-\mu^2+\mu^4) k^2 \chi^2) \hat{\epsilon}_1 \bigr) \mathtt{s}_3^{1/2} = 0 \,,
	\end{split}
\end{equation}
and the critical value of $\epsilon_1$ is $\epsilon_1^* = \hat{\epsilon}_1^* \epsilon_3$.
The central charge in terms of $\hat{\epsilon}_1^*$ is given by
\begin{equation}
	\begin{split}
		\mathbb{V}^{(3)} &= \frac{\bar{\nu}_{M5}^3 \epsilon_3^3 N^3}{24\chi^2 \xi^2} \Bigl\{ \mathtt{s}_3^{3/2} - \chi^3 \Bigl( \frac{\xi^2 (1-\mu^2)(k\chi + 2\mu\xi)}{2} - k^3 \chi^3 \Bigr) \hat{\epsilon}_1^*{}^3 \\
		& + 2\chi^2 \xi^2 (4 - 3\mu^2 \xi) \hat{\epsilon}_1^*{}^2 + 2\chi \xi^2 (k\chi + 6\mu \xi) \hat{\epsilon}_1^* - 8\xi^3 \Bigr\} \,.
	\end{split}
\end{equation}

\subsubsection{General geometries}\label{sec:M5II}

We now  discuss the general AdS$_5\times M_6$  solution with  no $\mathbb{Z}_2$ symmetry.  The fan \eqref{fanM6} corresponds now to a compact geometry.
The fan is the union of $2 d$ cones $(V^a,V^{a+1},V^{d+1})$ and $(V^a,V^{a+1},V^{d+2})$ corresponding to the fixed points of the torus action, that are specified by selecting a fixed point on $\Morb_4$ and simultaneously the North or South pole of the fibre $\bP^1$.

The equivariant volume is now given by
\be\label{volM6}
\evol = \sum_a \frac{\ex^{-\lambda_a \epsilon_1^a -\lambda_{a+1} \epsilon_2^a -\lambda_{d+1} (\epsilon_3-\epsilon^a_1-\epsilon^a_2) +\ldots}}{d_{a,a+1}\epsilon^a_1 \epsilon^a_2(\epsilon_3-\epsilon^a_1-\epsilon^a_2)} -  \sum_a \frac{\ex^{-\lambda_a \epsilon_1^a -\lambda_{a+1} \epsilon_2^a +\lambda_{d+2} (\epsilon_3-\epsilon^a_1-\epsilon^a_2) +\ldots}}{d_{a,a+1}\epsilon^a_1 \epsilon^a_2(\epsilon_3-\epsilon^a_1-\epsilon^a_2)}  \,,
\ee
where the dots at the exponents contain the higher times.

This expression can also be written as an integral over $\Morb_4$
\be
	\evol(\lambda_A,\lambda_{AB},\epsilon_I)=\int_{\Morb_4}\frac{\ex^{\tau_N^\bT}-\ex^{\tau_S^\bT}}{\epsilon_3+\sum_ac_1^\bT(L_a)}\:,
\ee
where we have defined the North pole equivariant form $\tau^\bT_N$ and South pole equivariant form $\tau^\bT_S$ as
\bea
	&\tau_N^\bT=\sum_A\lambda_A\:\cC_A^N+\sum_{A,B}\lambda_{AB}\:\cC_A^N\:\cC_B^N\:,\\
	&\tau_S^\bT=\sum_A\lambda_A\:\cC_A^S+\sum_{A,B}\lambda_{AB}\:\cC_A^S\:\cC_B^S\:,\\
	&\cC_a^N\,=\,\cC_a^S\,=\,c_1^\bT(L_a)\:,\qquad a=1,\ldots,d\:,\\[2mm]
	&\cC_{d+2}^S\,=\,-\,\cC_{d+1}^N\,=\,\epsilon_3+\sum_ac_1^\bT(L_a)\:,\\
	&\cC_{d+2}^N\,=\,\cC_{d+1}^S\,=0\:.
\eea
The flux equations are the following:
\be
\label{P1overM4flux}
	-\nu_{M5} M_A=\partial_{\lambda_A}\evol^{(2)}=\int_{\Morb_4}\frac{\cC_A^N\,\tau_N^\bT-\cC_A^S\,\tau_S^\bT}{\epsilon_3+\sum_ac_1^\bT(L_a)}\:.
\ee
For a generic fan, using the gauge transformations \eqref{gauge} and \eqref{gaugeCY} we can set all $\lambda_a=\lambda_{a,a}=\lambda_{a,a+1}=0$.
However, as already mentioned, for special fans, including $\bP^2$ and $\bP^1\times \bP^1$, one of the single times $\lambda_a$ remains unfixed.
As a difference with the $\bZ_2$ symmetric case,
an arbitrary $\lambda_a$ solves trivially the flux equations. Therefore we  set  $\lambda_{a,a}=\lambda_{a,a+1}=0$ and keep $\lambda_a$ with the understanding that the latter can be partially or totally gauged fixed to zero. The forms $\tau^\bT_N$ and $\tau^\bT_S$ with all variables can then be written as
\bea
	\tau_N^\bT=&\sum_a\lambda_{a}\,c_1^\bT(L_a)+\,\Big(\epsilon_3+\sum_ac_1^\bT(L_a)\Big)\Big(\wb\lambda_{d+1}+\sum_a\wb\lambda_{a,d+1}\,c_1^\bT(L_a)\Big)\:,\\
	\tau_S^\bT=&\sum_a\lambda_{a}\,c_1^\bT(L_a)\,-\Big(\epsilon_3+\sum_ac_1^\bT(L_a)\Big)\Big(\wb\lambda_{d+2}+\sum_a\wb\lambda_{a,d+2}\,c_1^\bT(L_a)\Big)\:,
\eea
where we have defined the $\wb\lambda$ variables as
\begin{equation}
	\begin{aligned}
		& \wb\lambda_{d+1}=\epsilon_3\lambda_{d+1,d+1}-\lambda_{d+1}\:, \qquad
		&& \wb\lambda_{d+2}=-\epsilon_3\lambda_{d+2,d+2}-\lambda_{d+2}\:,\\
		& \wb\lambda_{b,d+1}=\lambda_{d+1,d+1}-2\lambda_{b,d+1}\:, \qquad
		&& \wb\lambda_{b,d+2}=-\lambda_{d+2,d+2}-2\lambda_{b,d+2}\:.
	\end{aligned}
\end{equation}
Then equations (\ref{P1overM4flux}) become
\bea
	&- \nu_{M5} M_a=\sum_bD_{ab}(\bar\lambda_{b,d+1}+\bar\lambda_{b,d+2})\:,\\
	&- \nu_{M5} M_{d+1}=-\sum_{ab}D_{ab}\,\bar\lambda_{b,d+1}\:,\\
	&-\nu_{M5} M_{d+2}=\sum_{ab}D_{ab}\,\bar\lambda_{b,d+2}\:.
\eea
The expression for $\evol^{(3)}$ is
\bea \label{freeM6}
	\evol^{(3)} &= \frac16\int_{\Morb_4}\Big(\epsilon_3+\sum_ac_1^\bT(L_a)\Big)^2
		\Big[ \bigl(\Lambda^N\bigr)^3 + \bigl(\Lambda^S\bigr)^3 \Big] \\
		& + \frac12\int_{\Morb_4}\Big(\sum_a\lambda_{a}\,c_1^\bT(L_a)\Bigr)\Big(\epsilon_3+\sum_ac_1^\bT(L_a)\Big)
		\Big[ \bigl(\Lambda^N\bigr)^2 - \bigl(\Lambda^S\bigr)^2 \Big] \\
		& + \frac12\int_{\Morb_4}\Big (\sum_a\lambda_{a}\,c_1^\bT(L_a)\Big)^2
		\Big[ \Lambda^N + \Lambda^S \Big] \:,
\eea
where we defined
\begin{equation}
	\Lambda^N = \wb\lambda_{d+1}+\sum_a\wb\lambda_{a,d+1}\,c_1^\bT(L_a) \:,  \qquad
	\Lambda^S = \wb\lambda_{d+2}+\sum_a\wb\lambda_{a,d+2}\,c_1^\bT(L_a) \:.
\end{equation}

The  flux constraints are not enough to fix all the $\wb\lambda$, so the idea is again to extremize $\evol^{(3)}$ with respect to the remaining variables.
It is convenient to define $\wb\lambda_{b,+}$ and $\wb\lambda_{b,-}$ as
\be
	\wb\lambda_{b,\pm}=\wb\lambda_{b,d+1}\pm\wb\lambda_{b,d+2}\:,
\ee
so that all the $\wb\lambda_{b,+}$ are fixed (up to gauge transformations) by
\be
	- \nu_{M5} M_a=\sum_bD_{ab}\,\wb\lambda_{b,+}\:,
\ee
whereas the $\wb\lambda_{b,-}$ are only subject to the following constraint:
\be
\label{lambdaminusconstraint}
	\nu_{M5} (M_{d+1}+M_{d+2})=\,\sum_{ab}D_{ab}\,\wb\lambda_{b,-}\:.
\ee
The extremization conditions then are
\be\label{ext1}
	0=\frac{\partial\evol^{(3)}}{\partial\lambda_{a}}=\frac{\partial\evol^{(3)}}{\partial\wb\lambda_{d+1}}=\frac{\partial\evol^{(3)}}{\partial\wb\lambda_{d+2}}
		=\sum_a\rho^a\,\frac{\partial\evol^{(3)}}{\partial\wb\lambda_{a,-}}\:,\quad\forall\:\rho^a\text{ such that }\sum_{ab}D_{ab}\,\rho^b=0\:.
\ee
In general these equations do not look easy, but in the special case $M_{d+1}+M_{d+2}=0$ there is a simple solution:
we can set $\lambda_a=0$, $\wb\lambda_{d+1}=\wb\lambda_{d+2}$ and $\wb\lambda_{b,d+1}=\wb\lambda_{b,d+2}$
so that the equations with $\rho^a$ are trivially solved.
The rest of the computation reduces  to that of section \ref{genorb} for $\bZ_2$ symmetric geometries.
When $M_{d+1}+ M_{d+2}\ne0$ this simple solution is not possible because the constraint  (\ref{lambdaminusconstraint}) would not be satisfied.

\subsubsection{Examples of general geometries}

In this section we consider again the  examples based on $\mathbb{P}^2$, $\mathbb{P}^1\times \mathbb{P}^1$ and dP$_3$ and compare with the supergravity  solutions in \cite{Gauntlett:2006ai}, where the additional parameter $c$ is turned on.
Since the solutions in \cite{Gauntlett:2006ai} all correspond  to  a critical point at $\epsilon_1=\epsilon_2=0$, for simplicity in this section
we restrict again to configurations with this feature, which as we discussed requires a special choice of fluxes for the case of dP$_3$, while it is automatic for generic fluxes for $\mathbb{P}^2$ and
 $\mathbb{P}^1\times \mathbb{P}^1$.
The explicit value of the central charge has been written in \cite{Bah:2019rgq,Bah:2019vmq,BenettiGenolini:2023ndb}.

The free energy \eqref{freeM6} can be expanded in  a sum of integrals of equivariant Chern classes. Since the multiple intersections \eqref{multint} are homogeneous function of degree $p-2$ in $\epsilon_1$ and $\epsilon_2$, all the terms involving $D_{a_1,\ldots a_p}$ with $p>2$ in \eqref{freeM6} vanish
for  $\epsilon_1=\epsilon_2=0$, and the free energy simplifies to
  \bea 6 \evol^{(3)} &= \bar \lambda_{d+1}^3 \sum_{ab} D_{ab} + 3 \bar\lambda_{d+1}^2  \sum_{ab} D_{ab}(2 \epsilon_3 \bar\lambda_{a,d+1}+\lambda_a) \\&+
3  \bar \lambda_{d+1}  \sum_{ab} D_{ab}(\epsilon_3\bar\lambda_{a,d+1}+\lambda_a)(\epsilon_3 \bar\lambda_{b,d+1}+\lambda_b)  \\
&+ \bar \lambda_{d+2}^3 \sum_{ab} D_{ab} + 3  \bar\lambda_{d+2}^2  \sum_{ab} D_{ab}(2\epsilon_3\bar\lambda_{a,d+2} -\lambda_a)\\&+
3  \bar \lambda_{d+2}  \sum_{ab} D_{ab}(\epsilon_3\bar\lambda_{a,d+2}-\lambda_a)(\epsilon_3 \bar\lambda_{b,d+2}-\lambda_b)  \, .\eea
Using the flux constraints we can also write
 \bea \evol^{(3)} &= \bar \lambda_{d+1}^3 \sum_{ab} D_{ab} +  3 \bar\lambda_{d+1}^2 \epsilon_3 \nu_{M5} M_{d+1} + 3 \bar\lambda_{d+1}^2  \sum_{ab} D_{ab} \lambda_a^+ +
3  \bar \lambda_{d+1}  \sum_{ab} D_{ab}\lambda_a^+\lambda_b^+  \\
&+ \bar \lambda_{d+2}^3 \sum_{ab} D_{ab} - 3 \bar\lambda_{d+2}^2 \epsilon_3 \nu_{M5}  M_{d+2} + 3  \bar\lambda_{d+2}^2  \sum_{ab} D_{ab}\lambda_a^-+
3  \bar \lambda_{d+2}  \sum_{ab} D_{ab}\lambda_a^-\lambda_b^-  \, ,\eea
 where
 \be \lambda_a^+=  \epsilon_3 \bar\lambda_{a,d+1}+\lambda_a\, ,\qquad  \lambda_a^-=  \epsilon_3 \bar\lambda_{a,d+2}-\lambda_a \,  ,\ee
are constrained variables.

 We consider first the general case  ($\bP^2$, $\bP^1\times \bP^1$ and dP$_3$) with all fluxes associated to the fan of the KE$_4$ set equal to $N$. The fan and  intersection matrix are given in \eqref{fanKE} and \eqref{intersKE}.
 There are, in principle, $d-2$ independent fluxes on $\Morb_4$ that we can turn on but
in  the supergravity solution with KE metric they are equal and we first restrict to this case.
 The relations $\sum_A V_I^A M_A=0$
require $M_A=(N,\ldots,N,N_N,N_S)$ with $dN+N_N-N_S=0$ and we can parameterize $N_N=M- \frac d2 N$ and $N_S= M+\frac d2 N$, possibly allowing an half-integer $M$.
Given the symmetry of the problem, we take all $\lambda_a$ to be equal, and similarly for the $\bar\lambda_{a,d+1}$ and $\bar\lambda_{a,d+2}$. The condition $\sum_{abc} D_{abc} =0$ holds for these models, and therefore all linear terms in  $\epsilon_1,\epsilon_2$ in $\evol^{(3)}$ vanish, guaranteeing a critical point at $\epsilon_1=\epsilon_2=0$. The flux conditions are solved by
\be
\bar\lambda_{a,d+1}=\nu_{M5} \frac{\left ( M - \frac{ d N}{2}\right)}{d m_k}\,, \qquad \bar\lambda_{a,d+2}=- \nu_{M5} \frac{\left ( M +\frac{ d N}{2}\right)}{d m_k}\,,
\ee
where $\sum_{ab}D_{ab} =d m_k$ so that $m_k=3,2,1$ for $\bP^2$, $\bP^1\times \bP^1$ and dP$_3$, respectively.
Extremizing with respect to $\lambda_a$  and defining $\bar\lambda_{d+1}=\nu_{M5}(H+K)$ and  $\bar\lambda_{d+2}=\nu_{M5} (H-K)$ we find
\be  6 \nu_{M5}^{-3} \evol^{(3)} = 2 d m_k H^3 -\frac 32 d \epsilon_3^2 N^2 \frac{K^2}{m_k H} -6 d \epsilon_3 N H^2 +3 \epsilon_3 H (4 K M+ \frac{d}{2m_k}\epsilon_3 N^2) \, ,\ee
 which after extremization gives\footnote{Recall that to compare with section \ref{genorb} we need
to use the rescaling \eqref{half-full}.}
 \be \evol^{(3)} =\frac{d^2 \nu_{M5}^3 \epsilon_3^3 N^4}{12 m_k^2(d^2 N^2+12 M^2)^2}\left ((3 d^2 N^2-12 M^2)^{3/2}- d N(5 d^2 N^2-36 M^2)\right ) \, .\ee
An analogous formula for non-necessarily toric KE has recently appeared in \cite{BenettiGenolini:2023ndb}.

In the case $\mathbb{P}^1\times \mathbb{P}^1$ we can turn on two independent fluxes and have round metrics on the $\bP^1$s. We take the general assignment of fluxes
compatible with  $\sum_A V_I^A M_A=0$:
\bea M_A=(N_1,N_2,N_1,N_2,N_{N},N_{S})\, ,  \eea
where
$$2 N_1+2 N_2+N_{N}-N_{S}=0$$
and we can parameterize $N_N=M-N_1-N_2$ and $N_S= M+N_1+N_2$.
Using the gauge transformations \eqref{gauge} we can also reduce to the case
\bea  \lambda_{a+2}=\lambda_{a}\, ,\,
\bar\lambda_{a+2,d+1}=\bar\lambda_{a,d+1}\,, \,    \bar\lambda_{a+2,d+2}=\bar\lambda_{a,d+2} \, . \eea
Notice that, in this gauge, all the linear terms in  $\epsilon_1,\epsilon_2$ in $\evol^{(3)}$ vanishes since, as one can check,
\be \sum_{abc} D_{abc} l_a^{(1)}l_b^{(2)}l_c^{(3)} =0 \ee
provided the vectors $l_a^{(k)}$ satisfy $l_a^{(k)}=l_{a+2}^{(k)}$.
We can solve the flux constraints
\bea &2 \bar\lambda_{1,d+1}+2 \bar\lambda_{1,d+2}+\nu_{M5}N_1=0\, , \,\, &4 \bar\lambda_{1,d+1} +4 \bar\lambda_{2,d+1} +\nu_{M5}(-M +N_1+N_2)=0\, , \\
        &2 \bar\lambda_{2,d+1}+2 \bar\lambda_{2,d+2}+\nu_{M5}N_2=0    \, ,\,\, &4 \bar\lambda_{1,d+2} +4 \bar\lambda_{2,d+2} + \nu_{M5}(M +N_1+N_2)=0 \, ,\eea
         by
\bea &\bar\lambda_{2,d+1}=-\bar\lambda_{1,d+1}+\frac14\nu_{M5}(M-N_1-N_2)\, ,\\& \bar\lambda_{1,d+2}=-\bar\lambda_{1,d+1}-\nu_{M5}\frac{N_2}{2}\,, \\&  \bar\lambda_{2,d+2}=\bar\lambda_{1,d+1}+\frac14\nu_{M5}(-M-N_1+N_2)\, .\eea
Extremizing with respect to $\lambda_{1,2}$ and $\bar\lambda_{1,d+1}$ and defining $\bar\lambda_{d+1}=\nu_{M5} (H+K)$ and  $\bar\lambda_{d+2}=\nu_{M5} (H-K)$ we find
\be  6 \nu_{M5} ^{-3}\evol^{(3)} = 16 H^3 -3 \epsilon_3^2 N_1 N_2 \frac{K^2}{H} -12 \epsilon_3 (N_1+N_2) H^2 +3 \epsilon_3 H (4 K M+\epsilon_3 N_1 N_2) \, ,\ee
 which after extremization gives
 \bea  \evol^{(3)} &= \frac{\nu_{M5}^3 \epsilon_3^3 N_1^2 N_2^2 (4 N_1^2+4 N_1 N_2 +4 N_2^2 -3 M^2)^{3/2}}{6(4 N_1 N_2+3 M^2)^2} \\
 &-\frac{ \nu_{M5}^3\epsilon_3^3 N_1^2 N_2^2 (N_1+N_2) (8 N_1^2+4 N_1 N_2 +8 N_2^2 -9 M^2)}{6(4 N_1 N_2+3 M^2)^2} \, ,\eea
 reproducing  (5.7) of \cite{Bah:2019vmq}.

Finally, let us mention that in the case of dP$_3$ we can turn on four independent fluxes along the base plus one additional flux $M$, and the general extremization problem is intractable.
It is possible to solve it for the two special configurations of fluxes with enhanced symmetry discussed previously. We leave this as an instructive exercise for the reader.

 The case $\mathbb{P}^1\times \mathbb{P}^1$ has been interpreted in \cite{Bah:2019rgq,Bah:2019vmq} as a solution for M5 branes sitting at the orbifold $\bC^2/\bZ_2$ wrapped over one of the $\bP^1$. The interpretation follows by deriving the central charge from an anomaly polynomial computation. It would be very interesting to understand if our  general formula \eqref{freeM6} can be written as the  integral of the anomaly polynomial for some M5 brane theory wrapped over a two-cycle in $\Morb_4$ and give a field theory interpretation of the solution.

  \subsection{Comparison with other  approaches}\label{GGS}

It is interesting to compare with the recent approach based on Killing spinor bilinears in  M theory \cite{BenettiGenolini:2023kxp}.
The technique consists in considering a set of equivariantly closed differential forms which can be constructed from Killing spinor bilinears.
Three such forms have been explicitly constructed for AdS$_5\times M_6$ in \cite{BenettiGenolini:2023kxp} and
for AdS$_3\times M_8$ in \cite{BenettiGenolini:2023ndb,BenettiGenolini:2023yfe}. Our results in sections \ref{AdS3xM8} and \ref{sec:M5} partially overlap with those
in  \cite{BenettiGenolini:2023ndb,BenettiGenolini:2023yfe} and it is interesting to compare the two methods. We will show that they are actually equivalent, when they can be compared, although in a non-trivial way.

For both cases, AdS$_{11-k}\times M_{k}$ with $k=6,8$, the authors of \cite{BenettiGenolini:2023ndb,BenettiGenolini:2023yfe}  define an equivariant  $k$-form $\Phi$  whose higher-degree component is the warped volume of $M_k$ and the lowest component the third power of a special locally defined function $y$. Up to coefficients, the integral of $\Phi$ is the free energy, so we have
\be F= \int_{M_k} \Phi=
		\sum_\alpha\frac{y^3|_{\alpha} }{d_{\alpha}\:e^{\bT^{k/2}}|_{\alpha}}\:,
		\ee
where $\alpha$ are the fixed points of the geometry, and we recognize our expression for $\evol^{(3)}$ for M theory solutions%
\footnote{We are omitting a $(-1)^{k/2}$ sign in the expression for $\evol^{(3)}$. In this discussion we are ignoring all such overall numerical factors.}
 \be F=\evol^{(3)}=
		\sum_\alpha\frac{(\tau_{\alpha})^3 }{d_{\alpha}\:e^{\bT^{k/2}}|_{\alpha}}\:,
		\ee
upon identifying
\be y|_{\alpha}=\tau_\alpha\, . \ee

There exists also an equivariant four-form $\Phi^F$  whose higher-degree component is the M theory four-form and the lowest component the first power of the function $y$.
The flux quantization conditions give then
\bea\label{fc} M_{AB}& = \int_{M_8} \Phi^F c_1(L_A) c_1(L_B)=
		\sum_\alpha\frac{( c_1^{\mathbb{T}^4}(L_A)  c_1^{\mathbb{T}^4}(L_B) y )|_{\alpha} }{d_{\alpha}\:e^{\bT^4}|_{\alpha}}\:, \\
		M_{A}& = \int_{M_6} \Phi^F c_1(L_A) =
		\sum_\alpha\frac{( c_1^{\mathbb{T}^3}(L_A)   y )|_{\alpha} }{d_{\alpha}\:e^{\bT^3}|_{\alpha}}\:,
		\eea
for AdS$_3\times M_8$  and AdS$_5\times M_6$, respectively and it easy to see that these conditions are equivalent, up to coefficients,  to our \eqref{AdS3xM8_prescription} and \eqref{AdS5xM6_prescription} with the same identification $y|_{\alpha}=\tau_\alpha$.

Finally there exists another auxiliary form, a four-form $\Phi^{*F}$ in AdS$_3\times M_8$  and a two-form $\Phi^Y$ in AdS$_5\times M_6$, whose lowest component is the second power of the function $y$.

Consider first the AdS$_3\times M_8$ solutions with wrapped M5 branes of section  \ref{AdS3xM8}.
The vanishing of the $\Phi^{*F}$ flux along $S^4$ is used in \cite{BenettiGenolini:2023ndb,BenettiGenolini:2023yfe} to enforce a $\mathbb{Z}_2$ symmetry of the  solution by identifying $y^N|_a=-y^S|_a$, thus effectively cutting by half the number of fixed points. With the identification $y|_{a}=\tau_a$, our construction in section \ref{AdS3xM8} is then equivalent to the one in \cite{BenettiGenolini:2023ndb}.

Consider next the AdS$_5\times M_6$ solutions of section  \ref{sec:M5}. The approaches are complementary. While we consider toric orbifolds and the action of the full torus $\bT^3=U(1)^3$, the authors
of  \cite{BenettiGenolini:2023yfe} consider $\bP^1$ bundles over a smooth four-manifold $B_4$ and assume that the R-symmetry vector has no legs along $B_4$.
Let us observe that this assumption can fail in general. For a generic $B_4$ with abelian isometries there is no reason to expect that the R-symmetry does not mix with the isometries of $B_4$
and a full-fledged computation considering the torus action on $B_4$ is necessary. Also for the toric $B_4=\mathrm{dP}_3$ with a generic choice of fluxes we expect a mixing  with the isometries of $B_4$, as discussed in section \ref{genorb}.  Under this condition, the central charge given in \cite{BenettiGenolini:2023yfe} is not necessarily the extremum of the free energy.
Obviously, whenever the two approaches can be compared and the assumption in \cite{BenettiGenolini:2023yfe}  is satisfied, we find agreement.

From a technical point of view, this might be surprising. Recall indeed that the flux constraints do not completely fix the values of the times $\tau_a=y|_{a}$.
In our construction, we just extremize the free energy $F$ with respect to the remaining parameters. In \cite{BenettiGenolini:2023ndb,BenettiGenolini:2023yfe} instead, in a case-by-case analysis, the auxiliary form $\Phi^Y$ is used  to find additional conditions to fix the $y|_\alpha$. The two methods look superficially different, but we now show that they are effectively equivalent.

The extremization conditions with respect to the K\"ahler parameters that are not fixed by the flux constraints are written in \eqref{ext1}. The first three conditions
\be \frac{\partial \evol^{(3)}}{\partial \lambda_{A}} =0 \, , \qquad  A=a,d+1,d+2\, , \ee
 can also be rewritten as
\be\label{compcon} \sum_B V^B_i \frac{\partial \evol^{(3)}}{\partial \lambda_{BA}}(1+\delta_{AB}) =0 \, , \qquad i=1,2 \, , \qquad   A=a,d+1,d+2\, .\ee
Indeed
\bea
	 &\sum_B V^B_i \frac{\partial \evol}{\partial \lambda_{BA}}(1+\delta_{AB}) =
		-2\int_{M_6} \ex^{\tau^{\mathbb{T}^3}}  c_1^{\mathbb{T}^3}(L_{A}) \sum_B V^B_i c_1^{\mathbb{T}^3}(L_{B}) \\
	&\qquad= 2\epsilon_i \int_{M_6} \ex^{\tau^{\mathbb{T}^3}}  c_1^{\mathbb{T}^3}(L_{A}) =
		-2\epsilon_i \frac{\partial \evol}{\partial \lambda_{A}} \, ,
\eea
and taking the degree two component of this equation we see that all the conditions \eqref{compcon} collapse to the extremization of the free energy with respect to the parameter~$\lambda_{A}$.

The conditions \eqref{ext1} can be then written as
\be\label{ext2}
	0=\sum_b v^b_i\frac{\partial\evol^{(3)}}{\partial\lambda_{b,a}}(1+\delta_{ba})=\sum_b v^b_i\frac{\partial\evol^{(3)}}{\partial\wb\lambda_{b,d+1}}=\sum_b v^b_i\frac{\partial\evol^{(3)}}{\partial\wb\lambda_{b,d+2}}
		=\sum_a\rho^a\,\frac{\partial\evol^{(3)}}{\partial\wb\lambda_{a,-}}\:,
\ee
where $\rho^a$ is such that $\sum_{ab}D_{ab}\,\rho^b=0$.
Now, the equations
\bea (1+\delta_{bc})\frac{\partial\evol^{(3)}}{\partial\lambda_{b,c}} &= \sum_a (c_1^{\mathbb{T}^3}(L_b) c_1^{\mathbb{T}^3}(L_c))|_a\frac{  (\tau_a^N)^2-(\tau_a^S)^2}{d_{a,a+1}\epsilon^a_1 \epsilon^a_2(\epsilon_3-\epsilon^a_1-\epsilon^a_2)} \,, \\
\frac{\partial\evol^{(3)}}{\partial\lambda_{b,d+1}} &= \sum_a  c_1^{\mathbb{T}^3}(L_b)|_a\frac{ (\tau_a^N)^2}{d_{a,a+1}\epsilon^a_1 \epsilon^a_2} \,, \\
\frac{\partial\evol^{(3)}}{\partial\lambda_{b,d+2}} &= \sum_a  c_1^{\mathbb{T}^3}(L_b)|_a\frac{ (\tau_a^S)^2}{d_{a,a+1}\epsilon^a_1 \epsilon^a_2} \,,
\eea
given the identification $y|_{a}=\tau_a$ and the fact that $\Phi^Y$ has lowest component $y^2$, translate into the localization formulas for
\bea
\int_{\mathbb{P}^1_{bc}} \Phi^Y &= \int_{M_6} c_1(L_c) c_1(L_b) \Phi^Y \,, \\
\int_{D_b^N} \Phi^Y &= \int_{M_6} c_1(L_{d+1}) c_1(L_b) \Phi^Y \,, \\
\int_{D_b^S} \Phi^Y &= \int_{M_6} c_1(L_{d+2}) c_1(L_b) \Phi^Y \,,
\eea
respectively, where $\mathbb{P}^1_{bc}$ is the  fibre taken at the fixed points $D_c\cap D_b$ on the base ($b=c\pm 1$ necessarily) and the $D_b^{N,S}$ are the divisor on the base taken at the North and South pole of the fibre, respectively.  The extremization constraints are then equivalent to the following co-homological relations
\bea
0&=\sum_b v^b_i\int_{\mathbb{P}^1_{bc}} \Phi^Y=\sum_b v^b_i\int_{D_b^N} \Phi^Y=\sum_b v^b_i\int_{D_b^S} \Phi^Y\\
		&=\sum_a\rho^a\,\Big ( \int_{D_a^N} \Phi^Y - \int_{D_a^S} \Phi^Y\Big )\:,\quad\forall\:\rho^a\text{ such that }\sum_{ab}D_{ab}\,\rho^b=0\:.
\eea
The first three conditions are obvious: the cycles $\sum_b v^b_i \,\mathbb{P}^1_{bc}$, $\sum_b v^b_i \, D_b^{N,S}$ are trivial in homology. The last equation equates cycles sitting at the North and South pole. The corresponding fluxes of $\Phi^Y$  do not need to be equal   but they must be related. We know that $c_1(L_{d+2})=c_1(L_{d+1})+\sum_a c_1(L_a)$.\footnote{The $I=3$ condition of $\sum_A V^A_I c_1(L_A)=0$.} Then
\bea  \sum_a\rho^a\,\Big ( \int_{D_a^N} \Phi^Y - \int_{D_a^S} \Phi^Y\Big )=\sum_a\rho^a \int_{M_6} \Big ( c_1(L_{d+1})-c_1(L_{d+2}) \Big ) c_1(L_a) \Phi^Y \\=  -\sum_{ab}\rho^a \int_{M_6} c_1(L_b) c_1(L_a) \Phi^Y \propto \sum_{ab} \rho^a D_{ab} =0 \, .\eea
The last step follows by expanding $\Phi^Y$  in a sum of Chern classes, and by writing $\int_{M_6} c_1(L_b) c_1(L_a) \Phi^Y$ as a sum of triple intersections $D_{ABC}^{M_6}$ on $M_6$. But $D_{abc}^{M_6}=0$ and $D_{d+1,a,b}^{M_6}=D_{d+2,a,b}^{M_6}=D_{ab}$.\footnote{The triple intersections on $M_6$ are  easily computed as $D_{ABC}^{M_6}=\frac{\partial \evol^{(3)}}{\partial \lambda_A \partial\lambda_B \partial \lambda_C}$ from \eqref{volM6}.}

We see that our construction based on the equivariant volume naturally incorporates the localization approach  of \cite{BenettiGenolini:2023kxp,BenettiGenolini:2023ndb,BenettiGenolini:2023yfe}, with the advantage that all the geometrical constraints that must be imposed case-by-case in order to find the free energy in \cite{BenettiGenolini:2023kxp,BenettiGenolini:2023ndb,BenettiGenolini:2023yfe} appear naturally in our construction: they correspond to the extremization with respect to all parameters that remain after imposing the flux constraints. This avoids an analysis based on the specific topology of the background.

  \section{$\AdS_2$, $\AdS_3$ and $\AdS_4$ solutions in type II supergravities}\label{II theory}

 In this section we consider solutions in type II string theory with geometries that are fibrations over a four-dimensional orbifold $\Morb_4$. We consider the case of massive type IIA solutions with D4 brane flux, corresponding to D4 branes wrapped over $\Morb_4$ and the case of type IIB solutions with D3 brane flux. In all cases, we show  that the free energy can be obtained by extremizing the appropriate term in the equivariant volume.

  \subsection{$\AdS_2\times M_8$ solutions in massive type IIA}
\label{AdS2xM8}

In this section we turn our attention to D4 branes wrapped around a generic four-dimensional toric orbifold $\Morb_4$ \cite{Faedo:2022rqx,Couzens:2022lvg,Bomans:2023ouw}.
Specifically the brane system we study  corresponds to $\AdS_2\times M_8$ solutions in massive type IIA,
where $M_8$ is an $S^4$ fibration over $\Morb_4$. The geometry is similar to the case of M5 branes wrapped around $\Morb_4$ considered in section \ref{AdS3xM8} and we can borrow most of the computations. Here, due to the orientifold projection,\footnote{The brane system is actually D4 in the presence of D8, which generate the cosmological constant, and an orientifold plane O8 that cuts $S^4$ into half.}  the $\bZ_2$ projection used in section \ref{AdS3xM8}  is automatically implemented  and there is only one set of fixed points, at one of the poles of $S^4$. The geometry to consider is then  a CY$_4$, a $\bC^2$ fibration over $\Morb_4$ with toric fan generated by the vectors (\ref{C2overMorb_4_fan}).
As discussed in \cite{Martelli:2023oqk} and in the introduction, the prescription for D4 in massive type IIA is also similar to (\ref{AdS3xM8_prescription}), with different degrees of homogeneity:
\be
\label{AdS2xM8_prescription}
	\nu_{D4}\,(2-\delta_{AB})\,M_{AB}=-
		\frac{\partial}{\partial\lambda_{AB}}\evol^{(3)}(\lambda_A,\lambda_{AB},\epsilon_I)\,,
		\qquad F=\evol^{(5)}(\lambda_A,\lambda_{AB},\epsilon_I)\:.
\ee

The rest of the discussion is very similar to the  section \ref{AdS3xM8}.
We can write the flux equations as
\be
\label{AdS2xM8_flux_equation}
	\nu_{D4}\,M_{AB}=-\int_{\Morb_4}\frac{\cC_A\,\cC_B\big(\tau^\bT\big)^2}{2\,\cC_{d+1}\,\cC_{d+2}}=
		-\sum_a\frac{B^{(2)}_a\cdot\big(\cC_A\,\cC_B\big)|_{a}}{d_{a,a+1}\:\epsilon_1^a\:\epsilon_2^a}\:,
\ee
where the equivariant forms $\mathcal{C}_A$, $\tau^\bT$ and the $B^{(\alpha)}_a$ are defined respectively by (\ref{AdS3xM8_equivariant_forms})
and (\ref{Bdef}).
These equations are identical to the ones of section \ref{AdS3xM8}, with the only difference
being that $B^{(2)}_a$ takes the place of $B^{(1)}_a$. The solution can be read from
 (\ref{AdS3xM8_flux_equations_solution}) and (\ref{AdS3xM8_fluxes_result}):
\bea
\label{AdS2xM8_flux_equations_solution}
	&B^{(2)}_a=-\nu_{D4}\,N\:,\\
	&M_{AB}=N\sum_{c,d}\fkt_A^c\,\fkt_B^d\,D_{cd}\:,
\eea
with $\fkt_A^c$ given by (\ref{fktAc}).

Following prescription (\ref{AdS2xM8_prescription}), the solution (\ref{AdS2xM8_flux_equations_solution})
must be substituted in the expression for $\evol^{(5)}$, which depends on the $B^{(5)}_a$ as
\be
	\evol^{(5)}(\lambda_A,\lambda_{AB},\epsilon_I)=\sum_a\frac{B^{(5)}_a}{d_{a,a+1}\:\epsilon_1^a\:\epsilon_2^a}\:.
\ee
The relation between $B^{(5)}_a$ and $B^{(2)}_a$ is given in equation (\ref{Bchange}),
with the added complication that the exponents are half-integers and thus  we need to be careful about the signs:
\bea
	\,&B^{(5)}_a=\eta_a\,\frac{2^{\frac52}}{5!}\,\big|\nu_{D4}\,N\big|^{\frac52}\,
		\Big|\big(\cC_{d+1}\,\cC_{d+2}\big)|_{a}\Big|^{\frac32}\:,\\[2mm]
	\,&\eta_a=\sign\Big(\big(\cC_{d+1}\,\cC_{d+2}\big)|_{a}\Big)\cdot\sign\big(\tau_a\big)\:.
\eea
We note that the sign of $\big(\cC_{d+1}\,\cC_{d+2}\big)|_{a}$ is the same as the sign of $B^{(2)}_a$, and thus is fixed:
\be
	\sign\Big(\big(\cC_{d+1}\,\cC_{d+2}\big)|_{a}\Big)
		=\sign\left(\frac{(\tau_a)^2}{2\big(\cC_{d+1}\,\cC_{d+2}\big)|_{a}}\right)
		=\sign\big(-\nu_{D4}\,N\big)\equiv\,\sigma\:.
\ee
The sign of the $\tau_a$ however is not fixed by (\ref{AdS2xM8_flux_equations_solution}).
We can rewrite the equations $B^{(2)}_a=-\nu_{D4}\,N$ as
\be
	\tau_a=\sigma\,\eta_a\,\sqrt{-2\nu_{D4}\,N\big(\cC_{d+1}\,\cC_{d+2}\big)|_{a}}\:.
\ee
It is always possible to find $\lambda_A$ and $\lambda_{AB}$ that solve these equations, whatever the value of $\eta_a$ might be.

For the free energy we can write
\be F = \frac{2^{\frac52}}{5!} \big( -\nu_{D4} N\big)^{\frac52} \sum_a\frac{\eta_a \big( (\epsilon_3-\epsilon_4+(\fkt_a-1)\epsilon_1^a+(\fkt_{a+1}-1)\epsilon_2^a)
		(\epsilon_4-\fkt_a\epsilon_1^a-\fkt_{a+1}\epsilon_2^a)\big)^{\frac32}}{d_{a,a+1}\:\epsilon_1^a\:\epsilon_2^a}\:, \ee
thus reproducing the extremal function in \cite{Faedo:2022rqx}.\footnote{Compare formula (5.7) in \cite{Faedo:2022rqx} and set  $\varphi_1 =  \epsilon_4\, ,\varphi_2 = \epsilon_3-\epsilon_4\, , m_a\flp_1^a= \flt_a\, , m_a\flp_2^a=1- \flt_a\, ,W=0$, and set $\epsilon_3=2$ for simplicity of comparison. Our result for the free energy then matches theirs (up to an overall sign due to different conventions) upon choosing $\epsilon_3(-\nu_{D4})^{\frac52}=\frac{16\pi}{\sqrt{8-N_f}}$. Notice that in \cite{Faedo:2022rqx} the vectors $v^a$ are taken to be primitive, contrary to the conventions we are using in this paper. Our $v^a$ are their $\hat v^a$.}

The sign ambiguities  remain to be fixed by a more careful analysis.  For a convex fan, supersymmetry is preserved with a topological twist and we expect that all the $\eta_a$ have the same sign \cite{Faedo:2022rqx}. This could follow from a generalization of the following argument valid for the equivariant volume with single times only. The $\lambda_A$ determine the polytope
\be
	\cP=\{y_I\in\bR^4\:|\:y_IV_I^A\geq\lambda_A\}\:.
\ee
Naturally $\cP$ must be non-empty, so let us take $y_I\in\cP$.
If we contract the inequalities $y_IV_I^A\geq\lambda_A$ with $c_1^{\bT^4}(L_A)|_{a}$ we get
\be
	-y_I\epsilon_I\geq\tau_a\quad\forall a\in\{1,\ldots,d\}\:.
\ee
Given that $\cP$ is a resolved cone and that the equivariant volume is given by an integral over $\cP$
\be
	\evol(\lambda_A,\epsilon_I)=\int_\cP \dd^4y\:\ex^{-y_I\epsilon_I}\:,
\ee
then the exponent $-y_I\epsilon_I\geq\tau_a$ must be negative for convergence.
This implies $\tau_a\leq0$, and thus $\eta_a=-\sigma$. By choosing $\sigma=1$ we would find the result  of \cite{Faedo:2022rqx}.

The case of anti-twist requires taking  a non-convex fan for $\Morb_4$.
This can be  obtained by {\it formally} sending $v^a\rightarrow \sigma^a v^a$ everywhere, implying $\epsilon_1^a\rightarrow \sigma^a \epsilon_1^a$ and
$\epsilon_2^a\rightarrow \sigma^{a+1} \epsilon_2^a$. It was proposed in \cite{Faedo:2022rqx} that the correct assignment of signs is $\eta_a=-\sigma^{a}\sigma^{a+1}$,
and it would be interesting to understand this by a geometrical argument.

\subsection{AdS$_4\times M_6$ solutions in massive type IIA}
\label{subsec:AdS4xM6}

In this section we consider AdS$_4\times M_6$ solutions of massive type IIA supergravity,
which correspond to a system of D4 branes wrapped around a two-cycle inside a four-dimensional toric $\Morb_4$,
in the presence of D8 branes and an orientifold plane O8.
Explicit solutions of this type have been found in \cite{Passias:2018zlm}, with $M_6$ being a $\bP^1$ fibration over a four-dimensional manifold
that is either K\"ahler-Einstein or a product of Riemann surfaces, cut in half by the O8 plane.
The only toric manifolds that admit such metrics are $\bP^2$, $\bP^1\times\bP^1$ and dP$_3$: these are the cases we will be focussing on.

More precisely, we consider a half-geometry  modelled on a non-compact CY$_3$
corresponding to the canonical bundle over $\Morb_4$, with fan given by
\be
	 V^a=(v^a,1)\, ,\qquad  V^{d+1}=(0,0,1)\ \, , \qquad a=1,\ldots,d \,,
\ee
where $v^a$ are the vectors of the fan of $\Morb_4$.
This fan has the same structure as the ones in sections \ref{Z2} to \ref{nozeroeps:sec}, and for this reason the
discussion in this section will share some similarities with the former.
This half-geometry can accurately describe the solutions of \cite{Passias:2018zlm} when the parameters $\ell$ and $\sigma$ are set to zero.
We explain this point in more detail in appendix \ref{gravity_free_energy}, where we also compute the free energy of the solutions of
\cite{Passias:2018zlm} to be compared with the results of our approach.

Our prescription is the following:
\be
\label{AdS4xM6_prescription}
	\nu_{D4}\,M_A=-\frac{\partial }{\partial \lambda_A}\evol^{(3)}(\lambda_A,\lambda_{AB},\epsilon_I)\, ,
		\qquad  \Fext= \evol^{(5)}(\lambda_A,\lambda_{AB},\epsilon_I) \, ,\qquad \sum_A V_I^A M_A=0 \, .
\ee
The higher times are needed in order to find solutions to the flux constraints.
Similarly to the discussion of section \ref{sec:M5}, we will need to extremize the free energy with respect to any parameter that is not
fixed by the flux constraints.

Given the high-degree of symmetry of $\bP^2$, $\bP^1\times\bP^1$ and dP$_3$ we expect a critical point at $\epsilon_1=\epsilon_2=0$.
Indeed, it can be verified with a similar logic as equation (\ref{zero_epsilon_proof}) that the linear terms in $\epsilon_i$ in the expression of the free energy
vanish and thus $\epsilon_1=\epsilon_2=0$ is a critical point.

The flux equations are
\begin{equation}
	\begin{split}
		(I) \qquad  & \nu_{D4} M_a = -\frac12 \int_{\Morb_4} \frac{c_1^{\mathbb{T}}(L_a) \, \bigl(\tau^{\mathbb{T}}\bigr)^2}{\epsilon_3 + \sum_b c_1^{\mathbb{T}}(L_b)} \,, \\
		(II) \qquad & \nu_{D4} M_{d+1} = \frac12 \int_{\Morb_4} \bigl(\tau^{\mathbb{T}}\bigr)^2 \,,
	\end{split}
\end{equation}
where $\tau^\bT$ is defined as in (\ref{tau_for_CY3}).
For generic values of $\epsilon_i$ these equations are not independent:
since $\sum_a v^a_i M_a=0$ and $\sum_a v^a_i c_1^{\mathbb{T}}(L_a) =-\epsilon_i$, from $(I)$ we obtain
\begin{equation} \epsilon_i \int_{\Morb_4} \frac{ \bigl(\tau^{\mathbb{T}}\bigr)^2}{\epsilon_3 + \sum_b c_1^{\mathbb{T}}(L_b) } =0\, .\end{equation}
When $\epsilon_1$ and $\epsilon_2$ are not both zero this is a non-trivial relation that we can use to write
\begin{equation} \nu_{D4} \sum_a M_a = -\frac12 \int_{\Morb_4} \frac{\sum_a c_1^{\mathbb{T}}(L_a) \, \bigl(\tau^{\mathbb{T}}\bigr)^2}{\epsilon_3 + \sum_b c_1^{\mathbb{T}}(L_b)} =- \frac12 \int_{\Morb_4} \bigl(\tau^{\mathbb{T}}\bigr)^2 \,, \end{equation}
which is equation $(II)$.
Crucially,
this argument fails when $\epsilon_1=\epsilon_2=0$, which is the case we will be focussing on.
As we will see in this case equation $(II)$ becomes independent of $(I)$ and provides an additional constraint.

As already discussed in section \ref{genorb}, we have enough gauge freedom to set $\lambda_{a,a}=\lambda_{a,a+1}=0$.
For generic fans, it is also usually possible to gauge away the $\lambda_a$,
but this is not the case for the highly symmetric fans that we consider in this section.
For the $\bZ_2$ symmetric solutions studied in section \ref{genorb} it was always possible to find a critical point with $\lambda_a=0$ regardless,
as argued in appendix \ref{app:addendum}.
However the argument of appendix \ref{app:addendum} cannot be repurposed for the type IIA solutions of this section and
we are thus forced to keep the $\lambda_a$.
The equivariant form $\tau^\bT$ can then be parameterized as
\begin{equation}
	\tau^\mathbb{T} = \sum_a \lambda_a c_1^\mathbb{T}(L_a) +\Bigl( \epsilon_3 + \sum_a c_1^\mathbb{T}(L_a) \Bigr) \Bigl( \wb\lambda_{d+1} + \sum_b \wb\lambda_{b,d+1} c_1^\mathbb{T}(L_b) \Bigr) \,,
\end{equation}
where
\begin{equation}
	\wb\lambda_{d+1} = -\lambda_{d+1} + \lambda_{d+1,d+1} \epsilon_3 \,,  \qquad \wb\lambda_{a,d+1} = -2 \lambda_{a,d+1} + \lambda_{d+1,d+1} \,.
\end{equation}
Then for $\epsilon_1=\epsilon_2=0$ the flux equations become
\begin{equation}
\label{AdS4xM6_flux_2}
	\begin{split}
		(I) \qquad  & \nu_{D4} M_a = -\frac{\wb\lambda_{d+1}}{2} \Bigl( \wb\lambda_{d+1} \sum_b D_{ab} +2  \sum_b D_{ab} \Lambda_b\Bigr) \,, \\
	\sum_a\,(I)+	(II) \qquad & 0 = \sum_{ab} D_{ab} \Lambda_a\Lambda_b +2 \epsilon_3 \wb\lambda_{d+1} \sum_{ab} D_{ab} \wb\lambda_{a,d+1} \,,
	\end{split}
\end{equation}
where
\be
	\Lambda_a=\lambda_a+\epsilon_3\wb\lambda_{a,d+1}\, .
\ee
Notice that the second equation is not a consequence of the first, as we already anticipated.
The free energy restricted to $\epsilon_1=\epsilon_2=0$ is
\begin{equation} \label{AdS4xM6_free-e0}
	F=\evol^{(5)} = \epsilon_3^2\Big ( \frac{\wb\lambda_{d+1}^5}{20} \sum_{ab} D_{ab} + \frac{ \wb\lambda_{d+1}^4}{24} \sum_{ab} D_{ab} (3 \Lambda_a + \epsilon_3 \wb\lambda_{a,d+1})  +  \frac{ \wb\lambda_{d+1}^3}{12}  \sum_{ab} D_{ab} \Lambda_a\Lambda_b \Big ) \,.
\end{equation}
We can eliminate $\wb\lambda_{a,d+1}$ from the above expression by using the second flux constraint in (\ref{AdS4xM6_flux_2}) and find
\begin{equation}
	\begin{split}
	& F
 = \epsilon_3^2\Big ( \frac{\wb\lambda_{d+1}^5}{20} \sum_{ab} D_{ab} + \frac{ \wb\lambda_{d+1}^4}{8} \sum_{ab} D_{ab}  \Lambda_a   +  \frac{ \wb\lambda_{d+1}^3}{16}  \sum_{ab} D_{ab} \Lambda_a\Lambda_b \Big ) \,, \\
		  & \nu_{D4} M_a = -\frac{\wb\lambda_{d+1}}{2} \Bigl( \wb\lambda_{d+1} \sum_b D_{ab} +  2 \sum_b D_{ab} \Lambda_b\Bigr) \,.
	\end{split}
\end{equation}
The flux constraints are not sufficient to fix all parameters:
one parameter, say $\wb\lambda_{d+1}$, remains undetermined.
Our prescription is to extremize the free energy with respect to this leftover parameter.

Let us consider first the case of K\"ahler-Einstein base manifold, with all fluxes relative to two-cycles in the base equal, that is $M_a=N$.
The three cases of interest are then $\mathbb{P}^2$, $\mathbb{P}^1\times\mathbb{P}^1$ and dP$_3$.
We define the integers $M_k=\sum_{ab} D_{ab}$ and $m_k=\sum_b D_{ab}$,
which take values $M_k=(9,8,6)$ and $m_k=(3,2,1)$ for $\mathbb{P}^2$, $\mathbb{P}^1\times\mathbb{P}^1$ and dP$_3$ respectively.
Since the fluxes $M_a$ are all equal we can solve the flux equation by also setting all $\Lambda_a$ equal to each other, giving us
\begin{equation} \Lambda_a= -\frac{2 N \nu_{D4} +m_k \wb\lambda_{d+1}^2}{2 m_k \wb\lambda_{d+1}} \, .\end{equation}
The free energy as a function of $\wb\lambda_{d+1}$ is then
\begin{equation} F
 = \epsilon_3^2 M_k \Big ( \frac{\wb\lambda_{d+1}^5}{320} - \frac{ \nu_{D4} N \wb\lambda_{d+1}^3 }{16 m_k}    +  \frac{ \nu_{D4}^2N^2 \wb\lambda_{d+1}}{16 m_k^2} \Big ) \, ,\end{equation}
and extremizing it we find four solutions:
\bea\label{resI}
	F&=\pm \frac1{10} (3\sqrt{2}-4)\, \epsilon_3^2 \left ( \frac{\nu_{D4} N}{m_k}\right)^{5/2} M_k\:,\\
	F&=\pm \frac1{10}(3\sqrt{2}+4)\, \epsilon_3^2 \left (\frac{\nu_{D4} N}{m_k}\right)^{5/2} M_k \:.
\eea
The first solution, with a plus sign, reproduces the free energy of the massive type IIA supergravity solutions of \cite{Passias:2018zlm}
upon setting\footnote{Notice that the numerical values of $\nu_{D4}$ and $\epsilon_3$ here are
 different from those of the corresponding quantities in the previous section.} $\epsilon_3^2\:\nu_{D4}^{\frac52}=\frac{64\pi}{\sqrt{n_0}}$.
The  details about the computation of the free energy of the supergravity solutions are in appendix \ref{gravity_free_energy}.

Let us now consider the case of $\mathbb{P}^1\times\mathbb{P}^1$ with independent fluxes.
In this case the metric on each $\bP^1$ factor is round, but the two radii are different.
If we impose the condition $\sum_AV_I^AM_A=0$ then the fluxes can be parameterized as follows:
\begin{equation}
	M_a = (N_1, N_2, N_1, N_2) \,,  \qquad  M_{d+1} = -2(N_1 + N_2) \,.
\end{equation}
The flux constraints can then be solved by setting
\begin{equation}
	\Lambda_1 = \Lambda_3 = -\frac{\nu_{D4} N_2 + \wb\lambda_{d+1}^2}{2\wb\lambda_{d+1}} \,,  \qquad
	\Lambda_2 = \Lambda_4 = -\frac{\nu_{D4} N_1 + \wb\lambda_{d+1}^2}{2\wb\lambda_{d+1}} \,.
\end{equation}
The free energy takes the form
\begin{equation}
	F = \frac{\epsilon_3^2 \wb\lambda_{d+1}}{40} \bigl( \wb\lambda_{d+1}^4 - 5\nu_{D4} (N_1+N_2) \wb\lambda_{d+1}^2 + 5\nu_{D4}^2 N_1 N_2 \bigr) \,,
\end{equation}
and extremizing it with respect to~$\wb\lambda_{d+1}$ yields four solutions:
\begin{equation}\label{resII}
	\begin{split}
		F &= \pm\frac{\epsilon_3^2}{10} \bigl( \sqrt{8 + \mathtt{z}^2} - (2 + \mathtt{z}^2) \bigr) \sqrt{3 - \sqrt{8 + \mathtt{z}^2}} \, (\nu_{D4} N)^{5/2} \,, \\
		F &= \pm\frac{\epsilon_3^2}{10} \bigl( \sqrt{8 + \mathtt{z}^2} + (2 + \mathtt{z}^2) \bigr) \sqrt{3 + \sqrt{8 + \mathtt{z}^2}} \, (\nu_{D4} N)^{5/2} \,,
	\end{split}
\end{equation}
where for convenience we have introduced the parameterization
\be N_1 = (1+\mathtt{z})N, \qquad N_2 = (1-\mathtt{z})N,\qquad |\mathtt{z}| < 1\:.\ee
Once again the first solution, with a plus sign, reproduces the free energy of the supergravity solutions of \cite{Passias:2018zlm}
upon setting $\epsilon_3^2\:\nu_{D4}^{\frac52}=\frac{64\pi}{\sqrt{n_0}}$
(see appendix \ref{gravity_free_energy} for details).

\subsection{AdS$_3\times M_7$ solutions in type IIB}\label{sec:D3}

In this section we consider AdS$_3\times M_7$ solutions in type IIB, where $M_7$ is an $S^3/\bZ_p$ fibration over $B_4$,
which could potentially arise as the near-horizon limit of a system of D3 branes wrapped on a two-cycle in $B_4$.
Explicit solutions of this type have been found in \cite{Gauntlett:2006af,Gauntlett:2006ns} for K\"ahler-Einstein $B_4$
or products of K\"ahler-Einstein spaces.
The case of \emph{smooth} K\"ahler $B_4$ has been studied in \cite{Gauntlett:2019pqg} using the formalism of GK geometry and the GMS construction \cite{Gauntlett:2018dpc}.
The orbifold case has not been considered in the literature as of yet, so in this section we take $B_4$ to be a generic toric orbifold $B_4\equiv\Morb_4$ and we also allow a general dependence on all the equivariant parameters, including those on the base $\Morb_4$. As we already discussed, this is important to obtain the correct critical point for generic $\Morb_4$ without particular symmetries, even in the smooth case.
We also hope that our general formulas in terms of four-dimensional integrals will be useful to find a field theory interpretation of these solutions.

With odd-dimensional  $M_7$ we need to add one real dimension, the radial one, as familiar in holography. The relevant CY$_4$ geometry is given by
the fibration with $\Morb_4$ as the base and the K\"ahler cone over the Lens space as the fibre,
that is $\bC^2/\bZ_p\hookrightarrow\text{CY}_4\to\Morb_4$.
This CY$_4$ is toric and its fan is generated by the vectors
 \be
	V^a=(v^a,1,\fkt_a)\, ,\qquad V^{d+1} =(0,0,1,0)\, ,\qquad V^{d+2}=(0,0,1,p)\, ,
\ee
where as usual the vectors $v^a$ generate the fan of $\Morb_4$, $a=1,\dots,d$.

Our  prescription here  reduces to the GMS construction \cite{Gauntlett:2019pqg}, namely
\be
\label{AdS3xM7_prescription}
 \Fext= \evol^{(2)}(\lambda_A,\epsilon_I)\, ,  \qquad  \evol^{(1)} (\lambda_A,\epsilon_I)  = 0 \, , \qquad 	\nu_{D3}\,M_A=-\frac{\partial }{\partial \lambda_A}\evol^{(2)}(\lambda_A,\epsilon_I)\, ,
		\ee
where
\be \label{AdS3xM7_fluxes}
\sum_A V_I^A M_A =0 \, .
\ee

Here $\evol^{(2)}$ matches the ``supersymmetric action" introduced in \cite{Gauntlett:2018dpc} and  we know from \cite{Gauntlett:2018dpc} that there is no need to use higher times for these solutions. Notice that the second equation in \eqref{AdS3xM7_prescription}, which is consequence of the third and \eqref{AdS3xM7_fluxes},
 is the ``topological constraint'' in \cite{Gauntlett:2018dpc}.

When $p=1$ the CY$_4$ matches exactly the one of sections \ref{AdS3xM8} and \ref{AdS2xM8}.
The equivariant volume is computed in the same manner, with only minor corrections.
The one-to-one correspondence between the fixed points of CY$_4$ and $\Morb_4$ given by $(V^a,V^{a+1},V^{d+1},V^{d+2})\leftrightarrow(v^a,v^{a+1})$
still holds, but the orders of the orbifold singularities now differ by a factor of $p$:
\be
	d_{a,a+1,d+1,d+2}=p\,d_{a,a+1}\:.
\ee
The inward normals to the faces of the cone generated by $(V^a,V^{a+1},V^{d+1},V^{d+2})$ are now given by
 \bea
	&U^a=(p\,u^{a}_{1},0,0)\, ,\\
	&U^{a+1}=(p\,u^{a}_{2},0,0)\,,\\
	&U^{d+1}=\big((\fkt_a-p) u^{a}_{1}+(\fkt_{a+1}-p) u^{a}_{2}\,, p\,d_{a,a+1}, -d_{a,a+1}\big)\,,\\
	&U^{d+2}=(-\fkt_a  u^{a}_{1}-\fkt_{a+1} u_{a}^2\,,0,d_{a,a+1})\,.
\eea
From these we can derive the restriction of the equivariant Chern forms of CY$_4$ to the fixed points
\begin{equation}
	\begin{split}
		&c_1^{\bT^4}(L_b)|_{a}=-(\epsilon_1^a\,\delta_{a,b}+\epsilon_2^a\,\delta_{a+1,b})\,,\\[2mm]
		&c_1^{\bT^4}(L_{d+1})|_{a}=\frac{-(\fkt_a-p)\epsilon_1^a-(\fkt_{a+1}-p)\epsilon_2^a-p\,\epsilon_3+\epsilon_4}{p}\,,\\
		&c_1^{\bT^4}(L_{d+2})|_{a}=\frac{\fkt_a\epsilon_1^a+\fkt_{a+1}\epsilon_2^a-\epsilon_4}{p}\,,
	\end{split}
\end{equation}
and the respective restrictions to the base $\Morb_4$
\bea
\label{AdS3xM7_equivariant_forms}
	&\cC_a\,=\,c_1^\bT(L_a)\:,\qquad a=1,\ldots,d\:,\\[2mm]
	&\cC_{d+1}\,=\,\frac{-p\,\epsilon_3+\epsilon_4+\sum_a(\fkt_a-p)c_1^\bT(L_a)}{p}\:,\\
	&\cC_{d+2}\,=\,\frac{-\epsilon_4-\sum_a\fkt_ac_1^\bT(L_a)}{p}\:.
\eea
It is easy to verify that these forms satisfy $\sum_av_i^a\,c_1^\bT(L_a)=-\epsilon_i$ and $\sum_AV_I^A\,\cC_A=-\epsilon_I$.

The second degree homogeneous component of the equivariant volume
can be written as an integral on the base $\Morb_4$ as follows:
\be
	\bV^{\,(2)}(\lambda_A\,,\epsilon_I)=\int_{\bM_4}\,\frac{(\tau^\bT)^2}{2\,p\:\cC_{d+1}\:\cC_{d+2}}\:,\qquad
	\tau^\bT=\sum_A\lambda_A\,\cC_A\:.
\ee
The flux constraints then read
\be
\label{AdS3xM7_flux_equation}
	-\nu_{D3}\,M_A=\partial_{\lambda_A}\bV^{\,(2)}(\lambda_A\,,\epsilon_I)=
		\int_{\bM_4}\,\frac{\cC_A\:\tau^\bT}{p\:\cC_{d+1}\:\cC_{d+2}}=
		\sum_a\frac{B^{(1)}_a\cdot\big(\cC_A\big)|_{a}}{d_{a,a+1}\:\epsilon_1^a\:\epsilon_2^a}\:,
\ee
where $B_a^{(1)}$ is the restriction to the $a$-th fixed point of the form
\be
	B^{(1)}\equiv \frac{\tau^\bT}{p\:\cC_{d+1}\:\cC_{d+2}}\:.
\ee

The solution to equations (\ref{AdS3xM7_flux_equation}) takes the following form:
\be
\label{AdS3xM7_flux_equation_solution}
	\nu_{D3}^{-1}\,B^{(1)}_a=b(\epsilon_I)-\sum_bm_b\,c_1^\bT(L_b)|_{a}\:,
\ee
where the $m_b$ are such that $M_a=\sum_bD_{ab}\,m_b$.
Indeed, if we substitute this expression into the right-hand side of (\ref{AdS3xM7_flux_equation}) for $A\equiv b\in\{1,\ldots,d\}$ we obtain
\be
	\sum_a\frac{B^{(1)}_a\cdot\big(\cC_b\big)|_{a}}{d_{a,a+1}\:\epsilon_1^a\:\epsilon_2^a}=
		\nu_{D3}\int_{\Morb_4}\Big(b(\epsilon_I)-\sum_am_a\,c_1^\bT(L_a)\Big)\,\cC_b
		=-\nu_{D3}\sum_aD_{ab}\,m_a\:,
\ee
thus recovering
the left-hand side of (\ref{AdS3xM7_flux_equation}).
When $A=d+1,d+2$ from a similar computation we find that
\be
	M_{d+1}=\sum_a\frac{(\fkt_a-p)M_a}p\,,\qquad M_{d+2}=-\sum_a\frac{\fkt_aM_a}p\:,
\ee
which are precisely the values of $M_{d+1}$ and $M_{d+2}$ necessary to satisfy the relation expected from the fluxes,
$\sum_AV_I^AM_A=0$.

So far we have not specified the value of $b(\epsilon_I)$ in (\ref{AdS3xM7_flux_equation_solution}).
Using the gauge invariance \eqref{gaugeCY} we can fix three parameters $\lambda_{A}$.
Therefore only $d-1$ of the restrictions of $\tau^\bT$ to the fixed points are independent, which translates into a relation among the $B_a^{(1)}$
that we use to fix the value of $b(\epsilon_I)$.
This can be seen by observing that $\tau^\bT$ is an equivariant two-form and thus its integral over $\Morb_4$ vanishes, giving us
\be
	0=\int_{\Morb_4}\tau^\bT= \int_{\Morb_4} p\:\cC_{d+1}\,\cC_{d+2} B^{(1)} \, .
\ee
The value of $b(\epsilon_I)$ that satisfies this relation can then be written as
\be
	b(\epsilon_I)=\frac{\int_{\Morb_4}\cC_{d+1}\,\cC_{d+2}\sum_am_a\,c_1^\bT(L_a)}{\int_{\Morb_4}\cC_{d+1}\,\cC_{d+2}}\:.
\ee

We observe that
the reason why we had to turn on the higher times in the cases studied in the previous sections was related to the fact that $d-1$ independent
parameters were not enough to solve the flux constraints.
In the case considered in this section however the $d-1$ independent restrictions of $\tau^\bT=\sum_A\lambda_A\,\cC_A$
are sufficient and there is no necessity to include higher times.
Nonetheless, it is interesting to repeat the same computation of this section with the addition of higher times,
which we report in appendix \ref{app:triple times}.

The free energy is given by the second degree homogeneous component of the equivariant volume, which we write as
\be
	\Fext =\bV^{\,(2)}(\lambda_A\,,\epsilon_I)=\int_{\bM_4}\,\frac{(\tau^\bT)^2}{2\,p\:\cC_{d+1}\:\cC_{d+2}}=
		\sum_a\frac{B^{(2)}_a}{d_{a,a+1}\:\epsilon_1^a\:\epsilon_2^a}\:,
\ee
where the $B^{(2)}_a$ are the restrictions of the integrand to each fixed point.
The value of the K\"ahler moduli, and consequently of the $B^{(2)}_a$, is fixed by the flux constraints.
We can easily do this by employing the same strategy as formula (\ref{Bchange}) to relate the $B_a^{(2)}$ to the $B_a^{(1)}$:
\begin{align}
	&B^{(2)}_a\equiv\bigg(\frac{\big(\tau^\bT\big)^2}{2\,p\:\cC_{d+1}\:\cC_{d+2}}\bigg)\Big|_{a}=
		\frac p2\,\big(B^{(1)}_a\big)^2\,\big(\cC_{d+1}\,\cC_{d+2}\big)|_{a}\\\nn
	&=\frac{\nu_{D3}^2}{2p}\big(b(\epsilon_I)+m_a\epsilon_1^a+m_{a+1}\epsilon_2^{a}\big)^2
		\big(p\epsilon_3-\epsilon_4+(\fkt_a-p)\epsilon_1^a+(\fkt_{a+1}-p)\epsilon_2^a\big)
		\big(\epsilon_4-\fkt_a\epsilon_1^a-\fkt_{a+1}\epsilon_2^a\big).
\end{align}
We can also write the free energy as an integral over $\Morb_4$ as follows:
\be\label{freeM70}
	\Fext=\frac p2\,\nu_{D3}^2\int_{\Morb_4}\Big(b(\epsilon_I)-\sum_am_a\,c_1^\bT(L_a)\Big)^2\:\cC_{d+1}\,\cC_{d+2}\:.
\ee
Notice that the equation that fixes $b(\epsilon_I)$ can be written as
\be\label{constrM7}
	\int_{\Morb_4}\Big(b(\epsilon_I)-\sum_am_a\,c_1^\bT(L_a)\Big)\:\cC_{d+1}\,\cC_{d+2} = 0
\ee
and can be used to further rewrite the free energy as
\be\label{freeM7}
	\Fext=- \frac p2\,\nu_{D3}^2\int_{\Morb_4} \sum_am_a\,c_1^\bT(L_a)\Big(b(\epsilon_I)-\sum_am_a\,c_1^\bT(L_a)\Big)\:\cC_{d+1}\,\cC_{d+2}\:.
\ee
It would be very interesting to  understand if our  formulas  can be written as the  integral of the anomaly polynomial for some D3 brane theory wrapped over a two-cycle in $\Morb_4$ and thus providing a field theory interpretation of the solution.

\subsubsection{Examples: K\"ahler-Einstein  and  Hirzebruch surfaces}
\label{newtypeIIexamples}

We can check that our general formalism reproduces the know expressions for the toric cases $\mathbb{P}^2$, $\bP^1\times \bP^1$ and dP$_3$ with equal fluxes. The fan and  intersection matrix are given in \eqref{fanKE} and \eqref{intersKE}.
We take all the $M_a\equiv M$, $\mathfrak{t}_a\equiv \mathfrak{t}$ and $m_a\equiv m$ equal. We find $\sum_{ab}D_{ab} = d m_k$ and $M= m m_k$ with $m_k=3,2,1$ for $\bP^2$, $\bP^1\times \bP^1$ and dP$_3$, respectively. Since $\sum_{abc} D_{abc}=0$, there is no linear term in $\epsilon_{1,2}$ in $\evol^{(2)}$ which is extremized at $\epsilon_{1,2}=0$.
By expanding \eqref{constrM7} in integrals of Chern classes we find
\be
	b= \frac{m [(p \epsilon_3-\epsilon_4)\mathfrak{t} +\epsilon_4 (p - \mathfrak{t} )] }{\mathfrak{t} (p-\mathfrak{t})} \,
\ee
and
\be \label{FdP}
	\evol^{(2)} = - \nu_{D3}^2 \frac{d  M^2 [  \epsilon_3 \epsilon_4 p (p-3\flt) \flt + \epsilon_3^2 p^2 \flt^2  + \epsilon_4^2(p^2-3 p \flt + 3 \flt^2) ]}
		{2 p m_k (p -\flt) \flt } \, ,
\ee
which reproduces formula (5.6) in \cite{Gauntlett:2019pqg} with $\epsilon_4=\epsilon_3 b_2/2$.\footnote{In  \cite{Gauntlett:2019pqg} $N$ is the flux of the five-cycle fibred over $c_1/m_k$ in $\Morb_4$. To compare the formulas we need to identify $M = \frac{m_k N}{d}$ and $\mathfrak{t} = \frac{n_2}{m_k}$, which follows from (5.5). The formulas match for $\epsilon_3 \nu_{D3}=2\sqrt{6}$.}
This still needs to be extremized with respect to $b_2$.

As we already discussed in section \ref{sec:M5}, the critical point is generically at a non-zero value of $\epsilon_1$ and $\epsilon_2$, unless there is some extra symmetry in  the background and the fluxes. As an example where the critical point is not at $\epsilon_1=\epsilon_2=0$ we consider the case of the Hirzebruch surface $\Morb_4=\bF_k$ with fan
\begin{equation}
		v^1 = (1, 0) \,,  \qquad  v^2 = (-k, 1) \,,  \qquad  v^3 = (-1, 0) \,,  \qquad  v^4 = (0, -1) \,.
\end{equation}
The constraint $\sum_A V^A_I M_A=0$ leaves two independent fluxes on the base $\Morb_4$ and two fluxes associated with the fibre
\bea
&M_3 = M_1 - k \, M_2 \,,  \qquad  M_4 = M_2 \, ,\\[2mm]
&M_5=\frac{M_1(\fkt_1+\fkt_3-2p) +M_2(pk-2p +\fkt_2+\fkt_4-k \fkt_3)}{p} \, ,  \\
&M_6=-\frac{M_1(\fkt_1+\fkt_3) +M_2(\fkt_2+\fkt_4-k \fkt_3)}{p} \, .\eea
The vectors of the fan and the fluxes have a symmetry between the second and fourth entry, and therefore we expect that one of $\epsilon_i$ will be zero at the critical point.
Notice also that the physical fluxes depends only on two linear combinations of the $\fkt_a$. These are the combinations invariant under
\be\label{gg}  \fkt_a \rightarrow \fkt_a + \sum_{i=1}^2 \beta_i v_i^a \, .\ee
In the free energy \eqref{freeM7} this transformation can be reabsorbed in a redefinition of $\epsilon_4$ using $\sum_a v_i^a c_1(L_a)^\bT=-\epsilon_i$ and therefore the central charge depends only on the physical fluxes.  We also  solve $M_a=\sum_b D_{ab}m_b$, for example,  by $m_a=(0,0,M_2,M_1)$.\footnote{Any other choice would be equivalent. The equation $M_a=\sum_b D_{ab}m_b$ is invariant under  $m_a \rightarrow m_a + \sum_{i=1}^2 \gamma_i v_i^a$. This ambiguity can be reabsorbed in a shift of $b$ in \eqref{freeM70}.}
The constraint \eqref{constrM7} and the free energy \eqref{freeM7} can be expanded in a series of integral of Chern classes and expressed in terms of the intersections $D_{a_1\ldots a_k}$, which are homogeneous of order $k$ in $\epsilon_i$. We see then that $b$ and the free energy are homogeneous of degree one and two in all the $\epsilon_I$, respectively.
One can check explicitly that $F$ is extremized at $\epsilon_2=0$. The expressions are too lengthy to be reported so, for simplicity, we restrict to the case $M_2=M_1$. We also fix
$\fkt_a=(\fkt_1,\fkt_2,\fkt_1,\fkt_2)$ using \eqref{gg} for convenience. The free energy restricted to $\epsilon_2=0$ reads
\bea\label{F7} F=-\frac{\nu_{D3}^2 M_1^2 \cA}{8 p(p (\fkt_1+\fkt_2)-2 \fkt_1 \fkt_2)} \, ,\eea
where
\begin{small}
\bea &\cA = \epsilon_3^2 p^2[(k-2)\fkt_1-2 \fkt_2]^2+ 2 \epsilon_3\epsilon_4 p[(k-2)^2(p-2 \fkt_1)\fkt_1+4(p+(k-2)\fkt_1)\fkt_2-8 \fkt_2^2] \\
& +\epsilon_4^2[(k-4)^2 p^2 -4(k-3)(k-2) p \fkt_1 +4(k-2)^2 \fkt_1^2 + 4(k-6) p \fkt_2-8(k-2)\fkt_1\fkt_2 +16 \fkt_2^2] \\
&+4 \epsilon_1 k [\epsilon_3 p \fkt_1((k-1) p \fkt_1-(k-2) \fkt_1^2 - p \fkt_2) +\epsilon_4(-3(k-2)p\fkt_1^2 +2(k-2)\fkt_1^3 +p^2((k-3)\fkt_1+\fkt_2))]\\
&+4 \epsilon_1^2[ p^2 ((k^2+k-3) \fkt_1^2 +(k-4) \fkt_1 \fkt_2 -\fkt_2^2)+p \fkt_1(-(2 k^2+k-2)\fkt_1^2 +(10-3 k) \fkt_1 \fkt_2 + 4 \fkt_2^2)]\\
&+ \epsilon_1^2\fkt_1^2[k^2 \fkt_1^2 + (2k-4) \fkt_1 \fkt_2 - 4 \fkt_2^2]\, ,
\eea
\end{small}
which should be still extremized with respect to $\epsilon_1$ and $\epsilon_4$. One easily sees that the critical point is at a non-zero value of $\epsilon_1$.
This rectifies a result given in \cite{Gauntlett:2019pqg} where it was  assumed that the R-symmetry does not mix with the isometries of $\bF_k$.
The expression \eqref{F7} for $k=0$  is extremized at $\epsilon_1=0$ and it correctly reduces to the $\bP^1\times \bP^1$ result  \eqref{FdP} setting $\epsilon_1=0$ and $\fkt_1=\fkt_2$.

\section{Conclusions}

In this paper we have refined the proposal of \cite{Martelli:2023oqk}, that the geometry of an extensive class of supersymmetric solutions is captured
 by a universal quantity, depending only on the topology of the internal space and equivariant parameters associated with the expected symmetries of the solutions.
This quantity is  an extension of the equivariant volume, familiar from symplectic geometry, where we have introduced additional moduli dubbed higher times, which are
necessary to parameterize all the fluxes supported by a given topology. Although we have assumed from the outset that the spaces of interest are toric, we have indicated that
this assumption may be relaxed by considering for example ``non-convex'' geometries as well as configurations including a four-sphere, that are not toric geometries in the strict mathematical sense.
It is also possible to extend our construction to geometries with a number of expected abelian symmetries which is strictly less than half of the real dimension\footnote{The main difference is that
in these cases the localization formula involves fixed point sets that are not isolated points.}
 of the manifold/orbifold (or cone over it).
It is well known that in many situations the metric on the internal space
  (or the cone over it, in the odd-dimensional case) solving the supersymmetry equations may not be compatible with a K\"ahler or even symplectic structure. Nevertheless, the equivariant volume is a robust topological quantity, insensitive to the details of the metrics. Indeed, it may be regarded  as a gravitational analogue of anomalies in quantum field theory.
In all cases that we have analysed, we  extract an extremal function from the equivariant volume and our prescription for fixing the parameters on which it depends consists of extremizing over \emph{all} the parameters that are left undetermined by the flux quantization conditions.  This is consistent with the logic in the case of GK geometry \cite{Couzens:2018wnk} and indeed it is analogous to the paradigm of $a$-maximization in field theory \cite{Intriligator:2003jj}.  Geometrically, the existence of critical points to the various extremal functions that we proposed may be interpreted as providing necessary conditions to the existence of the corresponding supergravity solutions and indeed it would be very interesting
 to study when such conditions are also sufficient. In any case, if we assume that a solution exists, then our method calculates the relevant observables, yielding
 non-trivial predictions for the holographically dual field theories. It is worth emphasizing that in the procedure of extremization one should allow
 all the equivariant parameters not fixed by symmetries to vary,
 otherwise it is not guaranteed that the critical point found will be a true extremum of the gravitational action. We have demonstrated this point  in a number of explicit examples discussed in section
 \ref{nozeroeps:sec}
   as well as section \ref{newtypeIIexamples}.

  In this work we have focussed on setups involving internal geometries that are fibrations over four-dimensional orbifolds $\Morb_4$,
  that may be interpreted as arising from branes wrapping completely or partially
  $\Morb_4$.
  For example, the case of M5  branes completely wrapped on $\Morb_4$ yields a proof of  the gravitational block form of the trial central charge,
   conjectured in \cite{Faedo:2022rqx} (and derived in the field theory side in \cite{Martelli:2023oqk}).
   The case of M5 branes partially wrapped on a two-cycle inside $\Morb_4$ is still poorly understood from the field theory side, the best understood setup being
  the case of $\Morb_4=\Sigma_g \times S^2$, where $\Sigma_g$ is Riemann surface of genus $g$
    \cite{Bah:2019rgq}.
  The full internal space $M_6$ may then be viewed also as the fibration of the second Hirzebruch surface
   $\mathbb{F}_2\simeq S^2 \times S^2$ over the Riemann surface $\Sigma_g$, and interpreted as the backreaction of a stack of M5 branes at a (resolved)
  $\CC^2/\ZZ_2$ singularity, further wrapped on $\Sigma_g$, yielding insights about the dual four-dimensional field theories. In section \ref{nozeroeps:sec}
  we have discussed  the example of  $\Morb_4=\spindle \times S^2$, corresponding to M5 branes probing a  $\CC^2/\ZZ_2$ singularity, further wrapped on a spindle $\spindle$ and it would be interesting to confirm our predictions with a field-theoretic computation.
  It would also be nice to extend the methods of \cite{Bah:2019rgq} for computing anomalies to setups where the M5 branes wrap a two-cycle with non-trivial normal bundle in an $\Morb_4$.

In the context of type IIA supergravity, we have analysed the case of D4 branes completely wrapped on a general toric four-orbifold $\Morb_4$, proving the gravitational block form of the entropy function conjectured in \cite{Faedo:2022rqx}. It would be very interesting to reproduce this from a field theory calculation of the partition function of five-dimensional SCFTs on $S^1\times \Morb_4$, employing the method of \cite{Inglese:2023wky} for performing localization on orbifolds.
We have also analysed the case of D4 branes partially wrapped on a two-cycle inside $\Morb_4$, providing a dual field theoretic proposal for a class of solutions to massive type IIA supergravity, constructed in~\cite{Passias:2018zlm}.
Finally, we have also discussed the case of D3 branes partially wrapped on a two-cycle inside $\Morb_4$, corresponding to type IIB geometries of the form AdS$_3\times M_7$, making contact with the framework of fibred GK geometries studied in~\cite{Gauntlett:2019pqg}. In particular, we have improved some of the results previously obtained in~\cite{Gauntlett:2019pqg}, by revisiting some of the examples discussed there.
In this paper we have not discussed geometries associated to M2 and D2 branes (already briefly mentioned in  \cite{Martelli:2023oqk}), which are not naturally related to four-dimensional orbifolds $\Morb_4$, but we expect that for these  our method
will generalize straightforwardly. It would be very interesting to incorporate
 new classes of supersymmetric  geometries in our framework, such as for example AdS$_2\times M_8$ in type IIB in order to study entropy functions of AdS$_5$ black holes. It is tantalizing to speculate that our
  approach may be eventually extended to include geometries that do not necessarily contain AdS factors.

\section*{Acknowledgements}

We would like to thank Alessandro Tomasiello for useful discussions.
EC, FF and DM are supported in part by the INFN. AZ is partially supported by the INFN and the MIUR-PRIN contract 2017CC72MK003.

\appendix

\section{Fixing the K\"ahler moduli of $\AdS_5\times M_6$ solutions with $\bZ_2$ symmetry}
\label{app:addendum}

In this appendix we verify that for the Calabi-Yau geometry considered in section \ref{genorb} there is
a critical point of $\evol^{(3)}(\lambda_A,\lambda_{AB},\epsilon_I)$ with $\lambda_a=\lambda_{ab}=0$.
Even if the group of gauge transformations \eqref{gauge} has a sufficient number of parameters to potentially
gauge away all $\lambda_a$ and $\lambda_{ab}$, in orbifolds $\Morb_4$ with a small number of vectors in the fan there are often obstructions
that make this impossible.

In the following we will verify that the values of the K\"ahler moduli $\lambda_A$, $\lambda_{AB}$ given by
\be
\label{lambda_values}
	\begin{cases}
		\lambda_a=\lambda_{ab}=0\,,\\
		\wb\lambda_{a,d+1}\text{ such that }\sum_bD_{ab}\wb\lambda_{b,d+1}=-\wb\nu_{M5}\,M_a\,,\\
		\wb\lambda_{d+1}\text{ such that }\partial_{\,\wb\lambda_{d+1}}\evol^{(3)}=0\,,
	\end{cases}
\ee
are an extremum of $\evol^{(3)}$, under the constraints imposed by the flux equations
\be
	\partial_{\lambda_A}\evol^{(2)}(\lambda_A,\lambda_{AB},\epsilon_I)=-\wb\nu_{M5}\,M_A\:.
\ee
In practice, we will show that there exists a value for the Lagrange parameters $\rho_A$ such that the function
\be
	\cE=\evol^{(3)}+\sum_A\rho_A\big(\partial_{\lambda_A}\evol^{(2)}+\wb\nu_{M5}\,M_A\big)
\ee
has null derivatives with respect to $\lambda_A$, $\lambda_{AB}$. The equations that we will solve are then
\be
\label{extremization_with_Lagrange_params}
	\frac{\partial}{\partial\lambda_A}\,\cE\,(\lambda_A,\lambda_{AB},\epsilon_I,\rho_A)=0\,,
	\qquad\frac{\partial}{\partial\lambda_{AB}}\,\cE\,(\lambda_A,\lambda_{AB},\epsilon_I,\rho_A)=0\,,
\ee
where the $\lambda$ are given by (\ref{lambda_values}) while $\epsilon_1,\epsilon_2$ can take general values.
We will study the case $\epsilon_1=\epsilon_2=0$ separately.

\paragraph*{Case $(\epsilon_1,\epsilon_2)\ne(0,0)$.}
We claim that values of $\rho_A$ that solve (\ref{extremization_with_Lagrange_params}) exist and they are the solutions of the following linear system:
\bea
\label{Lagrange_param_equations}
	&\sum_A\rho_A\big(\cC_A\big)\big|_{b}=h_b\:,\qquad b=1,\ldots,d\, , \\
	&h_b\equiv-\frac12\,\Big[\Big(\epsilon_3 +\sum_a c_1^{\mathbb{T}}(L_a)\Big)^2
		\Big( \bar\lambda_{d+1} +\sum_a \bar\lambda_{a,d+1} c_1^{\mathbb{T}}(L_a)\Bigr)^2\Big]\Big|_{b}\:.
\eea
Indeed, when the above equations are satisfied we have
\bea
	&\partial_{\lambda_B}\sum_A\rho_A\big(\partial_{\lambda_A}\evol^{(2)}+\wb\nu_{M5}\,M_A\big)=
		\int_{\Morb_4}\frac{\sum_A\rho_A\,\cC_A\,\cC_B}{\epsilon_3+\sum_a c_1^{\mathbb{T}}(L_a)} \\
	&\qquad=-\frac12\int_{\Morb_4}\cC_B\Big(\epsilon_3 +\sum_a c_1^{\mathbb{T}}(L_a)\Big)
		\Big( \bar\lambda_{d+1} +\sum_b \bar\lambda_{b,d+1} c_1^{\mathbb{T}}(L_b)\Big)^2=-\,\partial_{\lambda_B}\evol^{(3)},
\eea
where we have used $\lambda_a=\lambda_{ab}=0$.
This gives us $\partial_{\lambda_A}\cE=0$.
The $\partial_{\lambda_{AB}}\cE=0$ equations can also be derived from (\ref{Lagrange_param_equations}) in a similar manner.

Let us now discuss the existence of solutions to the equations (\ref{Lagrange_param_equations}).
The restrictions of $\sum_A\rho_A\,\cC_A$ to the fixed points are not independent, they satisfy the following linear relation:
\be
	0=\int_{\Morb_4}\sum_A\rho_A\,\cC_A=\sum_b\frac{\sum_A\rho_A\big(\cC_A\big)|_{b}}{d_{b,b+1}\,\epsilon_1^b\,\epsilon_2^b}\:.
\ee
However, the
$h_b$ also satisfy the same linear relation, given that
the value of $\wb\lambda_{d+1}$ is set by the condition $\partial_{\,\wb\lambda_{d+1}}\evol^{(3)}=0$, which reads
\be
\label{h_b_constraint}
	0=\frac12\int_{\Morb_4}(\epsilon_3 +\sum_a c_1^{\mathbb{T}}(L_a))^2
		( \bar\lambda_{d+1} +\sum_a \bar\lambda_{a,d+1} c_1^{\mathbb{T}}(L_a))^2=-\sum_b\frac{h_b}{d_{b,b+1}\,\epsilon_1^b\,\epsilon_2^b}\:.
\ee
We can thus always eliminate one of the equations (\ref{Lagrange_param_equations}).
Considering that shifting $\rho_A\to\rho_A+\sum_I\alpha^IV_I^A$ with $\sum_I\alpha^I\epsilon_I=0$
leaves the left-hand side of (\ref{Lagrange_param_equations}) invariant, we can always gauge away $\rho_{d+1}$%
\footnote{We note that the $(\epsilon_1,\epsilon_2)\ne(0,0)$ hypothesis is needed to set $\rho_{d+1}=0$.}
and one of the $\rho_a$.
We are left with a system of $d-1$ equations in $d-1$ variables that generally is not singular and thus has a solution.

There is an edge case in which the system of equations must be further reduced: when there is $\wb a\in\{1,\ldots,d\}$ such that
$\epsilon_2^{\wb a}=0$ (and consequently $\epsilon_1^{\wb a-1}=0$).
Since $(\epsilon_1,\epsilon_2)\ne(0,0)$ and $\Morb_4$ is compact we must have $\epsilon_1^{\wb a}\ne0$, $\epsilon_2^{\wb a-1}\ne0$.
The $b=\wb a-1$ and $b=\wb a$ equations are ($\rho_{d+1}$ has been gauged away)
\be
	-\rho_{\wb a}\:\epsilon_2^{\wb a-1}=h_{\wb a-1}\,,\qquad-\,\rho_{\wb a}\:\epsilon_1^{\wb a}=h_{\wb a}\:.
\ee
In principle depending on the value of $h_{\wb a-1}$ and $h_{\wb a}$ the above equations might not have a solution.
However if we consider that $h_{\wb a-1}$ and $h_{\wb a}$ can only depend on $\epsilon_2^{\wb a-1}$ and $\epsilon_1^{\wb a}$ respectively,%
\footnote{By definition $h_{\wb a-1}$ and $h_{\wb a}$ are the restrictions of an equivariant form on the fixed points $\wb a-1$ and $\wb a$.}
and that in general $d_{\wb a-1,\wb a}\,\epsilon_1^{\wb a-1}=-d_{\wb a,\wb a+1}\,\epsilon_2^{\wb a}$,
then the only way for the right-hand side of (\ref{h_b_constraint}) to be finite is for
\be
	\frac{h_{\wb a-1}}{\epsilon_2^{\wb a-1}}=\frac{h_{\wb a}}{\epsilon_1^{\wb a}}\,,
\ee
which means that equations (\ref{Lagrange_param_equations}) are solvable without issue.

\paragraph*{Case $(\epsilon_1,\epsilon_2)=(0,0)$.}
It is not immediately clear whether the solutions to equations (\ref{Lagrange_param_equations}) are well-behaving in the limit $\epsilon_1,\epsilon_2\to0$.
However when $\epsilon_1$ and $\epsilon_2$ are zero the equations (\ref{extremization_with_Lagrange_params})
are quite simple and we can solve them directly.

For $\epsilon_1=\epsilon_2=0$ we have $c_1^\bT(L_a)=c_1(L_a)$ and thus
\be
	\int_{\Morb_4}c_1^\bT(L_{a_1})\ldots c_1^\bT(L_{a_k})=
	\begin{cases}
		D_{a_1a_2}\quad	&k=2\\
		0				&\text{otherwise}
	\end{cases}\:.
\ee
From this relation it easily follows that
\be
	\int_{\Morb_4}\frac{c_1^\bT(L_{a_1})\ldots c_1^\bT(L_{a_k})}{\epsilon_3 +\sum_b c_1^{\mathbb{T}}(L_b)}=
	\begin{cases}
		(\epsilon_3)^{-1}D_{a_1a_2}\quad	&k=2\\
		0							&k>2
	\end{cases}\:.
\ee
Using the above relations the extremization equations (\ref{extremization_with_Lagrange_params}) become
\be\label{limit_etremization_equations}\begin{cases}
	\partial_{\lambda_a}\text{ equation:}
		&\frac12\sum_bD_{ab}(\wb\lambda_{d+1}^2+2\,\epsilon_3\,\wb\lambda_{d+1}\wb\lambda_{b,d+1})
			+(\epsilon_3)^{-1}\sum_bD_{ab}\rho_b=0\\
	\partial_{\lambda_{ab}}\text{ equation:}
		&\frac12\,\epsilon_3\,\wb\lambda_{d+1}^2D_{ab}-\rho_{d+1}D_{ab}=0\\
	\partial_{\lambda_{a,d+1}}\text{ equation:}
		&-\sum_bD_{ab}(\epsilon_3^2\,\wb\lambda_{d+1}\wb\lambda_{b,d+1}+\epsilon_3\,\wb\lambda_{d+1}^2)
			-\sum_bD_{ab}(\rho_b-\rho_{d+1})=0\\
	\partial_{\lambda_{d+1,d+1}}\text{ equation:}
		&\begin{aligned}
			&\tfrac12\textstyle\sum_{a,b}D_{ab}(\epsilon_3^3\,\wb\lambda_{a,d+1}\wb\lambda_{b,d+1}+
				6\,\epsilon_3^2\,\wb\lambda_{d+1}\wb\lambda_{b,d+1}+3\epsilon_3\,\wb\lambda_{d+1}^2)\\
			&+\textstyle\sum_{a,b}D_{ab}(\rho_b-\rho_{d+1})=0
		\end{aligned}
\end{cases}\ee
The $\partial_{\lambda_{d+1}}$ equation was omitted because it is trivial:
$\partial_{\lambda_{d+1}}\sum_A\rho_A\big(\partial_{\lambda_A}\evol^{(2)}+\wb\nu_{M5}M_A\big)=0$
and $\partial_{\lambda_{d+1}}\evol^{(3)}=-\partial_{\,\wb\lambda_{d+1}}\evol^{(3)}=0$ because of (\ref{lambda_values}).

The solution to (\ref{limit_etremization_equations}) is%
\footnote{When we plug this solution into the left-hand side of the $\partial_{\lambda_{d+1,d+1}}$ equation we do not get
		zero straight away, but rather we get the same expression as $\partial_{\,\wb\lambda_{d+1}}\evol^{(3)}$, which is zero by (\ref{lambda_values}).}
\be
	\rho_a=-\epsilon_3^2\,\wb\lambda_{d+1}\wb\lambda_{b,d+1}-\frac12\,\epsilon_3\,\wb\lambda_{d+1}^2\:,\qquad
		\rho_{d+1}=\frac12\,\epsilon_3\,\wb\lambda_{d+1}^2\:,
\ee
and thus (\ref{lambda_values}) is the proper extremum of $\evol^{(3)}$ under the flux constraints.

\section{$\AdS_3\times M_7$ solutions with the addition of higher times}\label{app:triple times}

In this appendix we revisit the computation of section \ref{sec:D3}, now with the inclusion of second and triple times in the equivariant volume.
For the $\AdS_3\times M_7$ solutions we considered in section \ref{sec:D3} there was no need to add any higher times.
We will now show that it is still possible to perform the computation even when the equivariant volume is over-parameterized.
The extremization procedure for the parameters in excess plays a crucial role this time:
relations that where automatically verified when $\evol^{(2)}$ only included single times are now derived as extremization conditions.
This provides further evidence that extremization is the correct way to deal with any parameter $\lambda$ that is not fixed by the flux constraints.

The second degree homogeneous component of the equivariant volume with triple times  is given by
\bea
	&\bV^{\,(2)}(\lambda_A\,,\lambda_{AB}\,,\lambda_{ABC}\,,\epsilon_I)=\int_{\bM_4}\,\frac{(\tau^\bT)^2}{2\,p\:\cC_{d+1}\:\cC_{d+2}}\:,\\
	&\tau^\bT=\sum_A\lambda_A\:\cC_A+\sum_{A,B}\lambda_{AB}\:\cC_A\:\cC_B+\sum_{A,B,C}\lambda_{ABC}\:\cC_A\:\cC_B\:\cC_C\:.
\eea
We need to impose the following flux constraints:
\be
\label{triple_times_flux_equations}
	-\nu_{D3}\,M_A=\partial_{\lambda_A}\bV^{\,(2)}(\lambda_A\,,\lambda_{AB}\,,\lambda_{ABC}\,,\epsilon_I)=
		\int_{\bM_4}\,\frac{\cC_A\:\tau^\bT}{p\:\cC_{d+1}\:\cC_{d+2}}\:.
\ee
Proceeding in a similar way as we did in section \ref{sec:D3}, we will set all the $\lambda$ to zero except for $\lambda_{d+1,\,d+2}$ and $\lambda_{d+1,\,d+2,\,A}$.
This assumption is justified by the fact that in principle the group of gauge transformation for the single, double and triple times has
enough parameters to gauge away all the $\lambda$ except $\lambda_{d+1,\,d+2}$ and $\lambda_{d+1,\,d+2,\,A}$.%
\footnote{Note that $\lambda_{a,b}$, $\lambda_{a,b,d+1}$, $\lambda_{a,b,d+2}$ and $\lambda_{a,b,c}$ do not appear inside $\evol^{(2)}$
		unless $a,b,c\in\{\wb a,\wb a+1\}$ for some $\wb a\in\{1,\ldots,d\}$.}
At the end of this appendix we will quickly check that $\evol^{(2)}$ does indeed have a critical point for
$\lambda_{a,b}=\lambda_{a,b,d+1}=\lambda_{a,b,d+2}=\lambda_{a,b,c}=0$,
thus verifying the correctness of this choice of $\lambda$.
The flux constraints (\ref{triple_times_flux_equations}) now read
\be
	-\nu_{D3}\,M_a=\frac1p\int_{\bM_4}c_1^\bT(L_a)\Big(\lambda_{d+1,\,d+2}+\sum_A\lambda_{d+1,\,d+2,\,A}\:\cC_A\Big)=
		\sum_bD_{ab}\wb\lambda_b\:,
\ee
where
\be
	\wb\lambda_a=\frac{\lambda_{d+1,\,d+2,\,a}-\lambda_{d+1,\,d+1,\,d+2}}p+
		\frac{\fkt_a(\lambda_{d+1,\,d+1,\,d+2}-\lambda_{d+1,\,d+2,\,d+2})}{p^2}\:.
\ee
Up to gauge transformations, the $\wb\lambda_a$ are then fixed to be $\wb\lambda_a=-\nu_{D3}\,m_a$,
where the $m_a$ are such that $\sum_bD_{ab}m_b=M_a$.

We notice that the flux constraints did not fix all the $\lambda$, but rather there is one such parameter left:
\bea
	\bV^{\,(2)}=&\,\frac1{2p}\int_{\bM_4}\cC_{d+1}\:\cC_{d+2}\,\Big(\lambda_{d+1,\,d+2}+\sum_A\lambda_{d+1,\,d+2,\,A}\:\cC_A\Big)^2\\[2mm]
	=&\,\frac p2\,\nu_{D3}^2\int_{\bM_4}\cC_{d+1}\:\cC_{d+2}\,\Big(\wb\lambda-\sum_am_a\:c_1^\bT(L_a)\Big)^2\:,
\eea
where
\be
	\wb\lambda=\frac{\lambda_{d+1,\,d+2}-\epsilon_3\,\lambda_{d+1,\,d+1,\,d+2}}{p\,\nu_{D3}}+
		\frac{\epsilon_4\,(\lambda_{d+1,\,d+1,\,d+2}-\lambda_{d+1,\,d+2,\,d+2})}{p^2\,\nu_{D3}}\:.
\ee
Our procedure prescribes to fix the value of $\wb\lambda$ by extremizing $\bV^{\,(2)}$ with respect to it.
If we call $b(\epsilon_I)$ the extremal value of $\wb\lambda$, we
find that
\be
\label{AdS3xM7extremum}
	0=\frac{\partial}{\partial\wb\lambda}\evol^{(2)}=p\,\nu_{D3}^2\int_{\bM_4}\cC_{d+1}\:\cC_{d+2}\,\Big(b(\epsilon_I)-\sum_am_a\:c_1^\bT(L_a)\Big)\:.
\ee
Notably, the equation we obtain is the exact same as \eqref{constrM7}.
In the context of the computation without higher times, equation \eqref{constrM7} was a trivial relation,
a predictable consequence of the fact that there are only $d-1$ single times, but $d$ fixed points.
In the computation of this appendix the same relation is now derived as an extremization condition.

If we substitute (\ref{AdS3xM7extremum}) into $\evol^{(2)}$ we get
\be
\label{triple_times_result}
	\bV^{\,(2)}=-\frac p2\,\nu_{D3}^2\int_{\Morb_4}\cC_{d+1}\:\cC_{d+2}\,\Big(b(\epsilon_I)-\sum_am_a\:c_1^\bT(L_a)\Big)\sum_am_a\:c_1^\bT(L_a)\:,
\ee
which is the same as the main result of section \ref{sec:D3}.

We can quickly verify that the values of $\lambda$ that we have fixed are an extremum of $\evol^{(2)}$ by employing
the same strategy as appendix \ref{app:addendum}.
We can find the values of the Lagrange parameters $\rho_A$ such that the function
\be
	\cE(\lambda_A\,,\lambda_{AB}\,,\lambda_{ABC}\,,\epsilon_I,\rho_A)=
		\evol^{(2)}+\sum_A\rho_A\big(\partial_{\lambda_A}\evol^{(2)}+\nu_{D3}\,M_A\big)
\ee
has null derivatives with respect to $\lambda_A\,,\lambda_{AB}\,,\lambda_{ABC}$ by solving the following linear system:
\be
	\sum_A\rho_A\big(\cC_A\big)\big|_{b}=-p\,\nu_{D3}\Big[\cC_{d+1}\:\cC_{d+2}\,\Big(b(\epsilon_I)-\sum_am_a\:c_1^\bT(L_a)\Big)\Big]
		\Big|_{b}\:.
\ee
Using the same line of reasoning as in appendix \ref{app:addendum}, solutions to this system exist and thus (\ref{triple_times_result})
is the proper extremal value of $\evol^{(2)}$ (with respect to the extremization in~$\lambda$).

\section{$\AdS_4 \times M_6$ gravity solutions}
\label{gravity_free_energy}

In this appendix we study the family of $\AdS_4 \times M_6$ solutions to massive type IIA supergravity constructed in~\cite{Passias:2018zlm}. The internal space is a $\bP^1$ bundle over a four-dimensional compact manifold, $\bP^1 \hookrightarrow M_6 \rightarrow B_4$, where the base space can be either a K\"ahler-Einstein manifold $(B_4 = \KE_4)$ or the product of two Riemann surfaces $(B_4 = \Sigma_1 \times \Sigma_2)$. In the general class of solutions in~\cite{Passias:2018zlm}, the $\bP^1$ bundle is the projectivization of the canonical bundle over $B_4$, $\mathbb{P}(K\oplus \mathcal{O})$.
In what follows, we will focus on spaces with positive curvature and set to zero the constant parameter $\ell$ appearing in~\cite{Passias:2018zlm}. This last choice is motivated by our interest for systems with only D4 and D8 branes, therefore all fluxes, except for $F_{(0)}$ and $F_{(4)}$, must vanish.
In both configurations, the metric in the string frame is%
\footnote{Notice the different normalization of $\dd s_{M_6}^2$ with respect to~\cite{Passias:2018zlm}.}
\begin{equation}
	\dd s_\mathrm{s.f.}^2 = \ex^{2A} \bigl( \dd s_{\AdS_4}^2 + \dd s_{M_6}^2 \bigr) \,,
\end{equation}
where $\dd s_{\AdS_4}^2$ is the metric on $\AdS_4$ with unit radius. The details of the internal space, along with the expressions for the dilaton and the form fluxes, will be given case by case. The solutions in~\cite{Passias:2018zlm} corresponding to the geometries discussed in section \ref{subsec:AdS4xM6} are cut into half along the equator of the $\bP^1$ fibre due to the presence of an O8 plane.

\subsection{K\"ahler-Einstein base space}

We begin considering $B_4 = \KE_4$, in which case the metric on $M_6$ is given by (setting $\kappa=+1$ in~\cite{Passias:2018zlm})
\begin{equation}
	\dd s_{M_6}^2 = -\frac{q'}{4x q} \, \dd x^2 - \frac{q}{x q' - 4q} \, D\psi^2 + \frac{q'}{3q' - x q''} \, \dd s_{\KE_4}^2 \,,
\end{equation}
where
\begin{equation} \label{q(x)}
	q(x) = x^6 + \frac{\sigma}{2} x^4 + 4x^3 - \frac12 \,,
\end{equation}
with $\sigma$ a real parameter.
Here, we introduced $D\psi = \dd\psi + \rho$, where the one-form~$\rho$ is defined on $\KE_4$ and is such that $\dd_4\rho = -\mathcal{R}$, with $\mathcal{R}$ the Ricci form of $\KE_4$. The line element $\dd s_{\KE_4}^2$ is normalized such that its scalar curvature is $R_\KE=4$.
The background under exam corresponds to $\sigma>-9$, in which case the metric is smooth and well-defined given that $\psi$ is $2\pi$-periodic and $x\in[0,x_+]$, with $x_+$ the only positive root of $q(x)$. In $x=0$ the $S^1$ fibre parameterized by~$\psi$ does not shrink and here is located an O8-plane~\cite{Passias:2018zlm}.
The warp factor of the ten-dimensional metric is
\begin{equation}
	\ex^{2A} = L^2 \sqrt{\frac{x^2 q' - 4x q}{q'}} \,,
\end{equation}
with $L$ a real constant.
The dilaton reads
\begin{equation}
	\ex^{2\Phi} = \frac{72L^4}{f_0^2} \, \frac{x q'}{(3q' - x q'')^2} \biggl( \frac{x^2 q' - 4x q}{q'} \biggr)^{3/2} \,,
\end{equation}
where we find convenient to introduce the constant $f_0$ in order to parameterize the Romans mass
\begin{equation}
	F_{(0)} = \frac{f_0}{L^3} \,,
\end{equation}
and the four-form flux is given by
\begin{equation}
	F_{(4)} = -\frac{L \, f_0}{12} \biggl[ \frac{3x (x^6 - 5x^3 - \sigma x - 5)}{(1 - x^3)^2} \, \dd x \wedge D\psi \wedge \mathcal{R} + \frac{9x^5 + 5\sigma x^3 + 45x^2 + \sigma}{6(1 - x^3)} \, \mathcal{R} \wedge \mathcal{R} \biggr] \,.
\end{equation}
All the other fields, namely the two-forms $B_{(2)}$ and $F_{(2)}$, vanish.

The first step we take in the analysis is the quantization of the fluxes, which imposes
\begin{equation} \label{quantization}
	(2\pi\ell_s) F_{(0)} = n_0 \in \ZZ \,,  \qquad
	\frac{1}{(2\pi\ell_s)^3} \int_{\Sigma_4} F_{(4)} = N_{\Sigma_4} \in \ZZ
\end{equation}
for any four-cycle $\Sigma_4$ on $M_6$.
Letting $\Sigma_\alpha$ be a basis of two-cycles for $H_2(\KE_4,\ZZ)$, we take as a basis for $H_4(M_6,\ZZ)$ the set $\{C_\alpha,C_+\}$, where $C_\alpha$ are the four-cycles obtained by considering the fibration $\bP^1 \hookrightarrow C_\alpha \rightarrow \Sigma_\alpha$, and $C_+$ is a copy of the KE base space at $x=x_+$.
Performing the integrals, we obtain the fluxes
\begin{equation}
	\begin{split}
		& N_\alpha = \frac{\pi^2 L \, f_0}{6 (2\pi\ell_s)^3} \, \frac{x_+^2 (3x_+^3 + 2\sigma x_+ + 15)}{1 - x_+^3} \, m_k \, n_\alpha \,, \\
		& N_+ = -\frac{\pi^2 L \, f_0}{18 (2\pi\ell_s)^3} \, \frac{9x_+^5 + 5\sigma x_+^3 + 45x_+^2 + \sigma}{1 - x_+^3} \, M_k \,,
	\end{split}
\end{equation}
where we defined the integers
\begin{equation}
		n(\Sigma_\alpha) = \frac{1}{2\pi} \int_{\Sigma_\alpha} \mathcal{R} = m_k \, n_\alpha \,,  \qquad
		M_k = \frac{1}{4\pi^2} \int_{\KE_4} \mathcal{R} \wedge \mathcal{R} \,.
\end{equation}
$m_k$ is the Fano index of the $\KE_4$ and is the largest positive integer such that all of the $n_\alpha$ are integers. These integers take the values $m_k=(3,2,1)$ and $M_k=(9,8,6)$ for $\bP^2$, $\bP^1 \times \bP^1$ and dP$_3$, respectively.

For the rest of this subsection we will restrict to the case $\sigma=0$.
In order to understand this assumption, we first need to make contact with the equivariant volume extremization procedure.
The toric manifold $\KE_4$ is completely described by its fan $v^a$, which defines the toric divisors $\Sigma_a$ and their associated line bundles $L_a$. The set of divisors $D_a \subset M_6$ is naturally induced as the $\bP^1$ fibrations over $\Sigma_a$, to which we must add $D_{d+1}$, a copy of $\KE_4$ at the pole of the half $\bP^1$.
The corresponding integer fluxes are defined as
\begin{equation}
	M_A = \frac{1}{(2\pi\ell_s)^3} \int_{D_A} F_{(4)}
\end{equation}
and, for $a=1,\ldots,d$, they read
\begin{equation}
	M_a = \frac{\pi L \, f_0}{12 (2\pi\ell_s)^3} \, \frac{x_+^2 (3x_+^3 + 2\sigma x_+ + 15)}{1 - x_+^3} \times \int_{\Sigma_a} \mathcal{R} \,.
\end{equation}
Recalling that $\sum_a c_1(L_a) = c_1(T\KE_4) = \mathcal{R}/(2\pi)$, we obtain
\begin{equation}
	\sum_a \int_{\Sigma_a} \mathcal{R} = \sum_a \int_{\KE_4} \mathcal{R} \wedge c_1(L_a) = \frac{1}{2\pi} \int_{\KE_4} \mathcal{R} \wedge \mathcal{R} \,,
\end{equation}
which allows us to compute the sum
\begin{equation}
	\sum_a M_a = \frac{\pi^2 L \, f_0}{6 (2\pi\ell_s)^3} \, \frac{x_+^2 (3x_+^3 + 2\sigma x_+ + 15)}{1 - x_+^3} \, M_k \,.
\end{equation}
Identifying $M_{d+1}$ with $N_+$ we have
\begin{equation}
	\sum_A M_A = \sum_a M_a + N_+ = -\frac{\pi^2 L \, f_0 \, \sigma}{18 (2\pi\ell_s)^3} \, M_k \,,
\end{equation}
and consistency with the $I=3$ component of the third condition in~\eqref{AdS4xM6_prescription}, which reads $\sum_A M_A = 0$, imposes $\sigma=0$.

When $\sigma$ vanishes, the zeros of~\eqref{q(x)} can be computed analytically,
\begin{equation}
	x^3 = -2 \pm \frac{3}{\sqrt2}  \qquad  \implies  \qquad  x_+ = \biggl( \frac{3 - 2\sqrt2}{\sqrt2} \biggr)^{1/3} \,,
\end{equation}
and the fluxes simplify to
\begin{equation}
	N_\alpha = \frac{\pi^2 L \, f_0}{2 (2\pi\ell_s)^3} \, \biggl( \frac{3 + 2\sqrt2}{\sqrt2} \biggr)^{1/3} m_k \, n_\alpha \,,  \qquad
	N_+ = -\frac{\pi^2 L \, f_0}{2 (2\pi\ell_s)^3} \, \biggl( \frac{3 + 2\sqrt2}{\sqrt2} \biggr)^{1/3} M_k \,.
\end{equation}
In order for $N_\alpha$ and $N_+$ to be integers, as imposed by~\eqref{quantization}, we require
\begin{equation} \label{gravity_cond2}
	\frac{\pi^2 L \, f_0}{2 (2\pi\ell_s)^3} \, \biggl( \frac{3 + 2\sqrt2}{\sqrt2} \biggr)^{1/3} = \frac{N}{h} \,,
\end{equation}
where $N$ is an arbitrary integer and $h = \mathrm{hcf}(M_k,m_k)$. Specifically, $h=(3,2,1)$ for $\mathbb{P}^2$, $\mathbb{P}^1\times\mathbb{P}^1$ and dP$_3$, respectively.
On the other hand, the first condition of~\eqref{quantization} yields
\begin{equation} \label{gravity_cond1}
	\frac{f_0}{L^3} = \frac{n_0}{2\pi\ell_s} \,.
\end{equation}
Combining~\eqref{gravity_cond2} and~\eqref{gravity_cond1} we obtain the following quantization conditions on the parameters $L$ and $f_0$
\begin{equation}
	\begin{split}
		& L^4 = (2\pi\ell_s)^4 \, \frac{2^{4/3} (3 - 2\sqrt2)^{1/3}}{\pi^2 n_0} \biggl(\frac{N}{h}\biggr) \,, \\
		& f_0^2 = (2\pi\ell_s)^4 \, \frac{4 (3 - 2\sqrt2)^{1/2} \, n_0^{1/2}}{\pi^3} \biggl(\frac{N}{h}\biggr)^{3/2} \,.
	\end{split}
\end{equation}

The free energy of our $\AdS_4 \times M_6$ background with KE base space can be read off from the four-dimensional effective Newton constant $G_{(4)}$ as~\cite{Guarino:2015jca}
\begin{equation} \label{gravity_free-def}
	F = \frac{\pi}{2G_{(4)}} = \frac{16\pi^3}{(2\pi\ell_s)^8} \int \ex^{8A - 2\Phi} \, \vol(M_6) \,,
\end{equation}
which gives the general result
\begin{equation} \label{gravity_free-KE-int}
	F = \frac{1}{(2\pi\ell_s)^8} \, \frac{8\pi^6 L^4 f_0^2}{135} \, x_+^2 (9x_+^3 + 5\sigma x_+ + 45) \, M_k \,.
\end{equation}
In this computation we used the fact that, in our conventions, the K\"ahler form is $J_\KE = \mathcal{R}$, therefore the total volume of the $\KE_4$ can be determined from
\begin{equation}
	\Vol(\KE_4) = \frac12 \int_{\KE_4} \mathcal{R} \wedge \mathcal{R} = 2\pi^2 M_k \,.
\end{equation}
Setting $\sigma=0$ and substituting the expressions of $x_+$, $L$ and $f_0$ into~\eqref{gravity_free-KE-int}, the free energy then reads
\begin{equation} \label{gravity_free-KE}
	F = \frac{32\sqrt2 (3 - 2\sqrt2) \pi}{5n_0^{1/2}} \biggl(\frac{N}{h}\biggr)^{5/2} M_k \,,
\end{equation}
which agrees with the first equation in \eqref{resI} with a plus sign, taking into account that,  for our examples, $h=m_k$.

\subsection{$S^2 \times S^2$ base space}

We now move to the second case, $B_4 = S_1^2 \times S_2^2$, whose six-dimensional metric is (setting $\kappa_1=\kappa_2=+1$ in~\cite{Passias:2018zlm})
\begin{equation}
	\dd s_{M_6}^2 = -\frac{q'}{4x q} \, \dd x^2 - \frac{q}{x q' - 4q} \, D\psi^2 + \frac{q'}{x u_1} \, \dd s_{S_1^2}^2 + \frac{q'}{x u_2} \, \dd s_{S_2^2}^2 \,,
\end{equation}
where
\begin{equation}
	\begin{gathered}
		q(x) = x^6 + \frac{\sigma}{2} x^4 + 2(1 + t)x^3 - \frac{t}{2} \,, \\
		u_1(x) = 12x (1 - x^3) \,,  \qquad  u_2(x) = 12x (t - x^3) \,,
	\end{gathered}
\end{equation}
with $\sigma$ and $t$ real constants.
$D\psi = \dd\psi + \rho$, where $\rho$ is a one-form on $S_1^2 \times S_2^2$ such that $\dd_4\rho = -(\mathcal{R}_1 + \mathcal{R}_2)$, with $\mathcal{R}_i$ Ricci form of $S_i^2$, while each $\dd s_{S_i^2}^2$ is the metric on a two-sphere with unit radius.
The configuration of interest is realized when $t>0$ and $\sigma>-9\cdot4^{-1/3}(1+t)^{2/3}$, and in this region the metric is smooth and well-defined given that $\psi$ is $2\pi$-periodic and $x\in[0,x_+]$, with $x_+$ the only positive root of $q(x)$. Also in this case, we have an O8-plane in $x=0$.
The warp factor has the same expression as in the previous case, namely
\begin{equation}
	\ex^{2A} = L^2 \sqrt{\frac{x^2 q' - 4x q}{q'}} \,,
\end{equation}
whereas the dilaton is now given by
\begin{equation}
	\ex^{2\Phi} = \frac{72L^4}{f_0^2} \, \frac{q'}{x u_1 u_2} \biggl( \frac{x^2 q' - 4x q}{q'} \biggr)^{3/2} \,.
\end{equation}
The remaining non-vanishing fields are the Romans mass
\begin{equation}
	F_{(0)} = \frac{f_0}{L^3} \,,
\end{equation}
with $f_0\in\RR$, and the four-form flux
\begin{align}
	F_{(4)} &= -\frac{L \, f_0}{12} \biggl[ \frac{3x \bigl( x^6 -(t+4)x^3 - \sigma x - (2t + 3) \bigr)}{(1 - x^3)^2} \, \dd x \wedge D\psi \wedge \mathcal{R}_1 \nonumber \\
	& + \frac{3x \bigl( x^6 - (4t + 1)x^3 - \sigma t x - t (3t + 2) \bigr)}{(t - x^3)^2} \, \dd x \wedge D\psi \wedge \mathcal{R}_2 \\
	& - \frac{ 9x^8 + 5\sigma x^6 + 18(t+1)x^5 - 2\sigma (t+1)x^3 - 9(t^2 + 3t + 1)x^2 - \sigma t}{3(1 - x^3) (t - x^3)} \, \mathcal{R}_1 \wedge \mathcal{R}_2 \biggr] \nonumber \,.
\end{align}

In order to quantize the fluxes as in~\eqref{quantization}, we take as a basis for $H_4(M_6,\ZZ)$ the set $\{C_1,C_2,C_+\}$, where $C_i$ are the fibrations $\bP^1 \hookrightarrow C_i \rightarrow S_i^2$ (at a fixed point on the other sphere) and $C_+$ is a copy of $S_1^2 \times S_2^2$ at $x=x_+$.
The expressions of the three fluxes are
\begin{align}
	& N_1 = \frac{\pi^2 L \, f_0}{3 (2\pi\ell_s)^3} \, \frac{x_+^2 \bigl( 3x_+^3 + 2\sigma x_+ + 3(2t + 3) \bigr)}{1 - x_+^3} \,, \nonumber \\
	& N_2 = \frac{\pi^2 L \, f_0}{3 (2\pi\ell_s)^3} \, \frac{x_+^2 \bigl( 3x_+^3 + 2\sigma x_+ + 3(3t + 2) \bigr)}{t - x_+^3} \,, \\
	& N_+ = \frac{4\pi^2 L \, f_0}{9 (2\pi\ell_s)^3} \, \frac{ 9x_+^8 + 5\sigma x_+^6 + 18(t+1)x_+^5 - 2\sigma (t+1)x_+^3 - 9(t^2 + 3t + 1)x_+^2 - \sigma t}{(1 - x_+^3) (t - x_+^3)} \nonumber \,,
\end{align}
where we made use of the relation
\begin{equation}
	\frac{1}{2\pi} \int_{S_i^2} \mathcal{R}_i = \chi(S_i^2) = 2 \,.
\end{equation}

As before, we will restrict to configurations with $\sigma=0$, in which case the equation $q(x)=0$ can be solved analytically, giving
\begin{equation}
	x^3 = -(t+1) \pm \sqrt{\frac{(t+2)(2t+1)}{2}}  \quad  \implies  \quad  x_+ = \biggl( \sqrt{\frac{(t+2)(2t+1)}{2}} - (t+1) \biggr)^{1/3} \,.
\end{equation}
When $\sigma$ vanishes $N_+ = -2(N_1+N_2)$, therefore we will focus exclusively on the quantization of the fluxes $N_1$ and $N_2$, since the quantization of~$N_+$ follows immediately.
Setting $\sigma=0$, the fluxes simplify to
\begin{equation} \label{gravity_flux0}
	N_1 = \frac{\pi^2 L \, f_0}{(2\pi\ell_s)^3} \, \frac{x_+^2 (x_+^3 + 2t + 3)}{1 - x_+^3} \,,  \qquad
	N_2 = \frac{\pi^2 L \, f_0}{(2\pi\ell_s)^3} \, \frac{x_+^2 (x_+^3 + 3t + 2)}{t - x_+^3} \,,
\end{equation}
and taking their ratio we can immediately determine $t$
\begin{equation}
	t = \frac{\bigl[ \sqrt{9N_1^2 + 14N_1 N_2 + 9N_2^2} \pm 3(N_1 - N_2) \bigr]^2}{32N_1 N_2} \,.
\end{equation}
Since $t$ needs to be positive, $N_1$ and $N_2$ must have the same sign, \ie\ $N_1 N_2 > 0$; for the sake of simplicity, we will take both of them positive.
Taking the product of the fluxes~\eqref{gravity_flux0} and making use of~\eqref{gravity_cond1} we obtain
\begin{equation}
	\begin{split}
		& L^4 = (2\pi\ell_s)^4 \, \frac{1}{\pi^2 n_0} \biggl(\frac{2}{t}\biggr)^{1/2} \biggl( \sqrt{\frac{(t+2)(2t+1)}{2}} - (t+1) \biggr)^{1/3} (N_1 N_2)^{1/2} \,, \\
		& f_0^2 = (2\pi\ell_s)^4 \, \frac{n_0^{1/2}}{\pi^3}  \biggl(\frac{2}{t}\biggr)^{3/4} \biggl( \sqrt{\frac{(t+2)(2t+1)}{2}} - (t+1) \biggr)^{1/2} (N_1 N_2)^{3/4} \,.
	\end{split}
\end{equation}

The free energy of the $\AdS_4$ solution under exam can be computed performing the integral~\eqref{gravity_free-def} and takes the general expression
\begin{equation} \label{gravity_free-SS-int}
	F = \frac{1}{(2\pi\ell_s)^8} \, \frac{32\pi^6 L^4 f_0^2}{135} \, x_+^2 \bigl( 18x_+^3 + 10\sigma x_+ + 45(t+1) \bigr) \,,
\end{equation}
which, once all the ingredients are substituted, becomes
\begin{equation} \label{gravity_free-SS}
	\begin{split}
		F &= \frac{4\sqrt2 \pi}{5n_0^{1/2}} \Bigl( (N_1 + N_2) \sqrt{9N_1^2 + 14N_1 N_2 + 9N_2^2} - (3N_1^2 + 2N_1 N_2 + 3N_2^2) \Bigr) \\
		& \times \sqrt{3(N_1 + N_2) - \sqrt{9N_1^2 + 14N_1 N_2 + 9N_2^2}} \,.
	\end{split}
\end{equation}
Parameterizing the fluxes as $N_1 = (1+\mathtt{z})N$, $N_2 = (1-\mathtt{z})N$, with $|\mathtt{z}| < 1$, we obtain
\begin{equation}
	F = \frac{32 \pi}{5n_0^{1/2}} \, \bigl( \sqrt{8 + \mathtt{z}^2} - (2 + \mathtt{z}^2) \bigr) \sqrt{3 - \sqrt{8 + \mathtt{z}^2}} \, N^{5/2} \,,
\end{equation}
which agrees with  the first equation in \eqref{resII} with a plus sign.
Setting $\mathtt{z}=0$ we consistently retrieve the result~\eqref{gravity_free-KE} specified to the case $\mathbb{P}^1 \times \mathbb{P}^1$.

\bibliographystyle{JHEP}
\bibliography{biblioAdS5}

\end{document}